\documentclass[camera,letterpaper,nomarginnotes,nonarrowgutter]{jpaper}

\usepackage{fancyhdr}
\usepackage{datetime}

\usepackage{booktabs} 
\DeclareMathAlphabet{\mathcal}{OMS}{cmsy}{m}{n}
\usepackage{setspace}
\usepackage[italic]{mathastext}
\usepackage{array}
\usepackage{titlesec}
\usepackage{fancyhdr}
\usepackage[normalem]{ulem}
\usepackage{datetime} 
\usepackage{multirow}
\usepackage{multicol}
\usepackage{listings}
\usepackage{color}
\usepackage[font={small,bf}]{caption}
\usepackage{float}
\usepackage[noadjust]{cite}
\usepackage{tikz}
\usepackage{xcolor}
\usepackage[font={small,bf}]{subcaption}
\usepackage[noend]{algorithmic}
\usepackage[linesnumbered,ruled]{algorithm2e}
\usepackage{makecell}

\usepackage{pifont}
\usepackage{amssymb}
\usepackage{amsmath}

\usepackage[]{microtype}

\usepackage[utf8]{inputenc}
\usepackage[T1]{fontenc}

\usepackage{mathptmx} 

\usepackage[bookmarks=true,breaklinks=true,letterpaper=true,colorlinks,linkcolor=black,citecolor=black,urlcolor=black]{hyperref}

\usepackage{todonotes} 
\usepackage{marginnote} 
\usepackage{pifont}
\let\marginpar\marginnote
\usepackage{flushend}

\newcommand{\algcomment}[1]{\textcolor{blue}{//#1}}

\newcommand{\oms}[1]{{\color{black}#1}} 
\newcommand{\omt}[1]{{\color{black}#1}} 
\newcommand{\omf}[1]{{\color{black}#1}} 
\newcommand{\omsix}[1]{{\color{black}#1}} 
\newcommand{\omsev}[1]{{\color{black}#1}} 
\newcommand{\ome}[1]{{\color{black}#1}} 
\newcommand{\dc}[1]{{\color{black}#1}} 
\newcommand{\hy}[1]{{\color{black}{#1}}} 
\newcommand{\hys}[1]{{\color{black}{#1}}} 
\newcommand{\hyf}[1]{{\color{black}{#1}}} 
\newcommand{\hyfv}[1]{{\color{black}{#1}}} 
\newcommand{\hysix}[1]{{\color{black}{#1}}} 
\newcommand{\hysev}[1]{{\color{black}{#1}}} 

\makeatletter
\g@addto@macro{\normalsize}{%
  \setlength{\abovedisplayskip}{2pt plus 1pt minus 1pt}
  \setlength{\belowdisplayskip}{2pt plus 1pt minus 1pt}
  \setlength{\intextsep}{2pt plus 1pt minus 1pt}
  \setlength{\textfloatsep}{3pt plus 1pt minus 1pt}
  \setlength{\dbltextfloatsep}{3pt plus 1pt minus 1pt}
  \setlength{\skip\footins}{4pt plus 1pt minus 1pt}
}
\titlespacing\section{0pt}{8pt plus 4pt minus 2pt}{4pt plus 2pt minus 2pt}
\titlespacing\subsection{0pt}{6pt plus 4pt minus 2pt}{4pt plus 2pt minus 2pt}
\titlespacing\subsubsection{0pt}{4pt plus 4pt minus 2pt}{4pt plus 2pt minus 2pt}
\makeatother

\def\BibTeX{{\rm B\kern-.05em{\sc i\kern-.025em b}\kern-.08em
    T\kern-.1667em\lower.7ex\hbox{E}\kern-.125emX}}

\def\UrlBreaks{\do\/\do-\/\do.\/\do:}

\expandafter\def\expandafter\UrlBreaks\expandafter{\UrlBreaks
  \do\a\do\b\do\c\do\d\do\e\do\f\do\g\do\h\do\i\do\j
  \do\k\do\l\do\m\do\n\do\o\do\p\do\q\do\r\do\s\do\t
  \do\u\do\v\do\w\do\x\do\y\do\z\do\A\do\B\do\C\do\D
  \do\E\do\F\do\G\do\H\do\I\do\J\do\K\do\L\do\M\do\N
  \do\O\do\P\do\Q\do\R\do\S\do\T\do\U\do\V\do\W\do\X
  \do\Y\do\Z}
  
\definecolor{amber}{rgb}{1.0, 0.49, 0.0}
\definecolor{awesome}{rgb}{1.0, 0.13, 0.32}
\definecolor{dollarbill}{rgb}{0.52,0.73,0.4}
\definecolor{moegi}{rgb}{0.357, 0.537, 0.188}
\definecolor{burgundy}{rgb}{0.5, 0.0, 0.13}
\definecolor{ballblue}{rgb}{0.13, 0.67, 0.8}
\definecolor{ups-truck}{rgb}{0.53, 0.28, 0.21}
\definecolor{airforceblue}{rgb}{0.36, 0.54, 0.66}
\definecolor{cadmiumgreen}{rgb}{0.0, 0.42, 0.24}
\definecolor{darkcyan}{rgb}{0.0, 0.55, 0.55}
\definecolor{caribbeangreen}{rgb}{0.0, 0.8, 0.6}
\definecolor{flamingopink}{rgb}{0.99, 0.56, 0.67}
\definecolor{jazzberryjam}{rgb}{0.65, 0.04, 0.37}
\definecolor{mediumpersianblue}{rgb}{0.0, 0.4, 0.65}
\definecolor{coolblack}{rgb}{0.0, 0.18, 0.39}
\definecolor{bleudefrance}{rgb}{0.19, 0.55, 0.91}
\definecolor{ao}{rgb}{0.0, 0.0, 1.0}
\definecolor{babyblueeyes}{rgb}{0.63, 0.79, 0.95}
\definecolor{darkwarmgray}{rgb}{0.2,0,0}

\newcommand{\circled}[1]{{\tikz[baseline=(char.base)]{\node[shape=circle,inner sep=1.3pt,fill=black, text=white] (char) {\small \textbf{#1}};}}}

\newcommand*\circledletter[1]{\tikz[baseline=(char.base)]{
\node[font = {\small}, anchor=text, shape=circle, draw, inner sep=-0.5pt, minimum size=-1.1em] (char) {#1\strut};}}

\usepackage{tikz}

\DeclareRobustCommand\wcirc[1]{\tikz[baseline=(char.base)]{
           \node[shape=circle,draw,inner sep=1pt,fill=white, text=black] (char) {#1};}}

\newcommand{\squishlist}{
 \begin{list}{$\circ$}
  { \setlength{\itemsep}{0pt}
     \setlength{\parsep}{0pt}
     \setlength{\topsep}{0pt}
     \setlength{\partopsep}{0pt}
     \setlength{\leftmargin}{1em}
     \setlength{\labelwidth}{1em}
     \setlength{\labelsep}{0.5em} } }

\newcommand{\squishsublist}{
\begin{list}{$\rightarrow$}
 { \setlength{\itemsep}{0pt}
    \setlength{\parsep}{0pt}
    \setlength{\topsep}{-10em}
    \setlength{\partopsep}{-3pt}
    \setlength{\leftmargin}{1em}
    \setlength{\labelwidth}{1em}
    \setlength{\labelsep}{0.5em} } }

\newcommand{\squishend}{
  \end{list}  }

\newcommand{\head}[1]{\noindent\textbf{#1.}} 

\begin{document}
%
\title{GenPIP: In-Memory Acceleration of Genome Analysis \\ via Tight Integration of Basecalling and Read Mapping \vspace{2mm}}
\newcommand{\name}{GenPIP\xspace}
\newcommand{\entire}{genome analysis pipeline\xspace}
\newcommand{\chuncpp}{CP\xspace}
\newcommand{\earlyrej}{ER\xspace}



\newcommand{\affilETH}[0]{\small {$^1$}}
\newcommand{\affilbionano}[0]{{\small {$^2$}}}

\author{\vspace{-22pt}\\%
{Haiyu Mao\affilETH{}}\quad%
{Mohammed Alser\affilETH{}}\quad%
{Mohammad Sadrosadati\affilETH{}}\quad%
{Can Firtina\affilETH{}}\quad%
{Akanksha Baranwal\affilETH{}}\quad%
\\
{Damla Senol Cali\affilbionano{}}\quad%
{Aditya Manglik\affilETH{}}\quad%
{Nour Almadhoun Alserr\affilETH{}}\quad%
{Onur Mutlu\affilETH{}}\quad%
\vspace{0pt}\\%
\affilETH\emph{ETH Z{\"u}rich}%
\qquad\quad%
\affilbionano\emph{Bionano Genomics}%
\vspace{-12pt}}
\maketitle
\thispagestyle{plain}

\begin{abstract}
     Nanopore sequencing is a widely-used high-throughput genome sequencing technology that can sequence long fragments of a genome into raw electrical signals at low cost.
     Nanopore sequencing requires two computationally-costly processing steps for accurate downstream genome analysis. 
     The first step, basecalling, translates the raw electrical signals into nucleotide bases (i.e., A, C, G, T). 
     The second step, read mapping, finds the correct location of a read in a reference genome.
     In existing genome analysis pipelines, basecalling and read mapping are executed separately. We observe in this work that such separate execution of the two most time-consuming steps inherently leads to (1) significant data movement and (2) redundant computations on the data, slowing down the genome analysis pipeline.
     
     This paper proposes GenPIP, an in-memory genome analysis accelerator that tightly integrates basecalling and read mapping. GenPIP improves the performance of the genome analysis pipeline with two key mechanisms: (1) in-memory fine-grained collaborative execution of the major genome analysis steps in parallel; (2) a new technique for early-rejection of low-quality and unmapped reads to timely stop the execution of genome analysis for such reads, reducing inefficient computation.  Our experiments show that, for the execution of the genome analysis pipeline, GenPIP provides $41.6\times$ ($8.4\times$) speedup and $32.8\times$ ($20.8\times$) energy savings with negligible accuracy loss compared to the state-of-the-art software genome analysis tools executed on a state-of-the-art CPU (GPU). Compared to a design that combines state-of-the-art in-memory basecalling and read mapping accelerators, GenPIP provides $1.39\times$ speedup and $1.37\times$ energy savings.

\end{abstract}


%

\section{Introduction}
\dc{Long read genome sequencing technologies}~\cite{amarasinghe2020opportunities, murigneux2020comparison, wang2019efficient, cali2017nanopore} have significantly advanced the development of several genomic fields, such as personalized medicine~\cite{clark2019diagnosis,farnaes2018rapid,Ashley2016,flores2013p4,chin2011cancer,alkan2009personalized,ginsburg2009genomic}, forensic science~\cite{alvarez2017next,borsting2015next}, evolutionary biology~\cite{Prohaska2019,ellegren2016determinants,hoban2016finding,ellegren2014genome,romiguier2014comparative,Prado-Martinez2013}, and investigation of infectious disease outbreaks, especially during the COVID-19 pandemic~\cite{bloom2021massively,yelagandula2021multiplexed,wang2020initial,nikolayevskyy2016whole,qiu2015whole,gilchrist2015whole,meyer2022critical,gire2014genomic,alser2022covidhunter,le2013selected,lapierre2019micop}. 
Oxford Nanopore Technology (ONT)~\cite{cali2017nanopore} is one of the most widely-used \dc{long-read} sequencing technologies. ONT provides portable sequencing devices connected to a computer via a USB interface~\cite{wang2021nanopore,segerman2020most, amarasinghe2020opportunities,jain2016oxford}.
ONT devices generate long subsequences (called \emph{long reads}) based on the organism's DNA sequence~\cite{wang2021nanopore,de2021towards,amarasinghe2020opportunities,shafin2020nanopore,logsdon2020long,payne2019bulkvis,van2018third,ardui2018single,de2018nanopack,jain2018nanopore,rang2018squiggle,belser2018chromosome,pollard2018long,kchouk2017generations,weirather2017comprehensive,cali2017nanopore,jain2017minion,giordano2017novo,jain2016oxford,clarke2009continuous,alser2022molecules}.
Each read usually has a length ranging from a few hundreds to millions of base pairs~\cite{wang2021nanopore} (i.e., A, C, G, T nucleotide bases) but with a high sequencing error rate (10\% to 15\%~\cite{jain2018nanopore,ardui2018single,van2018third,kchouk2017generations,weirather2017comprehensive}). 

ONT devices \dc{sequence a genome} by detecting the fluctuations in electrical signals when bases of a DNA sequence pass through \dc{a nanoscale hole, called a nanopore}~\cite{jain2016oxford}.
\dc{A computational-processing step, called \emph{basecalling}~\cite{bonito,Guppy}, translates these} raw electrical signals into a sequence of nucleotide bases (i.e., a read). 
\dc{Basecalling} is  executed either inside the sequencing \oms{device}~\cite{bonito,Guppy} \hy{or using an external computer connected to 
the sequencing device via external connection links (e.g., USB, Ethernet)~\cite{jain2016oxford}}. 
A translated read is associated with a quality score for each base to reflect the accuracy of the translation.
After basecalling, reads are sent to a separate \oms{device} to \dc{perform} further analysis~\cite{minion}.
\dc{First,} \emph{\dc{read quality control}} \dc{detects and} filters out low-quality reads (i.e., \dc{reads} whose average quality score is \hy{lower than a threshold, indicating the accuracy of basecalling translation is low}) to avoid further computation \dc{on} unreliable reads. 
\hy{After read quality control, a computationally-costly \emph{read mapping} step identifies potential matching locations of reads against a reference genome (i.e., a high-quality representative sequence of a species)~\cite{mansouri2022genstore,firtina_blend_2021,alser2020accelerating,cali2020genasm,alser2020technology}.} 

In the entire genome analysis pipeline, basecalling and read mapping are \dc{two of the most time-consuming steps because they rely on computationally-intensive algorithms}~\cite{kim_fastremap_2022,firtina_aphmm_2022,alser2022molecules,cali_segram_2022,mansouri2022genstore,singh2021fpga,lou2020helix,firtina2020apollo, s_d_goenka_segalign_2020,angizi2020pim, alser2020accelerating,cali2020genasm,alser2020sneakysnake,nag2019gencache,kim2019airlift,kim2018grim,turakhia2018darwin,alser2017gatekeeper,xin2015shifted,xin2013accelerating}.
Basecalling commonly uses a deep neural network (DNN) to ensure high accuracy~\oms{\cite{bonito,Guppy}}. \dc{Read mapping} depends on dynamic programming (DP)-based algorithms~\cite{smith1981identification,needleman1970general} to find the \hy{potential matching locations} in the reference genome where the read can be aligned. 
\hy{Basecalling and read mapping are executed separately in different devices~\cite{alser2022molecules}. 
\oms{This} decoupled \oms{execution} of basecalling and read mapping causes three main issues. 
First, the data movement from the basecalling device to the read mapping device becomes a bottleneck.
Second, accelerators have been designed separately for basecalling~\cite{lou2020helix,lou2018brawl,wu2018fpga,alser2022molecules} and read mapping~\cite{mansouri2022genstore,khatamifard2021genvom,singh2021fpga,alser2020accelerating,jain2019accelerating,feng2019accelerating,gupta2019rapid,zokaee2018aligner,alser2022molecules,diab2022framework} \oms{to reduce the computational bottleneck, which exacerbates} the data movement bottleneck. 
Third, large execution time and energy consumption overheads \oms{ensue} due to the fact that a significant portion of \oms{the} basecalling output is \omt{\emph{not}} used in the subsequent analyses because of either low quality (10-20\% of the examined dataset~\cite{bowden2019sequencing}) and/or being different from the reference genome that is used for mapping (30-70\% of the examined datasets~\cite{lapierre2020metalign,ecoli}).}

\textbf{Our \dc{goal}} is to \hy{provide effective in-memory acceleration of the entire genome analysis pipeline while minimizing data movement and useless computation.} 
To this end, we propose \underline{\name}, a fast and energy-efficient in-memory acceleration \omt{system} for \omt{the} \underline{Gen}ome analysis \underline{PIP}eline\omt{, which we envision to be best implemented} inside the sequencing machine. 
\hy{The \textbf{key idea} of \name is to tightly integrate the two key steps of genome analysis (i.e., basecalling and read mapping) inside main memory to (1) minimize data movement by eliminating the need to store intermediate results and (2) minimize useless computation in the genome analysis pipeline that leads to unused outputs by providing timely feedback from read quality control and read mapping steps to the basecalling step.}
To realize \oms{our} key idea, we design an in-memory processing architecture for \name, equipped with two key techniques: (1) a chunk-based pipeline that provides fine-grained collaboration of basecalling and read mapping steps by processing reads at chunk granularity (i.e., a subsequence of a read, \oms{e.g.,} 300 bases)
and (2) an early-rejection technique \oms{that} predicts \omt{which reads} will not be useful downstream by analyzing \oms{multiple} chunks of the read, and then stops the execution of basecalling and read mapping for such reads. 

We compare \name with (1) the state-of-the-art software genome analysis tools on \dc{CPUs and GPUs} and (2) \dc{a combination of} state-of-the-art \dc{in-memory} basecalling~\cite{lou2020helix} and read mapping~\cite{chen2020parc} accelerators.   
Our experimental results show that \name~\dc{provides} $41.6\times$ ($8.4\times$) speedup and $32.8\times$ ($20.8\times$) energy \dc{savings} with negligible \dc{accuracy loss}, \dc{over the state-of-the-art software tools executed on a state-of-the-art} CPU (GPU). Compared \dc{to the} combination of prior \hys{in-memory} accelerators, 
\name delivers $1.39\times$ speedup and $1.37\times$ energy \dc{savings}.  

We make the following contributions:
\begin{itemize} [leftmargin=0pt, itemindent=10pt, itemsep=0pt]
    \item \hy{We observe that the \dc{combined} acceleration of \dc{multiple steps of the} genome analysis pipeline is \dc{critical} due to (1) \dc{large} data movement between \dc{multiple} genome analysis steps and (2) significant unnecessary computation \oms{due to} low-quality and unmapped reads. } 
    \item We propose \name as the \emph{first} in-memory accelerator for \omt{the} genome analysis pipeline, including basecalling, read quality control, and read mapping steps.
    \item We introduce two key mechanisms that \name employs: (1) fine-grained collaboration of two critical steps (i.e., basecalling and read mapping) using \oms{a} chunk-based pipeline, and  (2) timely prediction of low-quality and unmapped reads to stop the execution of basecalling and read mapping for such reads.
    \item \omt{We evaluate GenPIP and demonstrate that it provides significant performance and energy benefits over the state-of-the-art software and in-memory acceleration approaches.}
\end{itemize}

\section{Background and Motivation} ~\label{sec:background}
In this section, we first introduce the current genome analysis pipeline and its conventional execution environment. Second, we elaborate on \omt{the} shortcomings of current accelerator designs by studying the performance and energy bottlenecks in the \entire. 
Third, \omf{based} on our analysis, we \omt{describe our key goal in this work}.

\subsection{\omt{Nanopore} Genome Analysis Pipeline}~\label{sec:genomepipeline}
\begin{figure*}[!t]
    \centering
    \includegraphics[width=\linewidth]{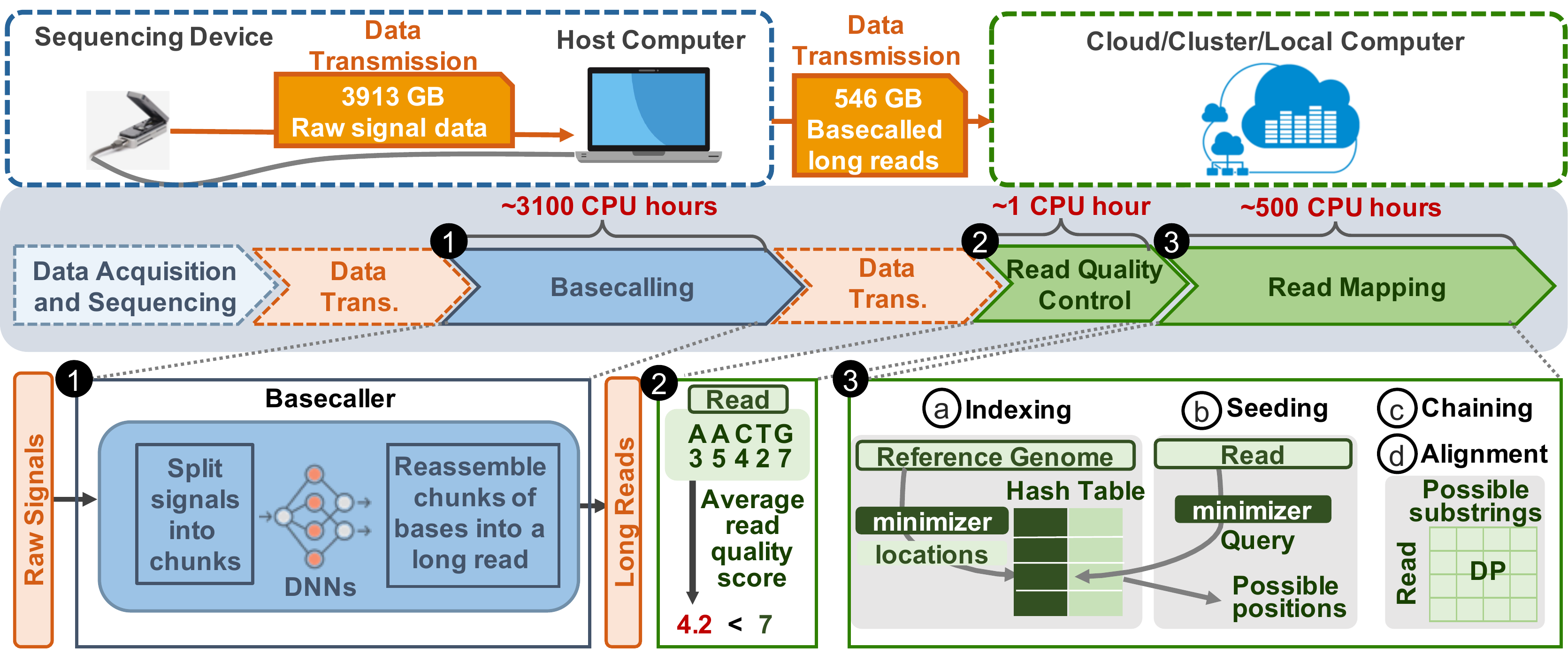}
    \caption{The genome sequencing and analysis pipeline. The basecalling step (\ding{202}) and the read mapping step (\ding{204}) are the two most time-consuming steps in the genome analysis pipeline. The read quality control step (\ding{203}) is a highly-recommended but optional step to reduce the workload of read mapping \omt{by eliminating unnecessary computation}. \omf{Dataset sizes and processing times are from~\cite{bowden2019sequencing}.}}
    \label{fig:analysis}
    \vspace{0.5em}
\end{figure*}

Oxford Nanopore Technology (ONT)~\cite{cali2017nanopore} is \omt{a} widely-used sequencing technology as it provides portable sequencing devices and offers \omt{much} higher sequencing speed than prior sequencing technologies~\cite{alser2022molecules,wang2021nanopore,segerman2020most, amarasinghe2020opportunities,jain2016oxford}. 
An ONT device generates long subsequences of DNA (called \emph{long reads}) by detecting the changes in electrical current signals when a DNA sequence passes through the device's nanopore~\cite{jain2016oxford} \hys{(called \omt{the} \textit{data acquisition and sequencing} step in genome sequencing)}.
\omf{The genome} analysis pipeline executes after genome sequencing to \omsix{identify} and analyze genomic features.
Figure~\ref{fig:analysis} shows an example of the ONT-based \omt{(i.e., nanopore)} genome sequencing and analysis pipeline. 
The pipeline includes two key steps, \omt{\emph{basecalling}} (\circled{1}) and \omt{\emph{read mapping}} (\circled{3}), and a highly-recommended but optional step, \omt{\emph{read quality control}}~\cite{alser2022molecules} (\circled{2}).


The first key step, \emph{basecalling}, receives the \omt{measured} electrical \omt{current} signals from \omt{an} ONT device \omf{and} then translates \omt{these signals} into \omt{nucleotide} bases (i.e., A, C, G, T).
State-of-the-art basecallers (e.g., Bonito~\cite{bonito}) use deep neural networks (DNNs) that provide high accuracy for translating electrical signals into bases~\cite{bonito,Guppy,Flappie,Scrappie,lou2020helix, 10.1093/gigascience/giy037,cali2017nanopore}. 
Basecallers report a \emph{read quality score} (i.e., the accuracy of the translation) for each base along with the translated base.
The basecaller first splits a long read in electrical-signal format \omt{(e.g., millions of signals)} into multiple \omt{smaller} \textit{chunks} (e.g., \hys{thousands of} signals per chunk) and then basecalls these chunks.
After basecalling all the chunks, the basecaller reassembles them back into a long read.


\hy{The second key step, \textit{read mapping}, maps} the basecalled read to \omt{a} reference genome (i.e., a high-quality representative genome sequence of a species).
Minimap2~\cite{li2018minimap2} is a state-of-the-art read mapping tool that performs read mapping mainly in four phases.
\omt{The first is a preprocessing step called \emph{indexing}} \hys{(\omsix{\circledletter{a}}}, shown at the bottom right of Figure~\ref{fig:analysis})\hyf{, which enable\omt{s} efficient queries to quickly find matches between the subsequences of a reference genome and reads.}
\omt{Indexing} is performed \omt{offline and} \textit{only} once for each reference genome.
In \omt{the} indexing \omt{step}, Minimap2 generates minimizers~\cite{roberts_reducing_2004,schleimer2003winnowing} (i.e., representative subsequences) from the reference genome, 
and insert\omt{s} them \omt{into} a key-value hash table, where minimizers are the keys and their locations in the reference genome are the values.
Second, Minimap2 performs \emph{seeding} \hys{(\omf{\circledletter{b}})} to generate minimizers from a basecalled read and query \omt{the generated minimizers} in the hash table to quickly find
matching regions between \omf{the} reference genome 
and the read sequence~\cite{firtina_blend_2021}.
Third, Minimap2 executes \emph{chaining} \hys{(\omf{\circledletter{c}})}~\cite{li2018minimap2} to identify the \emph{candidate regions} in the reference genome that have a high similarity with the read based on the matching minimizers and \omf{distances} between \omf{the read and the reference genome}.
Chaining is a dynamic programming (DP) approach \hys{\cite{alser2021technology,eddy2004dynamic}} that assigns a \emph{chaining score} for a chain of matching minimizers 
based on the distances (i.e., gaps) between these minimizers. As the chaining score increases, the similarity between the corresponding region in the reference genome and read sequence increases.
Fourth, Minimap2 performs \emph{sequence alignment} \hys{(\omf{\circledletter{d}})} to quantitatively identify the similarity between the read and each candidate \omt{region} in the reference genome.
\hyf{Sequence alignment calculates \omt{an} \emph{alignment score} to quantitatively \omt{represent} the difference between the two sequences. To calculate the alignment score, sequence alignment uses a computationally-expensive DP algorithm~\cite{alser2022molecules,alser2020technology} that performs approximate string matching between two sequences.} 

\textit{Read quality control (RQC)} \omsix{is a highly-recommended~\cite{alser2022molecules} but optional step that takes place after basecalling but before read mapping.}
RQC filters out low-quality reads generated by the basecaller to (1) improve the overall accuracy of \omt{the entire genome} analysis pipeline, \omt{and} (2) reduce the computation and memory overheads associated with processing such low-quality reads in \omt{later steps (e.g., read mapping)}.
First, RQC calculates the average quality score of a read by summing the quality score\omt{s} of  \omt{each of the read's bases} and then dividing \omt{this sum} by the number of bases.
Second, RQC uses a threshold to categorize the reads into low-quality and high-quality groups, and filters out the low-quality reads before performing read mapping. 
For example, several prior works consider a read with an average quality score of less than \emph{7} as a low-quality read that \omt{is not useful} in further analysis \dc{step}s~\cite{fukasawa2020longqc,leger2019pycoqc}.

In the genome analysis pipeline, \hys{basecalling and read mapping steps are usually executed on different machines~\cite{minionrequirements}.
The computer used for basecalling \omf{is} 
usually located in the \emph{wet lab} with the sequencing device~\cite{minion}. \omf{Later} analysis steps are executed on \omf{completely separate (and \omsix{usually physically distant})} machines located in the \emph{dry lab}.} 
In the entire pipeline, basecalling and read mapping are two of the most time-consuming computational steps~\cite{kingsmore2020measurement, clark2019diagnosis,farnaes2018rapid}.
In the \omf{real \omsix{system} study} shown in~\cite{bowden2019sequencing} \omt{and \omf{pictorially}  demonstrated in Figure~\ref{fig:analysis} (middle)}, the basecalling step takes $\sim3100$ CPU hours and the read mapping step takes $\sim500$ CPU hours. 
This motivates \omt{system} designers to accelerate two key computational steps, basecalling and read mapping.
\omt{Next}, we describe the \omt{shortcomings} of prior accelerator design\omt{s} \omt{in the context of the entire} genome analysis pipeline. 

\subsection{State-of-the-art Solutions}\label{sec:accelerators}
Several works propose \omt{hardware} accelerators for  basecalling~\cite{lou2020helix,lou2018brawl,wu2018fpga} or read mapping~\cite{mansouri2022genstore, khatamifard2021genvom, singh2021fpga, alser2020technology, alser2020accelerating, alser2020sneakysnake, angizi2020pim, s_d_goenka_segalign_2020, cali2020genasm, nag2019gencache, jain2019accelerating,feng2019accelerating,gupta2019rapid,turakhia2018darwin,zokaee2018aligner}.  
Among these accelerators, non-volatile memory (NVM)-based processing in memory (PIM) accelerators offer \omt{high} performance \omt{and efficiency} since NVM-based PIM provides in-situ and highly-parallel computation \omt{support} for matrix-vector multiplications (MVM)~\cite{ankit2019puma,ankit2020panther,yuan2021forms,huang2021mixed,longlive,song2018graphr,imani2019floatpim,mittal2019survey,gupta2019nnpim,lergan,mao2020lrgan} and string matching operations~\cite{karam2015emerging,imani2016masc,li2016nvsim,deng2016multi,chang20173t1r,yantir2017approximate,yin2017design,imani2017multi,yin2018ultra,wang2018novel,gnawali2018low,behnam2018r,tan2019experimental,arakawa2019multi,Zhao2019RFAccA3,halawani2019reram,li2020analog,graves2020memory,yin2020fecam,xiu2021capacitive}, \omt{two major operations \omf{used} in \omf{the} genome analysis pipeline.} MVM is the main operation in DNN-based \omt{basecallers}~\cite{bonito,Guppy} and string matching is the main operation in read mapping\omt{~\cite{li2018minimap2}}. 

\begin{figure}[htbp]
\centering
\includegraphics[width=\linewidth]{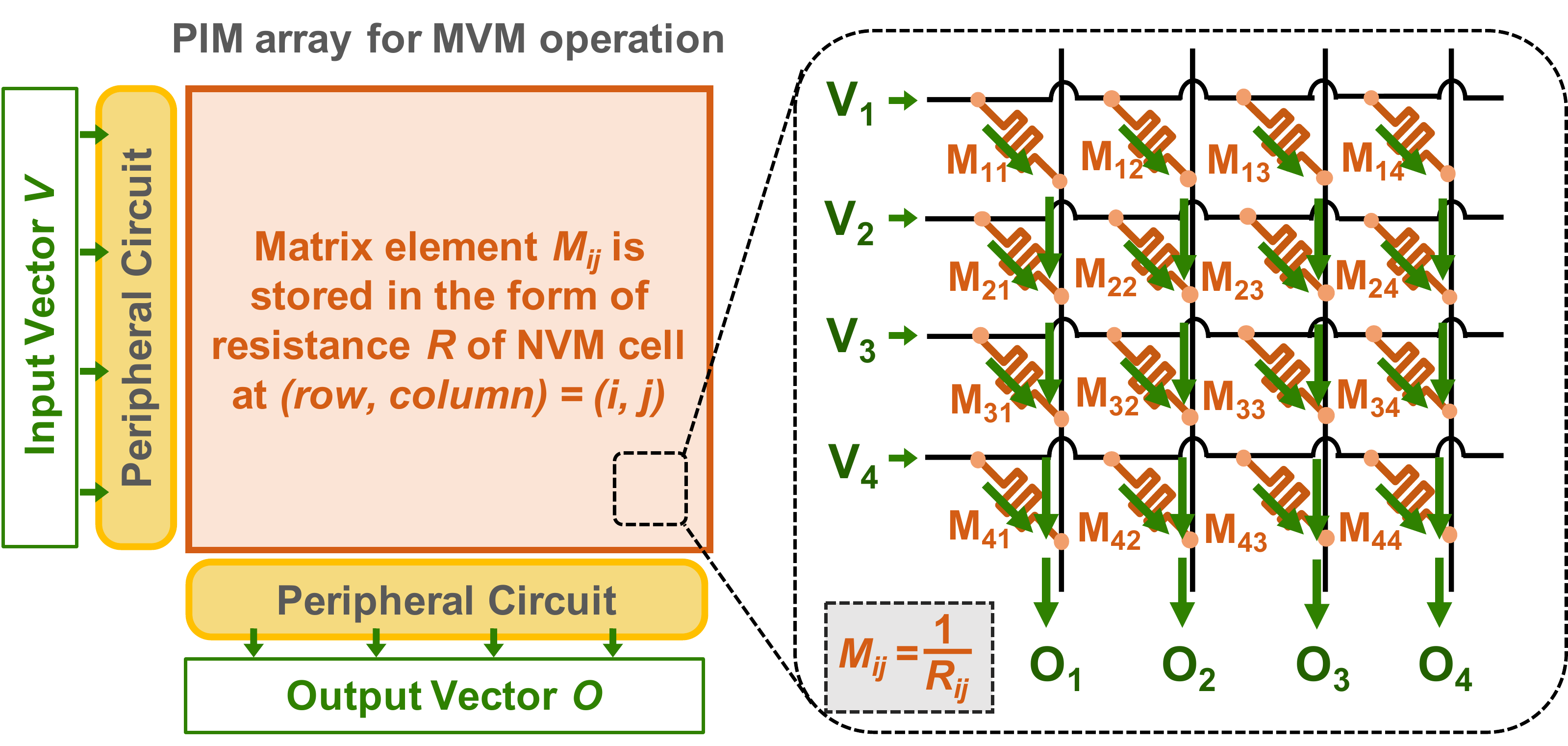}
\caption{The basic structure of an NVM-based PIM array \omt{designed for computing an} MVM operation.} \label{fig:mvmarray}
\end{figure}

\textbf{NVM-based PIM Array for MVM Operation\omt{s}.} Helix~\cite{lou2020helix} is the state-of-the-art basecalling accelerator \omt{that exploits} \hys{an NVM-based} PIM array designed for \omt{efficiently performing} MVM operation\omt{s}. Figure~\ref{fig:mvmarray} shows the basic structure of an NVM-based PIM array designed for \omt{the} MVM operation~\cite{chi2016prime}.
The NVM-based PIM array performs \hys{in-situ} MVM \hys{operation by applying \omf{1)} voltages ($V$ represents the input vector) on the wordlines of the array that stores the matrix ($M$) and \omf{2)} sensing the output vector ($O$) on the bitlines}.
\hys{In the MVM operation ($O = V \times M$), the PIM array uses the resistance ($R$) of each NVM cell to represent the corresponding element of matrix $M$ ($M_{i,j} = 1/R_{i,j}$).}
\hys{Based on Kirchhoff's Law, the currents sensed on the bitlines represent $O$.} 
Using the PIM array, an MVM operation can be \omf{performed} inside the NVM array in nearly a single NVM read cycle if the matrix fits in the PIM array.
\vspace{4mm}

\begin{figure}[htbp]
\centering
\includegraphics[width=0.8\linewidth]{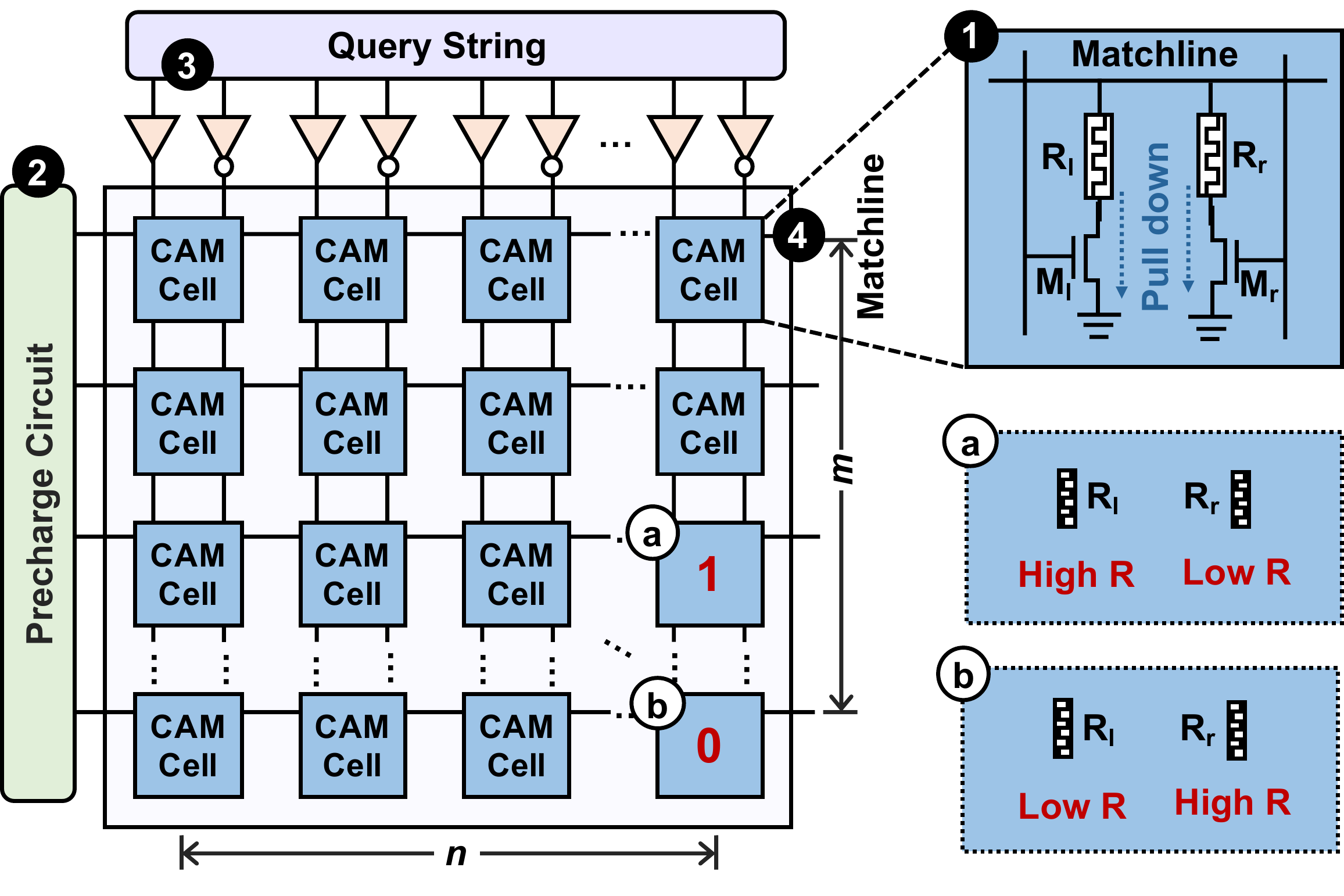}
\caption{\hyf{An example NVM-based CAM array for string matching.}} \label{fig:cam}
\end{figure}

\textbf{NVM-based PIM Array for String Matching Operation\omt{s}.}  Content addressable memory (CAM) is often leveraged for accelerating string matching operations. 
\hyf{PARC~\cite{chen2020parc} is the state-of-the-art work that accelerates the
computationally-expensive chaining step in read mapping 
using \omt{an} NVM-based CAM.}
Figure~\ref{fig:cam} 
\hyf{shows an example NVM-based CAM used for string matching. The CAM array consists of $m \times n$ CAM cells that house \emph{m} reference strings, each of which is \emph{n-bit} long. Each CAM cell stores \emph{one} bit and has two programmable resistors ($R_{l}$ and $R_{r}$) and two transistors ($M_{l}$ and $M_{r}$) (Figure~\ref{fig:cam}\circled{1}). To store 1 (or 0) in a CAM cell, $R_{l}$ and $R_{r}$ are programmed to high and low (or low and high) resistance, respectively (Figure~\ref{fig:cam}\omsix{\circledletter{a}\circledletter{b}}). 

The NVM-based CAM array is able to query the existence of an \emph{n-bit} string in parallel across all \emph{m} rows. First, the CAM array precharges the matchline signals to \emph{high} voltage (\circled{2}). 
Second, each bit in the input string and its \omsix{complement} drive the gate voltages of $M_{l}$ and $M_{r}$ transistors of the CAM cells in the corresponding column, respectively (\circled{3}). 
Third, each CAM cell compares its stored bit to the corresponding bit in the input string. If these two bits are different, the pull down network in the CAM cell is turned on and the matchline becomes "0". Otherwise, the matchline keeps its precharged \emph{high} voltage. We elaborate on this operation using an example. Assume that the bit stored in a CAM cell is "1", which means $R_{l}$ and $R_{r}$ are high and low resistance \hysix{(\circledletter{a})}, respectively. Having "1" in the corresponding bit of the input string implies that transistors $M_{l}$ and $M_{r}$ are \emph{on} and \emph{off}, respectively. Hence, none of the pull down circuits are active in this CAM cell since 1) the left circuit cannot drain current due to the high resistance value of $R_{l}$, and 2) the right circuit cannot also, due to the off transistor $M_{r}$. As a result, matchline keeps its \emph{high} voltage indicating that it is a match in this CAM cell. However, having "0" in the corresponding bit of the input string turns on the right pull down circuit and discharges the matchline signal ($R_{r}$ is low resistance and $M_{r}$ is on). 
Fourth, if all bits of the input string match with all corresponding CAM cells in a row, the matchline will remain high, indicating an \omsix{\emph{exact match}} between the input string and the reference string stored in the CAM array (\circled{4}).} 



\subsection{Limitations of the State-of-the-art \omt{Accelerators}}\label{sec:limitations}
Although \omt{state-of-the-art works} accelerate the basecalling and read mapping steps \omt{separately}, \omt{no prior system is} designed to support \omt{multiple key steps of} the \entire in a single accelerator design. This leads to two important limitations: \omt{(1)}  \omt{Different} steps in the \entire are separated from each other \omt{and executed in different devices}, \omt{which results} in \omt{in a large amount of} data movement between the \dc{step}s. 
\omt{(2)} A considerable portion of computation done in the \entire is likely to be useless due to \hys{low-quality or unmapped reads.} \omf{Next, we describe each of these limitations in more detail.}

\hys{First, \omf{executing the \omsix{genome analysis} steps separately from each other} generates a large amount of data movement between the machine that \omf{performs basecalling, and the machines perform} read quality control and read mapping\omf{,} as shown in the example in Figure~\ref{fig:analysis}.
Such data movement introduces two main issues\omf{:}
1) \omf{A} large amount of intermediate data (e.g., 3913 GB raw signal dataset and 546 GB basecalled reads\omf{, based on real datasets \omsix{analyzed} in}~\cite{bowden2019sequencing}) needs \omf{to be stored in} large memory or storage \omf{structures}. 
2) Transferring data \omf{between different machines that execute the different steps}
is \omf{both} time-consuming and energy hungry\omf{, and it} significantly bottlenecks \omf{both} the performance and energy efficiency of the entire genome analysis pipeline.
When machines use \omsix{state-of-the-art} accelerators (Section~\ref{sec:accelerators}), data movement between \omf{different} machines becomes an even \omf{larger} bottleneck as computation time reduces with fast yet separate accelerators.}

\hys{Second, a considerable amount of useless data \omf{that flows through the \omsix{genome} analysis pipeline} wastes computation and memory resources.
Even though read quality control (Section \ref{sec:genomepipeline}) filters \omf{out} the low-quality reads, these reads have already \omf{been} processed by the expensive basecalling step \omt{(because basecalling happens \omf{earlier} in a separate machine)}.}
\hys{To quantitatively show the amount of low-quality reads, we perform a descriptive \omf{statistical analysis} on the Escherichia coli (i.e., E. coli) genome dataset~\cite{ecoli}.}
We make a key observation that a large number of reads (20.5\% in~\cite{ecoli}) are basecalled but eventually discarded, including very long reads.
Besides the low-quality reads, some high-quality reads cannot be mapped (called \emph{unmapped reads}) to the reference genome due to high dissimilarity~\cite{li2018minimap2}. \hys{To quantitatively show the amount of unmapped reads, we map E. coli reads~\cite{ecoli} to the reference genome and \omf{find that} 10\% of \omf{all} reads \omf{are} unmapped.} \omf{Thus, a total of 30.5\% of all reads in the E. coli dataset are useless. Such a} \hys{large amount of useless reads motivates us to reject such reads as soon as possible \omf{(ideally even before they go through basecalling)} to reduce the computation and memory \omsix{overheads caused by them}.}

\subsection{Potential Benefits}\label{sec:potential}
\omt{We would like to quantitatively demonstrate the potential benefits of overcoming the two limitations we identify in prior works~\hys{\cite{lou2018brawl,wu2018fpga,mansouri2022genstore, alser2020technology, alser2020accelerating, singh2021fpga, alser2020sneakysnake, angizi2020pim, s_d_goenka_segalign_2020, cali2020genasm, turakhia2018darwin, nag2019gencache, jain2019accelerating,feng2019accelerating,zokaee2018aligner,gupta2019rapid,khatamifard2021genvom,lou2020helix,chen2020parc}}. To this end, we devise a study 
to compare performance of the following four systems using the E. coli dataset \omsix{we describe in Section~\ref{sec:limitations}}:\footnote{\omf{We provide our methodology in Section~\ref{sec:evaluation_setup}}}}
\begin{enumerate}[leftmargin=0pt, itemindent=46pt, itemsep=0pt]
    \item[System A.] \textbf{Current \omsix{practice.}} \omf{This system} separately \omf{executes} the state-of-the-art open-source basecalling and read mapping software, Bonito~\cite{bonito} and Minimap2~\cite{li2018minimap2}. \omf{Each software is executed on a separate machine, a state-of-the-art GPU machine~\cite{gpu} for Bonito~\cite{bonito} and a state-of-the-art CPU server~\cite{cpu} for Minimap2~\cite{li2018minimap2}}. 
    \hys{Reads whose average quality score is less than 7 are discarded after basecalling but before read mapping.}
    \item[System B.] \textbf{\omt{State-of-the-art} accelerators.} The basecalling and \omsix{read mapping steps} 
    are executed in \omf{separate} NVM-based PIM accelerators\hys{, Helix~\cite{lou2020helix} and PARC~\cite{chen2020parc}}. The read quality control step is executed in 
    \omf{a} \hys{state-of-the-art} CPU~\cite{cpu}.
    \item[System C.] \textbf{\omf{Accelerators with no} data movement overhead.} 
    \hyf{This system is an idealized version of System B where we ideally eliminate all data movement between separate NVM-based accelerators and \omsix{the} CPU. We demonstrate this ideal system to show the potential benefit of eliminating data movement between separate accelerators and CPUs executing different parts of the genome analysis pipeline.}
    \hysix{We assume there is \emph{no} data movement between these NVM-based PIM accelerators and} \omsev{the} \hysix{CPU by removing the latency of data movement in our analysis.}
    \item[System D.] \textbf{No data movement and no useless reads.} \omf{This system is an even more ideal version of System C. Here\omsix{,} we ideally eliminate useless and unmapped reads even before they are basecalled. As such, useless and unmapped reads do not have any overhead in the pipeline.}
\end{enumerate}

\begin{figure}[htbp]
    \centering
    \includegraphics[width=\linewidth]{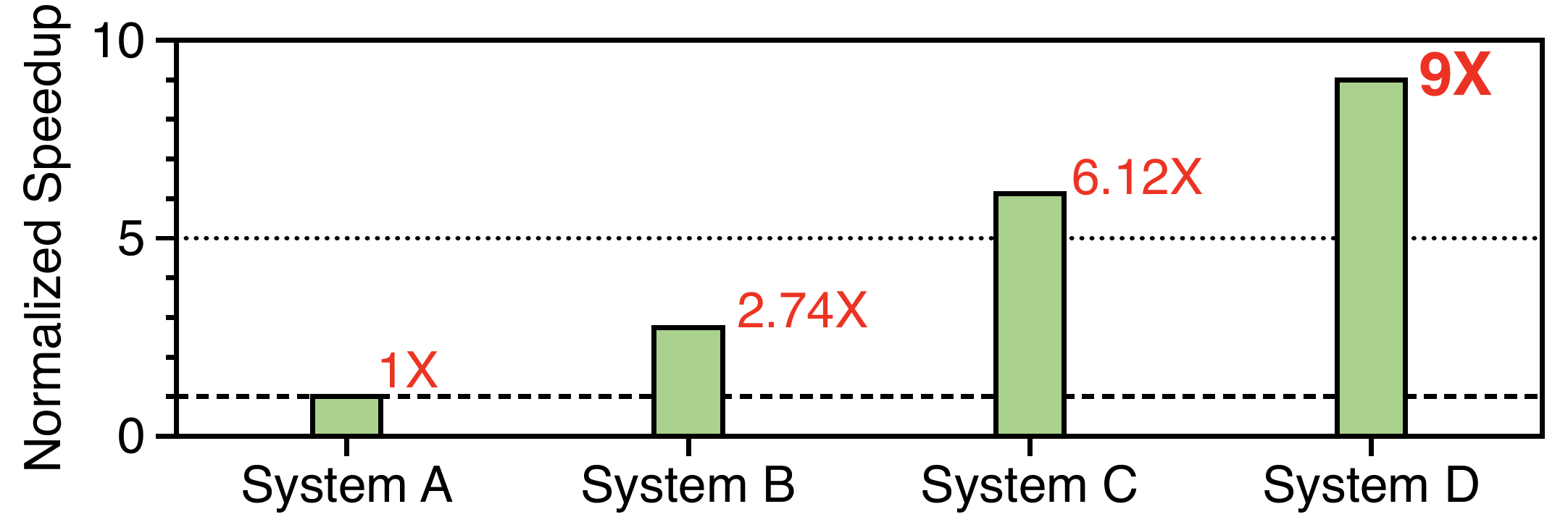}
    \caption{\omt{Performance comparison between four different systems.}}
    \label{fig:motivation}
    \end{figure}

Figure~\ref{fig:motivation} shows the speedup of using \omt{Systems} B, C, and D \omt{normalized to the performance of System A.}
We make two observations. First, \omf{both System A and System B are} \hys{bottlenecked by data movement and useless \omf{reads}. \hyf{System} C and \hyf{System} D provide $2.23 \times$ and $3.28 \times$ speedup over System B, respectively, \omf{by eliminating these  bottlenecks}. Second, there is a significant potential (as System C and System D show) to accelerate the current \omsix{practice} \omt{(System A)} by tightly integrating the \omf{basecalling and read mapping} accelerators to reduce \omt{both} data movement and \omt{useless computation due to} useless reads.}

\subsection{Our Goal}
\omt{We aim} to \omt{1)} reduce the data movement in the genome analysis pipeline effectively and \omt{2)} avoid processing useless reads \omt{as early as possible in the genome analysis pipeline}.
To this end, we tightly integrate the computation of \hys{basecalling, read quality control, and read mapping steps} inside the memory. \omt{Doing so provides two major} opportunities for optimizing \omf{the} \entire~ \omf{holistically}: (1) 
Consuming intermediate data items as soon as they are generated. This eliminates the need for both storing intermediate data items in main memory and storage \omt{and transferring them via slow and \omf{energy-hungry} interconnects}. (2) Providing timely feedback from read quality control and read mapping \omf{steps} to terminate basecalling as soon as possible \hys{when the read is \omf{determined to be} \omsix{useless (i.e.,} low-quality or unmapped\omsix{)}}.

\section{\name: Overview}~\label{sec:design}
In this section, we present \name, a fast and energy-efficient in-memory \hyf{system} for \omsix{holistically} accelerating \omsix{the} genome analysis pipeline. \omsix{We} envision \omsix{\name} to be best implemented inside the sequencing machine. 
\hyf{The key idea of \name is to tightly integrate the two key steps of genome analysis (basecalling and read mapping) inside main memory to (1) minimize data movement by eliminating the need to store intermediate results and (2) minimize useless computation in the basecalling step that leads to unused outputs by providing timely feedback from read quality control and read mapping steps to the basecalling step.}
\hyf{\name is equipped with} two key \hyf{techniques}: (1) chunk-based pipeline (\chuncpp) and (2) early rejection technique (\earlyrej). 
\hyf{\chuncpp is a chunk-based pipeline that provides fine-grained collaboration of basecalling, read quality control, and read
mapping steps by processing reads at chunk granularity (i.e.,
a subsequence of a read, e.g., 300 bases)}.
\hy{\hysix{\name applies \earlyrej on top of \chuncpp to}} predict reads that will not be useful downstream by \hyf{sampling the quality of a number of chunks in each read} 
and stop the execution of \chuncpp for low-quality or unmapped reads. 
\hy{\earlyrej} includes two sub-techniques\omf{:} (1) rejection based on \emph{quality score}, which executes after basecalling but before read mapping, and (2) rejection based on \hyf{\emph{chaining score}, which executes during read mapping}. \hyf{The rest of this section explains \name's two key mechanisms (\chuncpp and \earlyrej)}

\subsection{Chunk-based \omf{P}ipeline \hy{(\chuncpp)}}~\label{sec:pipeline}
\begin{figure}[htbp]
    \centering
    \includegraphics[width=\linewidth]{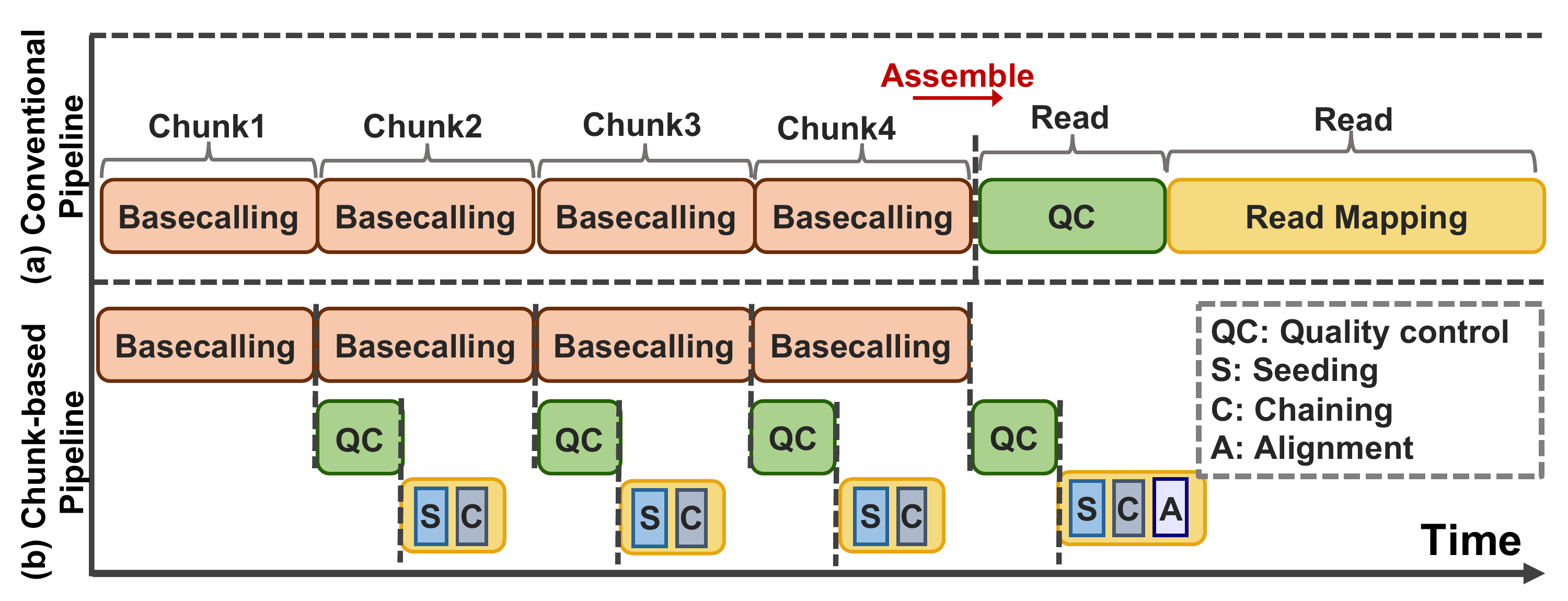}
    \caption{\omf{Conventional  pipeline (a) vs. the chunk-based pipeline (\chuncpp) of \name (b).}}
    \label{fig:chunk_pipeline}
\end{figure}

\chuncpp processes reads \omf{at the} granularity of a chunk (i.e., a subsequence of a read, e.g., 300 
bases, instead of the complete read sequence, e.g., 90k bases) to increase parallelism and the utilization of computation resources by overlapping the execution of \hyf{different steps in the \entire}.
Figure~\ref{fig:chunk_pipeline} \hyf{compares} \chuncpp to the \hyf{conventional} pipeline.
\hys{In the \hyf{conventional} pipeline (\omf{Figure~\ref{fig:chunk_pipeline}(a)}),} basecalling is executed \omf{at} the granularity of a chunk, while subsequent read quality control and read mapping steps are executed \omf{at} the granularity of a read (i.e., assembled by \omsix{basecalling of} \hyf{tens to hundreds} of chunks). 
\omf{We observe that} most of the computations performed in the read quality control and read mapping steps do \omsix{\emph{not}} require the information of an entire basecalled read.
For example, once a chunk is basecalled, we can calculate its average quality score, perform seeding (query the minimizers of this chunk), and perform chaining with the possible locations in this chunk. \omf{While this chunk is going through quality control and read mapping, the next chunk is basecalled.}  
In other words, a significant part of \dc{read quality control} and read mapping steps can be performed \omf{concurrently with} basecalling.
\hyf{After the last chunk goes through read quality control and read mapping, \chuncpp merges the results of all chunks in a read and \omsix{outputs} the read as the input to the sequence alignment step.}

\hyf{We illustrate the \chuncpp~mechanism using an example, assuming  a read of length $2c$ has two chunks, each of which has a length of $c$. We explain the independent \omf{and concurrent} execution of read quality control and basecalling in detail.}
The \hyf{conventional} pipeline calculates the average read quality score ($AQS_{read}$) \hysix{by calculating the average value of the quality scores of all bases in the read (i.e., $q_{1}, q_{2}$, ..., $q_{2c}$)} after the \omsix{\emph{entire}} read is basecalled (Equation~\ref{equ:aqso}), \omsix{yet} the $SQS$ of a chunk (i.e., sum of the quality scores of its bases) ($SQS_{first}$) can be calculated as soon as \omsix{\emph{that particular chunk}} is basecalled (Equation~\ref{equ:aqsp1}). 
After basecalling the next chunk, \chuncpp calculates the $SQS$ of this chunk and merges it with the result of the previous one (Equation~\ref{equ:aqsp}) \hyf{to calculate the $AQS_{read}$ of the entire read.}
\begin{equation}
    AQS_{read} = (q_{1} + q_{2} + ... + q_{c} + q_{c+1} + q_{c+2} + ... + q_{2c})/ 2c
    \label{equ:aqso}
\end{equation}
\begin{equation}
    SQS_{first} = q_{1} + q_{2} + ... + q_{c}
    \label{equ:aqsp1}
\end{equation}
\begin{equation}
    AQS_{read} = (SQS_{first} + q_{c+1} + q_{c+2} + ... + q_{2c})/2c
    \label{equ:aqsp}
\end{equation}
\hyf{Similarly, we use our example to explain the independent and concurrent execution of \omsix{the} seeding and chaining steps. As soon as the seeding step obtains a set of minimizer hits in the first chunk, the chaining step can work on the output of seeding step while the seeding step can obtain a set of minimizer hits in the second chunk. In the end, the chaining step combines the results from the two chunks.}

\omf{As we described} above, by \hys{tightly integrating} the basecalling, read quality control, and read mapping steps inside the sequencing machine, we can pipeline the execution of these steps \omf{at} the granularity of a chunk. Based on this insight, we propose a chunk-based pipeline, \hy{called \chuncpp}, that executes the partial computations of \dc{read quality control}, \hyf{seeding, and chaining} once a chunk is basecalled.
Figure~\ref{fig:chunk_pipeline}(b) \omf{shows} our \hy{\chuncpp} design.  As the figure shows\omsix{,} chunk-based basecalling, read quality control, and a part of read mapping (seeding and chaining) are pipelined. \hyf{The chunk-based execution flow 
not only saves time \omsix{via} pipeline\omf{d} execution (\omsix{by} overlapping the execution of several steps), but also reduces the \hyf{need for storing intermediate data as each pipeline step can quickly consume the \omsix{small amount of} output that is produced by the previous step.}}

\subsection{Early Rejection Based on Chunks \hy{(\earlyrej)}}~\label{sec:rejection}
\begin{figure}[htbp]
    \centering
    \includegraphics[width=\linewidth]{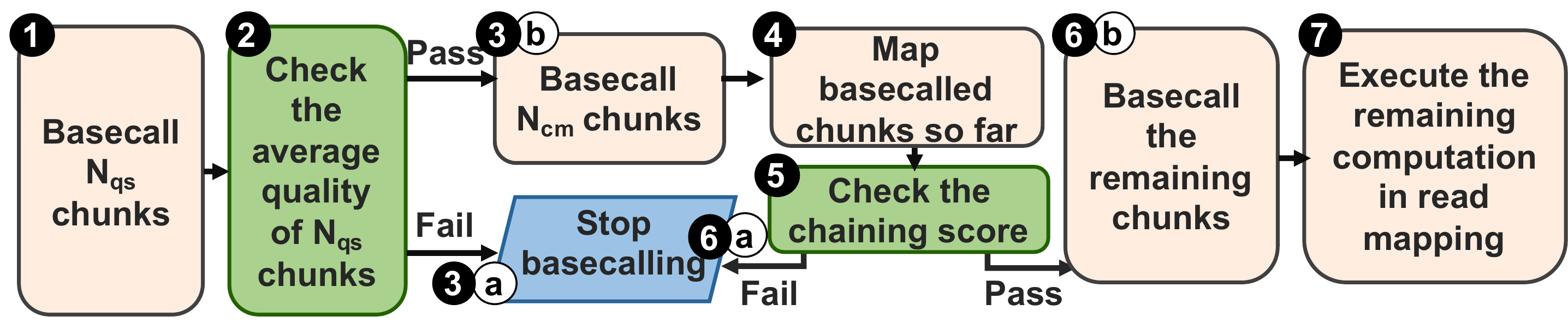}
    \caption{Overview of \omsix{the} early rejection \omsix{(\earlyrej) technique in the genome analysis pipeline (the green boxes \ding{203}\ding{206} are \omsev{the} two early-rejection steps \omsev{we introduce}).}}
    \label{fig:rejection}
\end{figure}

\hys{The goal of \earlyrej is to predict and eliminate low-quality and unmapped reads from 
both basecalling and read mapping steps. \omf{Doing so} can significantly \omf{lower} the execution time \omf{of the entire genome analysis pipeline.}}
\hys{To achieve this goal, \omsix{the key idea of \earlyrej is to use} information \omf{about} several basecalled chunks in a read to predict the \omsix{quality and} usefulness of \omsix{the} read.} \hysix{\name applies the \earlyrej technique on top of \chuncpp.}
Figure~\ref{fig:rejection} shows the overview of \hy{\earlyrej}.
\hys{Instead of basecalling \emph{all} \hyf{$N_{total}$} (\hysix{e.g., tens to hundreds})
chunks \omsix{in a read} \omsev{and} then check\omf{ing} the average quality score of the \emph{entire} read \omsix{(as done in conventional systems)}, 
\earlyrej~\omsix{first checks} the average quality score of \omsix{only \emph{a small number} of} \omf{(i.e., $N_{qs}$)} chunks basecalled by \chuncpp (Figure~\ref{fig:rejection}\circled{1}\circled{2}).
If the read fails this chunk-based quality score check, then \earlyrej stops basecalling \omsix{the remaining} chunks \omf{in the read} and discards \omf{the} read \omsix{(\circled{3}\circledletter{a})}. 
Otherwise, \chuncpp basecalls \omsix{some} more chunks \hyf{(i.e., $N_{cm}$)} in \omf{the} read \omsix{(\circled{3}\circledletter{b}}) and \hysix{ then maps the basecalled chunks \omsix{so far} \hyf{(i.e., $N_{qs} + N_{cm}$ \omsix{chunks}) (\circled{4}). \earlyrej checks the chaining score of} the $N_{qs} + N_{cm}$ basecalled chunks} (\circled{5}) (\hysix{i.e., \omsev{it predicts}} the likelihood of mapping \hyf{the read to the reference genome}). If the read fails the chunk-based \hyf{chaining score} check, \earlyrej stops basecalling \omsix{the remaining chunks} and discards \omf{the} read \omsix{(\circled{6}\circledletter{a})}.
Otherwise, \chuncpp basecalls the \omf{remaining} chunks \hyf{(i.e., $N_{total} - (N_{qs} + N_{cm})$ \omsix{chunks})} in \omf{the} read (\circled{6}\circledletter{b}) and executes the remaining computation in read mapping (\circled{7}).} 

\omsix{As such,} \earlyrej involves two filtering steps: (1) rejection based on the quality score of \hyf{$N_{qs}$} chunks (\circled{2}) and (2) rejection based on the \hyf{chaining score} of \hyf{$N_{qs} + N_{cm}$} chunks (\circled{5}).
We \omf{describe} these two filtering steps in detail.

\subsubsection{\omf{Early} Rejection Based on \omsix{Chunk} Quality \omsix{Scores}}~\label{sec:rejq}
\begin{figure}[htbp]
    \centering
    \includegraphics[width=\linewidth]{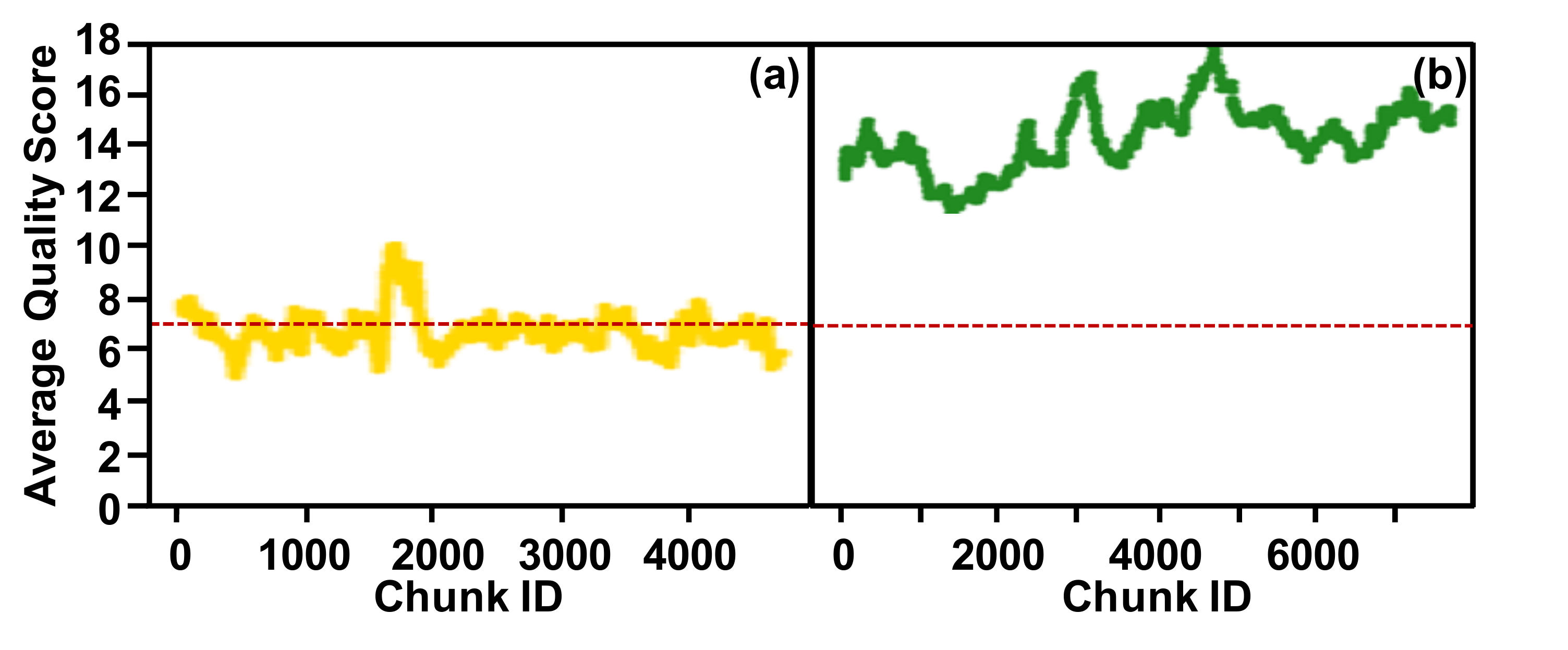}
    \caption{The quality scores of the chunks in two representative reads\omsix{: (a) a low-quality read and (b) a high-quality read}.}
    \label{fig:chunkq}
\end{figure}

\hyf{\omsix{The early} rejection technique based on the quality score of chunks relies on how \omsix{accurately} it can estimate the quality of the entire read \omsix{by} checking the quality of a \omsix{small number of \omsev{(i.e., $N_{qs}$)}} sampled chunks. We \omsix{first investigate} whether or not it is possible to \omsev{accurately} estimate the quality of the entire read using a \omsix{small number of} chunks. 
To this end, we study chunk quality scores from both low-quality reads and high-quality reads in \omsix{the} E. coli~\cite{ecoli} dataset (Section~\ref{sec:limitations}) using \omsix{a} chunk size \omsix{of} 300 bases.
As a representative example, Figure~\ref{fig:chunkq} shows the 
chunk quality scores in a low-quality read (Figure~\ref{fig:chunkq}(a)) and a high-quality read (Figure~\ref{fig:chunkq}(b)).}

We make three key observations:
(1) The range of quality scores for the chunks extracted from high-quality reads (e.g., ranging from 11 to 18 in Figure~\ref{fig:chunkq}(b)) is \omf{greatly} higher than that from low-quality reads (e.g., ranging from 4 to 10 in Figure~\ref{fig:chunkq}(a)).
\hyf{(2) \omf{A single} chunk\omf{'s} quality score is not enough to predict the read quality score because there are many chunks whose quality scores are larger than 7 (\omsix{which is} a common threshold used to distinguish between low-quality and high-quality reads\omsev{,} as shown in~\cite{fukasawa2020longqc,leger2019pycoqc}) \omf{in a low-quality read.}} 
(3) \omsev{Consecutive} chunks' quality scores are usually close to each other, indicating that sampling consecutive chunks \omsix{may not be} representative enough to estimate the quality of \omsix{an} entire read.
We conclude that early rejection based on \omsix{the} quality score \omsix{of chunks} should sample a \omsix{small number of} \omsev{\emph{non-consecutive}} chunks to accurately \omsix{guess whether or not a read is low-quality}.

Leveraging our key conclusion, we propose an early rejection technique based on \omsix{the} quality score \omsix{of chunks}, \omsix{called Quality-Score-based Rejection ($QSR$)}. $QSR$ 1) calculates the average quality score of \omsev{a set of non-consecutively} \hyf{sampled} chunks (i.e., $N_{qs}$ chunks) in a read, and 2) \omsix{predicts} \hyf{the entire read} as either low quality or high quality\omsix{, by comparing} the calculated average score \omsev{of the $N_{qs}$ chunks} \omsix{to the quality score threshold ($\theta_{qs}$).}

\hys{Algorithm~\ref{alg:rejc} illustrates the procedure of \omsix{the} $QSR$ \omf{technique}.} \hyf{$QSR$ 1) samples $N_{qs}$ chunks \omsix{that are} evenly distributed in a read, 2) calculates the sum of \omsev{the} quality \omsix{scores} of these sampled chunks (lines 1-3), and 3) uses the average quality score of these sampled chunks (line 4) to predict whether or not the quality score of the read is higher than \omsix{the quality score} threshold, $\theta_{qs}$ (lines 5-9).}

We determine the number of \omsix{sampled} chunks ($N_{qs}$) by \omsix{a} one-time preprocessing of \omsix{the chunk} quality scores of the reads of a species (see Section~\ref{sec:sensitivity} for more detail). For example, we experimentally observe that sampling \omsix{only} two chunks is enough for \omsix{accurate} read quality prediction in E. coli~\cite{ecoli}.


\begin{algorithm}[h]
    \footnotesize
    \SetAlgoLined
  \KwIn{
  the original read: $read_{original}$; \\
  length of the original read: $N$; \\
  chunk size: $C$\\
  number of chunks \omf{needed} for $QSR$: $N_{qs}$; \\ 
  \omsix{quality score threshold}: $\theta_{qs}$;}
  \KwOut{$rejection$}
  \For{i=0;i$ \textless N_{qs}$;i++}
  {
    sum\_sample\_score += quality score of the chunk \hysix{located at $\lfloor i/(N_{qs}-1) \rfloor \times \lfloor N/C \rfloor$ in $read_{original}$} \algcomment{sum the quality scores of evenly-sampled chunks in the read}
  }
  average\_score = sum\_sample\_score/$N_{qs}$\;
  \eIf{average\_score $<$ $\theta_{qs}$} 
  {
    return $rejection=$ TRUE\;
  }{
   return $rejection=$ FALSE\;
  }
  \caption{Quality-Score-based Rejection \hyf{(QSR)}}\label{alg:rejc}
\end{algorithm}

\subsubsection{\omf{Early} Rejection Based on Chunk Mapping}
~\label{sec:reja}
\hys{The key idea of \omsix{the} \omsev{chunk-mapping-based} early rejection \omsix{technique} is that a read probably cannot be mapped to the reference genome if enough consecutive chunks in this read cannot be mapped to the reference genome (i.e., the chaining score of the minimizers in these chunks is lower than a threshold).} 
\hyf{Unfortunately, 
mapping short chunks provides \omsev{too large a} list of possible mapping locations.}
\hyf{To predict whether or not a read can be mapped to the reference genome, \omsix{our} technique needs larger chunk sizes.}

\omsix{W}e propose a chunk-mapping-based \omf{early} rejection mechanism, $CMR$, that is based on three key steps:
(1) $CMR$ basecalls a number of \omsev{(\omsix{$N_{cm}$})} continuous chunks.  
\hyf{(2) $CMR$ combines the $N_{cm}$ continuous chunks into a larger chunk (e.g., \omsix{by} combining five continuous 300-base chunks to \omsix{create} a \omsix{larger} 1500-base chunk).}
(3) $CMR$ maps the large chunk to the reference genome and checks the chaining score. 
If the chaining score is lower than a threshold $\theta_{cm}$ (\omf{indicating} that this chunk is significantly different from the reference genome), $CMR$ rejects the read \omf{and stops} basecalling it.
\omsix{We determine the value of $N_{cm}$ via a} one-time preprocessing \omsix{of} the reads of a species. \hyf{For example, we experimentally find that combining \omsix{five} consecutive \omsix{300-base} chunks can effectively \omsix{predicts} the \omsix{mapping} behavior of the read\omsix{s} \omsix{in the} E. coli dataset~\cite{ecoli} (see Section~\ref{sec:sensitivity} for more detail).}

\section{\name: \omsix{Architecture \&} Implementation}~\label{sec:hardware}
\begin{figure}[t]
    \centering
    \includegraphics[width=\linewidth]{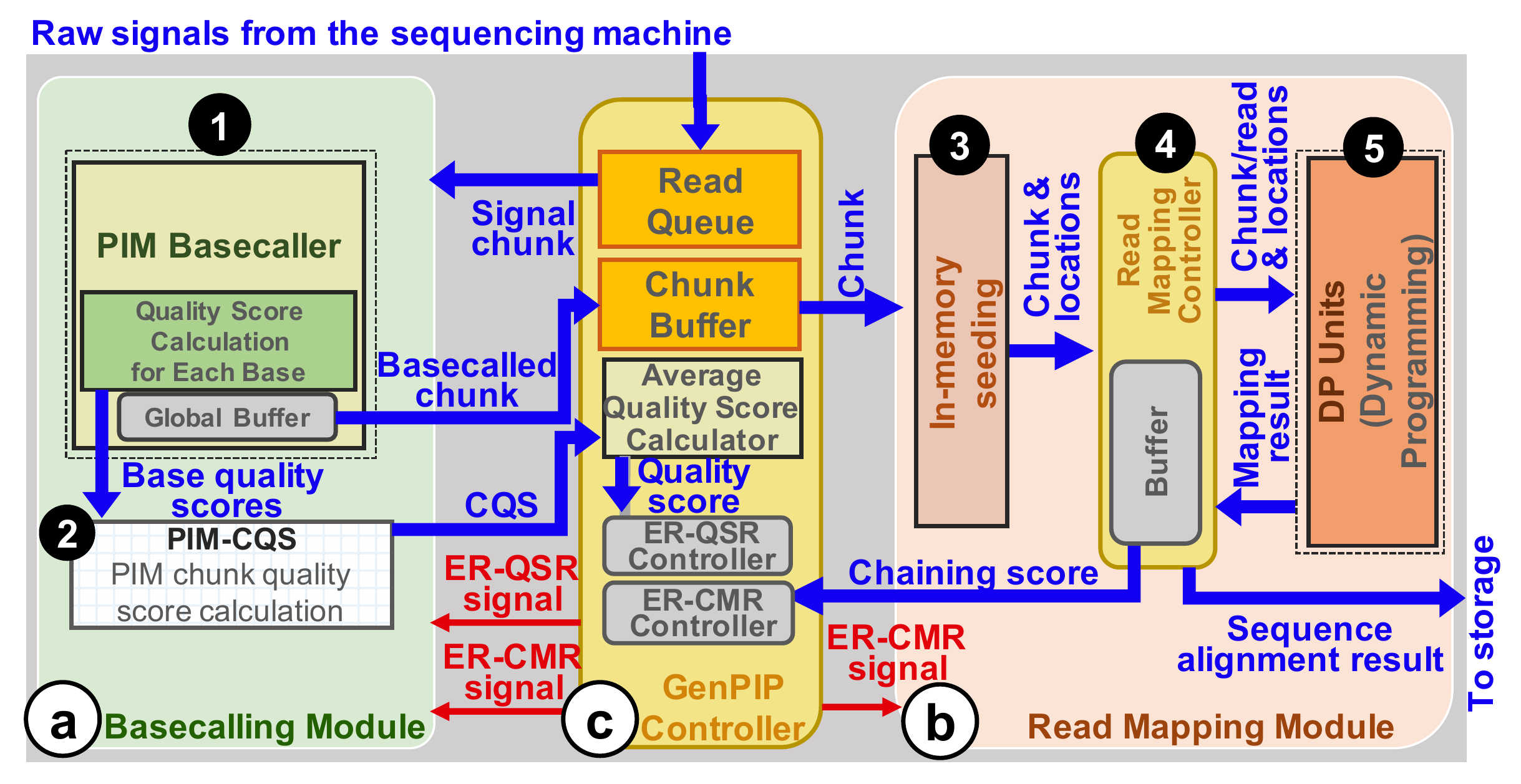}%
    \caption{Architecture overview of \name. \wcirc{a} The basecalling module. \wcirc{\scriptsize{b}} The read mapping module. \wcirc{c} The \name controller.}%
    \label{fig:overview}
\end{figure}

In this section, we describe the architecture and implementation of GenPIP. Figure~\ref{fig:overview} shows the overview of \omsix{the} \name architecture.
\hysix{\name architecture has three modules: the basecalling module (Figure~\ref{fig:overview}\circledletter{a}), the read mapping module (\circledletter{b}), and the \name controller (\circledletter{c}).
The basecalling module (\circledletter{a}) has two main units: 1) a PIM basecaller similar to prior work~\cite{lou2020helix} (\circled{1}) and 2) a new PIM accelerator for chunk quality score calculation (PIM-CQS \circled{2}). The read mapping module (\circledletter{b}) has three main units: 1) a new PIM accelerator for \omsev{the} seeding step (\circled{3}), 2) the read mapping controller (\circled{4}), and 3) dynamic programming units for chaining and alignment steps similar to prior work~\cite{chen2020parc}. The \name controller (\circledletter{c}) aims to 1) control the execution of \chuncpp and 2) issue early rejection commands using \earlyrej.} 
In this section, we first explain how \omsix{the} \name architecture implements \chuncpp and \chuncpp+~\earlyrej by providing a detailed walkthrough over \name components (Section~\ref{sec:DW}). We then explain the details of \name's new components in Sections~\ref{sec:controller}-\ref{sec:seeding}.

\subsection{Detailed Walkthrough}
\label{sec:DW}
\noindent\textbf{\omsix{Chunk-Based Pipeline (\chuncpp)} in \omsix{the} \name Architecture.} 
\omsev{We first describe the operation of \chuncpp without \earlyrej.} First, the \name controller (\circledletter{c}) receives raw electrical signals from the sequencing machine and stores these signals in the \emph{read queue}. 
Second, the \name controller sends the raw signals to \hysix{the PIM basecaller (\circled{1}) in} the basecalling module (\circledletter{a}) chunk by chunk. 
Third, the PIM basecaller translates each chunk into nucleotide bases using a deep neural network. 
For \omsix{the} PIM basecaller, \name uses a similar architecture as Helix~\cite{lou2020helix}, the state-of-the-art NVM-based PIM accelerator for basecalling. 
\hysix{The PIM basecaller performs the inference of basecaller's neural network via an NVM-based PIM array for matrix-vector computation (as described in Section~\ref{sec:accelerators}) and calculates the quality score for \emph{each base} after the inference.}
The PIM basecaller stores the basecalled chunks in \omsix{a} global buffer.  
Fourth, \hysix{after a chunk is basecalled}, PIM-CQS calculates the chunk quality score (CQS) \hysix{by \omsev{summing} the quality scores of the chunk's bases}.  
Fifth, the basecalling module sends the \hysix{basecalled} chunk together with its CQS to the \name controller. 

Sixth, the \name controller stores the basecalled chunk inside the \emph{chunk buffer}, and forwards the chunk to the read mapping module (\circledletter{b}).
Seventh, the read mapping module first performs the seeding operation on each basecalled chunk to \omsev{identify} the possible match locations in the reference genome. 
For \omsix{the} seeding step, we \omsev{design} a new PIM-based accelerator, \omsev{the} in-memory seeding component (\circled{3}), to enable fast and energy efficient seeding (\omsev{see} Section~\ref{sec:seeding}). 
The seeding component sends \hysix{a list of} the possible match locations to the read mapping controller (\circled{4}). 
Eighth, the read mapping controller sends the \omsev{chunk} and \omsev{its} possible match locations to the DP units to perform chaining (\circled{5}). 
For chaining, \name uses a similar design as PARC~\cite{chen2020parc}, the state-of-the-art NVM-based PIM accelerator for chaining. PARC customizes an NVM-based CAM array to implement a DP algorithm used for chaining and alignment. 
Ninth, the read mapping controller stores the chaining results in its \emph{buffer}. 
After finishing the chaining step for all chunks \omsev{in a read}, the read mapping controller \hysix{compares the chaining score of the entire read with a threshold ($\theta_{cm}$).} 
If the chaining score is lower than this threshold, the read mapping controller stops the read mapping \omsev{for this read}. 
Otherwise, the read mapping controller assembles the entire basecalled read and sends the read to DP units to execute the sequence alignment step. 
For sequence alignment, \name uses the same hardware units used for chaining while modifying its penalty score calculation, similar to PARC~\cite{chen2020parc}. 
\hysix{Tenth, the read mapping controller sends the mapping result to the storage after sequence alignment.}

\noindent\textbf{\chuncpp Working in Tandem with \earlyrej in \omsix{the} \name Architecture.} 
\omsev{We describe how \earlyrej is integrated \ome{with} \chuncpp.} As discussed in Section~\ref{sec:rejection}, \earlyrej includes two sub-techniques, \earlyrej-$QSR$ and \earlyrej-$CMR$. \name supports \earlyrej-$QSR$ using \hysix{\omsev{the} PIM-CQS unit in the basecalling module (\circledletter{a}\circled{2}) and the \name controller (\circledletter{c})}. 
The \hysix{PIM-CQS unit} (1) receives the first \hysix{$N_{qs}$} basecalled chunks \hysix{(depending on the dataset; see Section~\ref{sec:sensitivity}),} (2) calculates the quality score of the basecalled chunks, \hysix{and (3) sends the chunk quality scores to the \name controller (\circledletter{c})}.
\hysix{The \name controller (1) calculates the average quality score of the $N_{qs}$ chunks by using the average quality score calculator unit and (2) compares the average quality score with the threshold ($\theta_{qs}$) by using the \earlyrej-$QSR$ controller unit.} If the quality score is \omsix{lower} than the threshold (\omsix{i.e., if the read is predicted to be low-quality}), the \name controller sends \hysev{the ER--$QSR$ signal to the basecalling module} to terminate \omsev{basecalling on} the current read, and starts processing the next read. 
To support \earlyrej-$CMR$, the read mapping controller (\circled{4}) (1) combines $N_{cm}$ chunks (depending on the dataset; see Section~\ref{sec:sensitivity}) to \omsix{create} a larger chunk, (2) sends the \omsix{large} chunk for chaining, and (3) sends the chaining score of the \omsix{large} chunk to the \earlyrej-$CMR$ controller inside the \name controller. \earlyrej-$CMR$ controller compares the chaining score with the threshold ($\theta_{cm}$). If the chaining score is \hysix{lower} than the threshold (\omsix{i.e., if the read is predicted to be unmapped}), \hysev{the \name controller sends the ER-$CMR$ signal to the basecalling module and read mapping module to terminate \chuncpp processing for the current read}.

\subsection{ The \name Controller}~\label{sec:controller}
The \name controller (\circledletter{c}) (1) communicates with the basecalling module  (\circledletter{a}) and the read mapping module (\circledletter{b}) to control the chunk-based execution of \omsev{the} genome analysis pipeline, 
(2) issues early-rejection signals to basecalling and read mapping modules, and (3) merges the quality \omsev{scores} of basecalled chunks. 
The \name controller has \hysix{five key units\omsix{:} read queue, chunk buffer, average quality score calculator, \earlyrej-$QSR$ controller, and \earlyrej-$CMR$ controller. We explain each unit in \omsev{more} detail.}

\noindent\textbf{Read Queue.} 
The \name controller uses the read queue to store raw electrical signals. The sequencing machine enqueues raw electrical signals to this queue, and the controller dequeues them for \omsix{the} basecalling process. \name sizes this buffer as large as needed to store the longest signal (which is around 6 MB~\cite{amarasinghe2020opportunities,alser2022molecules}). \name 
can use different memory technologies to build the read queue. However, the memory technology should provide high write endurance, low read/write energy consumption, low read/write latency, and high density. We find \omsix{that} eDRAM~\cite{edram} is an example memory technology that provides a good tradeoff across \omsix{these} optimization goals. \omsix{Related work~\cite{lou2020helix,shafiee2016isaac}} also \omsix{uses} eDRAM-based buffers for the same reason.

\noindent\textbf{Chunk Buffer.} 
The \name controller uses the chunk buffer to store the basecalled chunks. The chunk buffer \omsix{keeps} the basecalled chunks until the end of sequence alignment \omsix{process for an entire read}, unless \earlyrej terminates the process of the read. 
In \name, the chunk buffer is able to house 2.3 million bases\omsix{, which} is the longest read length based on \ome{state-of-the-art} nanopore sequencing \omsev{technology}~\cite{amarasinghe2020opportunities}. \name uses \omsix{the} eDRAM technology for \omsix{the} chunk buffer\omsix{, for the same reasons as it does for the} read queue. 

\noindent\textbf{Average Quality Score (AQS) Calculator.} 
The \name controller uses the \emph{AQS} calculator unit to calculate the average quality score of \hysix{either} \omsix{an entire} read or \hysix{$N_{qs}$ chunks for \earlyrej-$QSR$}. The \emph{AQS} calculator unit has a buffer that keeps the sum of the quality scores of the chunks it has received so far. 
Once the \emph{AQS} unit receives all basecalled chunks \omsix{for the read}, it divides the final calculated sum by the total number of chunks to calculate the average quality score for the entire read. 

\noindent\textbf{\earlyrej-QSR Controller.} This unit receives the average quality score of $N_{qs}$ chunks from \omsev{the} \emph{AQS} calculator unit and compares it with the threshold ($\theta_{qs}$) to predict whether or not the read is low-quality. \hysix{If so, the \name controller issues \earlyrej-$QSR$ signal to the basecalling module to stop \chuncpp~\omsev{processing} for the predicted low-quality read.}



\noindent\textbf{\earlyrej-CMR Controller.} This unit receives the chaining score of a \emph{large} chunk (assembled with $N_{cm}$ chunks) from the read mapping module and compares it with the threshold ($\theta_{cm}$) to predict whether or not the read is unmapped. \hysix{If so, the \name controller issues \omsev{the} \earlyrej-$CMR$ signal to stop \chuncpp~\omsev{processing} for the \omsev{predicted-unmapped} read.}

\subsection{Timely Early Rejection}~\label{sec:scorecal}
This section explains how we implement \omsix{the} \earlyrej technique in \name to predict low-quality and unmapped reads \omsix{in a timely fashion}, and stop the execution of \omsix{the} \entire~on such reads. We describe the implementation of early rejection based on chunk quality scores (\earlyrej-$QSR$) and early rejection based on chunk mapping (\earlyrej-$CMR$) in Sections \ref{sec:qsr-imp} and \ref{sec:cmr-imp}, respectively.
\subsubsection{\earlyrej-QSR Implementation}
\label{sec:qsr-imp}
\omsix{As described in Section~\ref{sec:rejq}, t}he goal of $QSR$ is to calculate the quality score of a \omsix{small number of} sampled basecalled chunks ($N_{qs}$ \hysix{chunks}) and compare the result with the threshold of $QSR$ ($\theta_{qs}$). \name implements this technique partly in the basecalling module (\circledletter{a}) and partly in the \name controller (\circledletter{c}).  
Inside the basecalling module, we add a new unit, \hysix{PIM-CQS  (\circled{2}),} to calculate a chunk's quality score \omsix{by} \omsev{summing} the quality scores of its bases. 
\hysix{PIM-CQS} is an NVM-based PIM array \omsix{that performs the} MVM operation (as \omsix{described} in Section~\ref{sec:accelerators}).
\hysix{We use the PIM-CQS unit to perform the summation of the quality scores of the bases in a chunk by (1) storing the quality scores of bases in a column and (2) inputting} an all-1 vector so that a dot product becomes a simple addition. 
The basecalling module sends the \hysix{results (i.e., chunk quality scores)} to the \name controller. 
The \name controller calculates the average quality score of the sampled chunks and compares it with the threshold ($\theta_{qs}$). 
The \name controller \hysix{sends \omsev{the ER-$QSR$ signal} to the basecalling module to terminate basecalling on the current read} if the calculated average quality score is lower than $\theta_{qs}$.

\subsubsection{\earlyrej-CMR Implementation}
\label{sec:cmr-imp}
\omsix{As described in Section~\ref{sec:reja}, t}he goal of $CMR$ is to check the chaining score of a larger chunk, which is assembled \omsix{by} combining a \omsix{small number of} consecutive chunks \omsix{$N_{cm}$ chunks}, to estimate whether or not the entire read is unmapped. We implement $CMR$ partly inside the read mapping module (\circledletter{b}) and partly inside the \name controller (\circledletter{c}). 
Inside the read mapping module, the read mapping controller \hysix{(\circled{4}) enqueues the basecalled chunks in its buffer after the seeding step.} 
Once the controller has $N_{cm}$ chunks in the buffer, it assembles a larger chunk and sends the \omsix{large} chunk to the chaining step. The read mapping controller sends the chaining score to the \earlyrej-$CMR$ controller inside the \name controller. The \earlyrej-$CMR$ compares the chaining score of the \omsix{large} chunk with \omsix{the} $\theta_{cm}$ threshold. 
If the chaining score is lower than the threshold, the \name controller \hysix{sends \omsev{the ER-$CMR$} signal to the basecalling module and the read mapping module} to terminate the execution of \chuncpp~\hysix{on the current \omsev{predicted-unmapped} read}.

\subsection{\omsev{In-Memory} Seeding}~\label{sec:seeding}
\begin{figure}[!h]
    \centering
    \includegraphics[width=\linewidth]{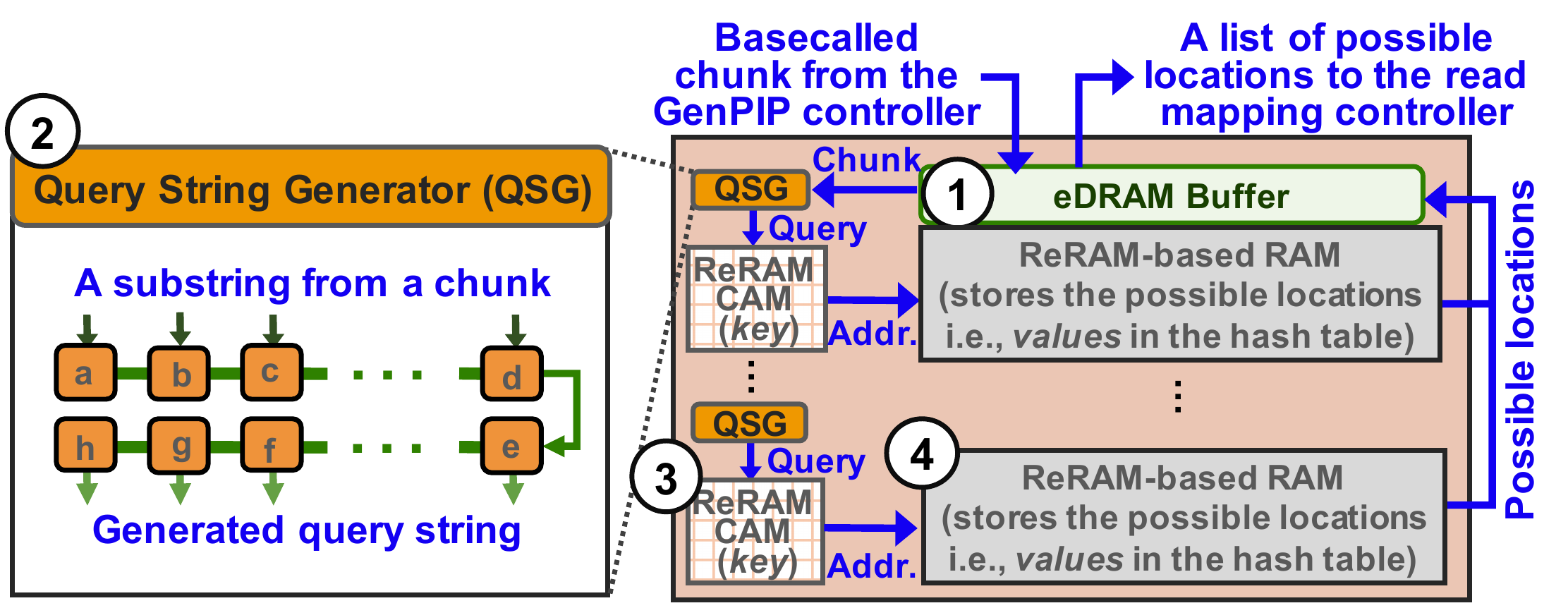}%
    \caption{Microarchitecture of the in-memory seeding \omsev{accelerator}.}%
    \label{fig:seeding}
\end{figure}

\omsix{As described in Section~\ref{sec:genomepipeline}, t}he seeding component aims to generate query strings (e.g., minimizers) from a basecalled chunk and queries them in the
hash table to quickly find matching regions between the reference genome and the chunk. 
\omsix{To this end, we design a new in-memory seeding accelerator \hysix{(Figure~\ref{fig:overview}\circled{3}} to speed up the process of seeding so that it can keep up with other components of \name.} 
\hysix{Figure~\ref{fig:seeding} shows the components of the \omsev{in-memory seeding accelerator.}} \omsev{There are} four main components\omsix{:} an eDRAM buffer (Figure~\ref{fig:seeding}\circledletter{1}), the query string generator (\circledletter{2}),  ReRAM-based CAM (\circledletter{3}) and RAM arrays (\circledletter{4}). 
First, the \name controller writes a basecalled chunk into the seeding unit's eDRAM buffer (\circledletter{1}). 
Second, the seeding unit moves a substring from the chunk to the query string generator (\circledletter{2}). 
Third, the query string generator uses each substring to generate multiple query strings \omsix{by} shifting the substring one base at a time. 
Fourth, the seeding module queries each query string using \omsev{the} ReRAM-based CAM and RAM arrays. \name stores multiple reference strings inside the CAM array (as the \emph{keys}, \circledletter{3}), and the locations of the reference strings in the reference genome inside the ReRAM-based RAM \omsix{array} (as the \emph{values}, \circledletter{4}). 
The implementation of ReRAM-based CAM \omsix{array} is similar to what we explain in Section~\ref{sec:accelerators}. If the query string matches one reference string in the ReRAM-based CAM, \hysev{the ReRAM-based CAM outputs the address ($Addr.$) to access  the corresponding values (i.e., the possible match locations) stored inside the ReRAM-based RAM.}
The seeding unit then reads out the list of possible match locations in the reference genome for that particular reference string and stores the possible locations in the eDRAM buffer. 
Fifth, the seeding unit outputs the possible match locations of the chunk to the read mapping controller unit. 

\section{Evaluation Methodology}
\label{sec:evaluation_setup}
\noindent\head{Performance, Power, and Area Analysis}
We implement an in-house simulator to evaluate the performance, energy consumption, and \omsix{handware} area \omsix{overhead} of \name. Since \name includes several components, we embed the latency, power, and area values for each \name component in our simulator. To calculate these values, we use different tools depending on the technology of the component.
We use Verilog HDL to implement the logic components in \name. To estimate the area and power consumption of logic components, we synthesize our HDL implementation using the Synopsys Design Compiler~\cite{kurup2012logic} with a 32nm process technology node at 1.6 GHz clock frequency. To model the performance, energy, and area of our ReRAM-based RAM and CAM arrays, we use the state-of-the-art models for non-volatile memories, NVSim~\cite{dong2012nvsim} and NVSim-CAM~\cite{li2016nvsim}, respectively.
We use CACTI 6.5~\cite{muralimanohar2007optimizing} to model the performance, energy, and area of the embedded DRAM (eDRAM).
For Helix and PARC accelerators, we use the performance, power and area results reported in the original works~\cite{lou2020helix, chen2020parc}.

\head{Comparison points}
Our goal is to 1)~compare \name with the state-of-the-art CPU/GPU implementations \omsix{of software genome analysis tools} and \omsev{state-of-the-art} PIM accelerators, and 2)~show the benefits of integrating the key mechanisms of \name (\chuncpp and \earlyrej) to these CPU/GPU implementations \omsix{and accelerators}. To this end, we evaluate the following systems:

\begin{itemize}[leftmargin=0pt, itemindent=10pt, itemsep=0pt]
    \item \texttt{CPU:} We use the CPU-based state-of-the-art basecaller, Bonito~\cite{bonito}, and the CPU-based read mapper, minimap2~\cite{li2018minimap2} executed on an Intel® Xeon® Gold 5118 \omsev{CPU~\cite{cpu}} at 2.3 GHz, with 192 GB DDR4 memory.
    \item \texttt{CPU-\chuncpp:} \texttt{CPU} integrated with \chuncpp (\omsix{the} chunk-based pipeline technique we describe in Section~\ref{sec:pipeline}). 
    \item \texttt{CPU-GP:} \texttt{CPU} integrated with both \chuncpp and \earlyrej (GP stands for \name).
    \item \texttt{GPU:} We use the GPU implementation of Bonito~\cite{bonito} as \omsix{the} basecaller executed on an NVIDIA GeForce RTX 2080 Ti GPU and the CPU implementation of minimap2 as \omsix{the} read mapper.
    \item \texttt{GPU-\chuncpp:} \texttt{GPU} integrated with \chuncpp.
    \item \texttt{GPU-GP:} \texttt{GPU} integrated with both \chuncpp and \earlyrej.
    \item \texttt{PIM:} To compare to a single PIM-based accelerator that executes both basecalling and read mapping steps, we connect two state-of-the-art PIM-based accelerators for 1)~basecalling, Helix~\cite{lou2020helix}, and 2)~read mapping, PARC~\cite{chen2020parc}. We \emph{optimistically} make the following three assumptions. (1) There is no latency and energy overhead for data movement between \omsix{the} basecalling and read mapping steps when connecting these accelerators. (2) There are processing elements executing the read quality control step without any \omsix{performance} overhead. (3) There is enough memory to store the intermediate data.
    \item \texttt{\name-\chuncpp:} Our \name design \emph{only} equipped with \chuncpp. 
    \item \texttt{\name-\chuncpp-QSR:} Our \name design with both \chuncpp and \emph{only} $QSR$ in \earlyrej. 
    \item \texttt{\name:} The \omsix{full} \name design using  both \chuncpp and \earlyrej~\omsix{(i.e., both $QSR$ and $CMR$)}. 
\end{itemize}

\head{Datasets}
Table~\ref{table:dataset} shows the details of the datasets we use in our evaluations. For all of our experiments, we evaluate \name using the datasets that are representatives of small and large genomes to \omsix{cover} the commonly used genome sizes in genome analysis. As a small genome, we use a publicly available dataset of the Escherichia coli (\emph{E. coli}) genome.\footnote{The \emph{E. coli} dataset is available at: \url{http://lab.loman.net/2016/07/30/nanopore-r9-data-release/}} As a large genome, we use a human genome dataset of the NA12878 sample. The human dataset can be accessed through ENA~\cite{ena} or NCBI~\cite{ncbi} with accession PRJEB30620. Both \emph{E. coli} and human genomes are sequenced using Oxford Nanopore Technologies (ONT) with R9-based chemistry~\cite{jain2017minion}. This chemistry provides sequencing data with around 80-85\% sequencing accuracy~\cite{senol2017nanopore}, which is slightly lower than the most recent chemistry (R10.4) that provides around 95-99\% accuracy~\cite{sereika2022oxford}. We include these less accurate datasets in our experiments to show the robustness \omsix{and effectiveness} of \name~\omsix{in the presence of considerable} sequencing \omsix{in}accuracy.

\begin{table}[h!]
\small
\begin{center}
\vspace{1em}
\caption{Details of datasets used in the evaluation.}\label{table:dataset}
\vspace{-0.5em}
\begin{tabular}{@{}|l|r|r|@{}}\hline
\textbf{Dataset Details} & \textbf{\emph{E. coli}~\cite{ecoli_dataset}} & \textbf{\emph{Human}~\cite{human_dataset}}\\\hline \hline
\textbf{Mean read length}    & 9,005.90                            & 5,738.30                            \\ \hline
\textbf{Mean read quality}   & 7.9                                 & 11.3                                \\ \hline
\textbf{Median read length}  & 8,652                               & 6,124                               \\ \hline
\textbf{Median read quality} & 9.3                                 & 12.1                                \\ \hline
\textbf{Number of reads}     & 58,221                              & 449,212                             \\ \hline
\textbf{Total bases}         & 524,330,535                         & 2,577,692,011                       \\ \hline
\end{tabular}
\vspace{-1em}
\end{center}
\end{table}

\section{Results}~\label{sec:result}
In this section, we present the experimental results of \name, including 1)~the performance of \name compared to the baseline systems \omsix{(Section~\ref{sec:performance})}, 2)~the energy \omsev{consumption} of \name compared to the baseline systems \omsix{(Section~\ref{sec:energy})}, 3)~the sensitivity analysis of \name ~\omsix{(Section~\ref{sec:sensitivity})}, and 4)~the area and power analysis of the \name architecture \omsix{(Section~\ref{sec:area})}.

\subsection{Performance Analysis}\label{sec:performance}
\begin{figure}[h!]
    \centering
    \includegraphics[width=\linewidth]{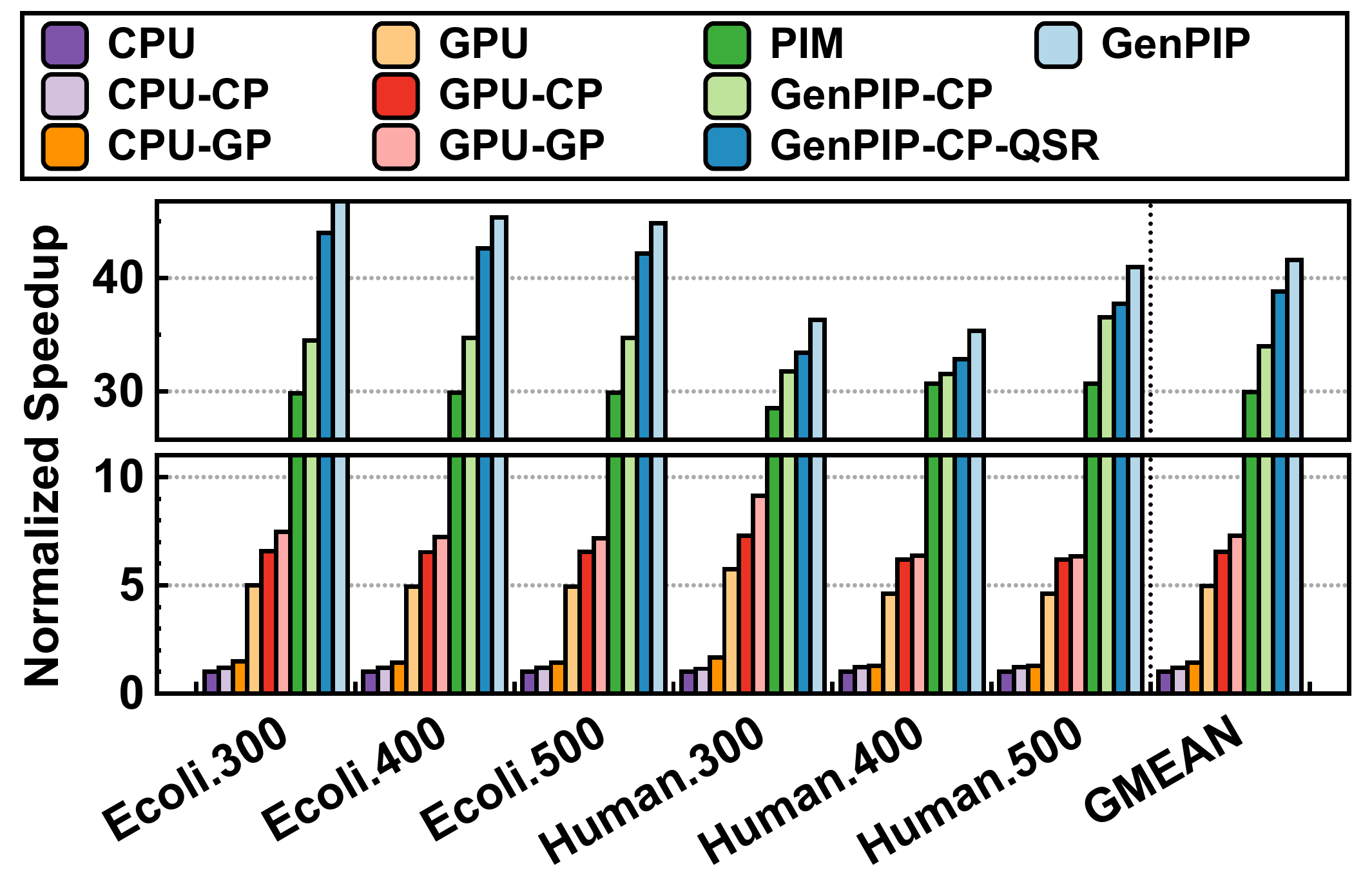}
    \caption{\omsix{Speedups of various systems normalized to \texttt{CPU}} (300, 400, and 500 in the x-axis represent the three chunk sizes used in the evaluation).}
    \label{fig:speedup}
\end{figure}

To study the effect of \name on accelerating \omsix{the} genome analysis pipeline, we measure the performance \omsix{of} \name and the baseline systems. Figure~\ref{fig:speedup} shows the performance of \name compared to the CPU, GPU, and PIM-based systems that we explain in Section~\ref{sec:evaluation_setup} (results are normalized to the performance of the \texttt{CPU} system).
We use chunk \omsix{sizes} of 300, 400, and 500 bases in the evaluation \omsix{of} two datasets, E. coli and human (300 is the suggested \omsix{chunk size} by state-of-the-art basecallers~\cite{bonito,farnaes2018rapid}).

We make four key observations. 
First, \name provides $41.6\times$, $8.4\times$, and $1.39\times$ \omsix{speedup} compared to the \texttt{CPU}, \texttt{GPU}, and \texttt{PIM} systems, respectively. 
\name~\omsix{achieves} such speedups as it 1)~efficiently enables the fine-grained collaboration of the basecalling and \omsix{the} read mapping steps \hyfv{via \omsix{the} \chuncpp technique and 2)~reduces useless computation via \omsix{the} \earlyrej technique.} 
Second, we observe \omsix{that} the \texttt{\name-\chuncpp}, \texttt{\name-\chuncpp-QSR}, and \texttt{\name} systems \omsix{provide} $1.16\times$, $1.32\times$, and $1.39\times$ \omsix{speedup} compared to the \omsix{idealized} PIM-based accelerator (\texttt{PIM}) that integrates the state-of-the-art basecalling and read mapping accelerators with optimistic assumptions. These speedups \omsix{identify} the main benefits of the key mechanisms of \name that tightly integrate the basecalling and read mapping steps rather than simply connecting two PIM-based accelerators that perform basecalling and read mapping separately \omsix{(even with idealized assumptions, as we did for \texttt{PIM})}.
Third, \hyfv{1)~\texttt{CPU-\chuncpp} and~\texttt{CPU-GP} provide  $1.20\times$ and $1.42\times$ \omsix{speedup} compared to \texttt{CPU}, and 2)~\texttt{GPU-\chuncpp} and \texttt{GPU-GP} provide $1.32\times$ and  $1.46\times$ \omsix{speedup} compared to \texttt{GPU}.}
Implementing the \chuncpp and \earlyrej techniques significantly improves performance in CPUs and GPUs as these techniques are effective \omsev{at} reducing data movement and useless computation \omsix{in} any system.
Fourth, \omsix{\name's} performance \omsix{benefits do} not change significantly \omsev{with chunk size}.
We conclude that 1)~\chuncpp and \earlyrej techniques significantly improve the overall performance of genome analysis over the state-of-the-art CPU- and GPU-based approaches, and 2)~\name~\omsix{outperforms} the optimistic integration of the state-of-the-art PIM-based read mapping and basecalling accelerators.

\subsection{Energy Efficiency Analysis}~\label{sec:energy}
\begin{figure}[h!]
    \centering
    \includegraphics[width=\linewidth]{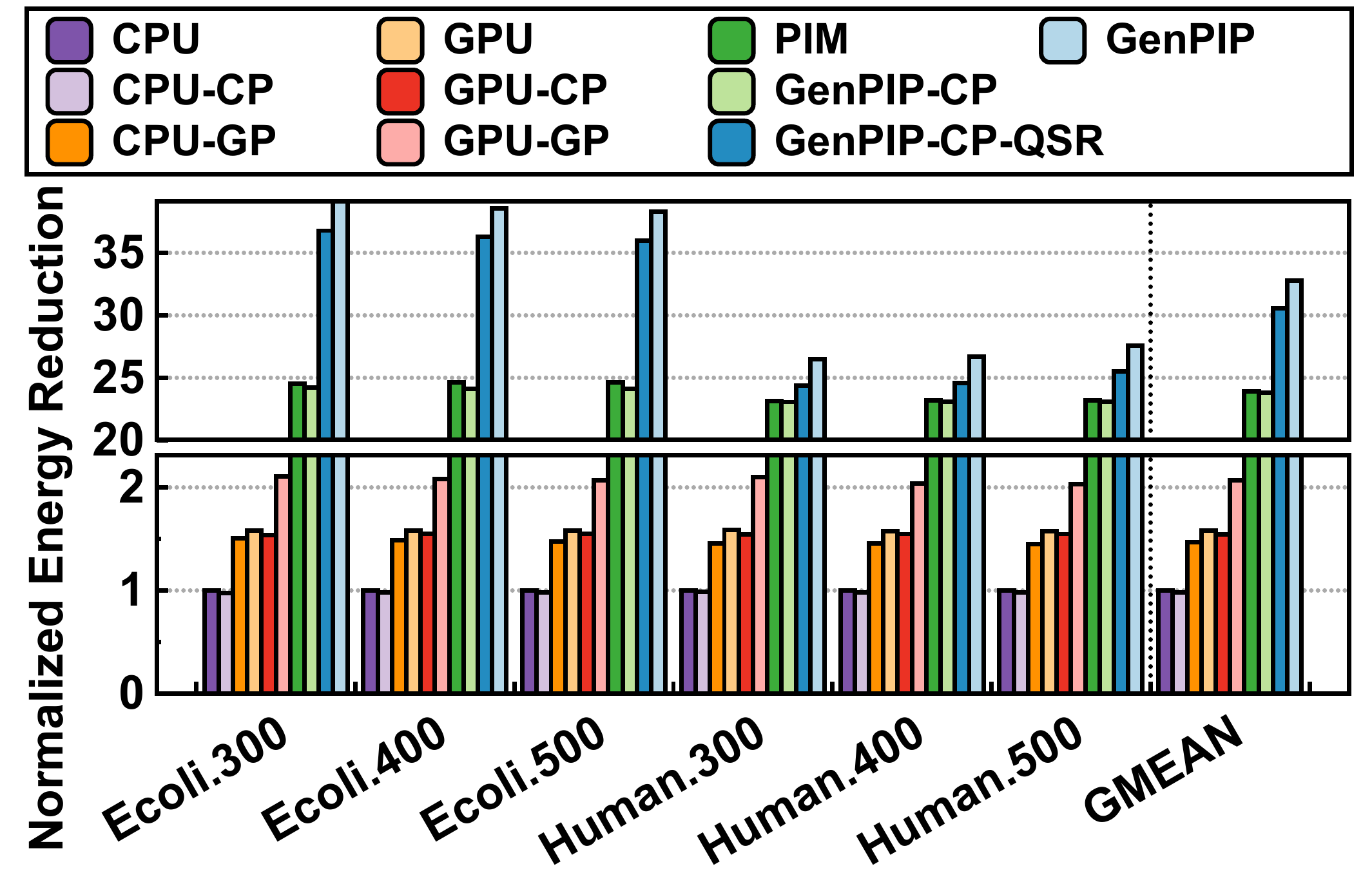}
    \caption{\omsix{Energy reduction of various systems normalized to \texttt{CPU}} (300, 400, and 500 in the x-axis represent the three chunk sizes used in the evaluation).}
    \label{fig:energy}
\end{figure}

To study the energy efficiency of \name, we measure the energy consumption of \name and the baseline systems. 
Figure~\ref{fig:energy} shows the energy savings of each evaluated system normalized to the energy consumption of the \texttt{CPU} system. We make three key observations. 
First, \name~\omsix{provides} $32.8\times$, $20.8\times$, and $1.37\times$ energy \omsix{reduction}, compared to \texttt{CPU}, \texttt{GPU}, and \texttt{PIM} systems, respectively. These energy savings are in line with the performance improvements that \name provides by reducing \hyfv{1) the data movement between the basecalling and read mapping steps} \hyfv{and 2)}~the useless computation \omsix{due to} the low-quality reads and unmapped reads. 
\hyfv{Second}, \name~\omsix{reduces energy by} $1.07\times$ and $1.37\times$ than \texttt{\name-\chuncpp-QSR} and \texttt{\name-\chuncpp}, which shows that filtering based on \emph{both} read quality score and chunk mapping is \omsix{important} to improve the overall energy savings of \name. 
Third, similar to the performance results, the energy consumption of the evaluated systems is robust to chunk sizes. \omsev{We conclude that \name is very effective at reducing energy compared to state-of-the-art CPU, GPU, and PIM systems.}

\subsection{Sensitivity Analysis}~\label{sec:sensitivity}
In this section, we study the sensitivity of the number of \omsix{sampled} chunks on \omsix{the} effectiveness of \earlyrej-$QSR$ (Section~\ref{sec:qsr-analysis}) and \earlyrej-$CMR$ (Section~\ref{sec:cmr-analysis}). To this end, we calculate two metrics\omsix{:} \emph{rejection ratio} and \emph{false negative ratio}. \omsev{\emph{Rejection ratio}} is ratio of rejected reads (via \earlyrej-$QSR$ or \earlyrej-$CMR$) over all reads. \omsev{\emph{False negative ratio}} is the ratio of \omsix{\emph{incorrectly}} rejected reads over all rejected reads. 

\subsubsection{Effect of \omsev{the} Number of \omsix{Sampled} Chunks on \earlyrej-QSR}
\label{sec:qsr-analysis}
\begin{figure}[h]
    \centering
    \vspace{0em}
    \includegraphics[width=\linewidth]{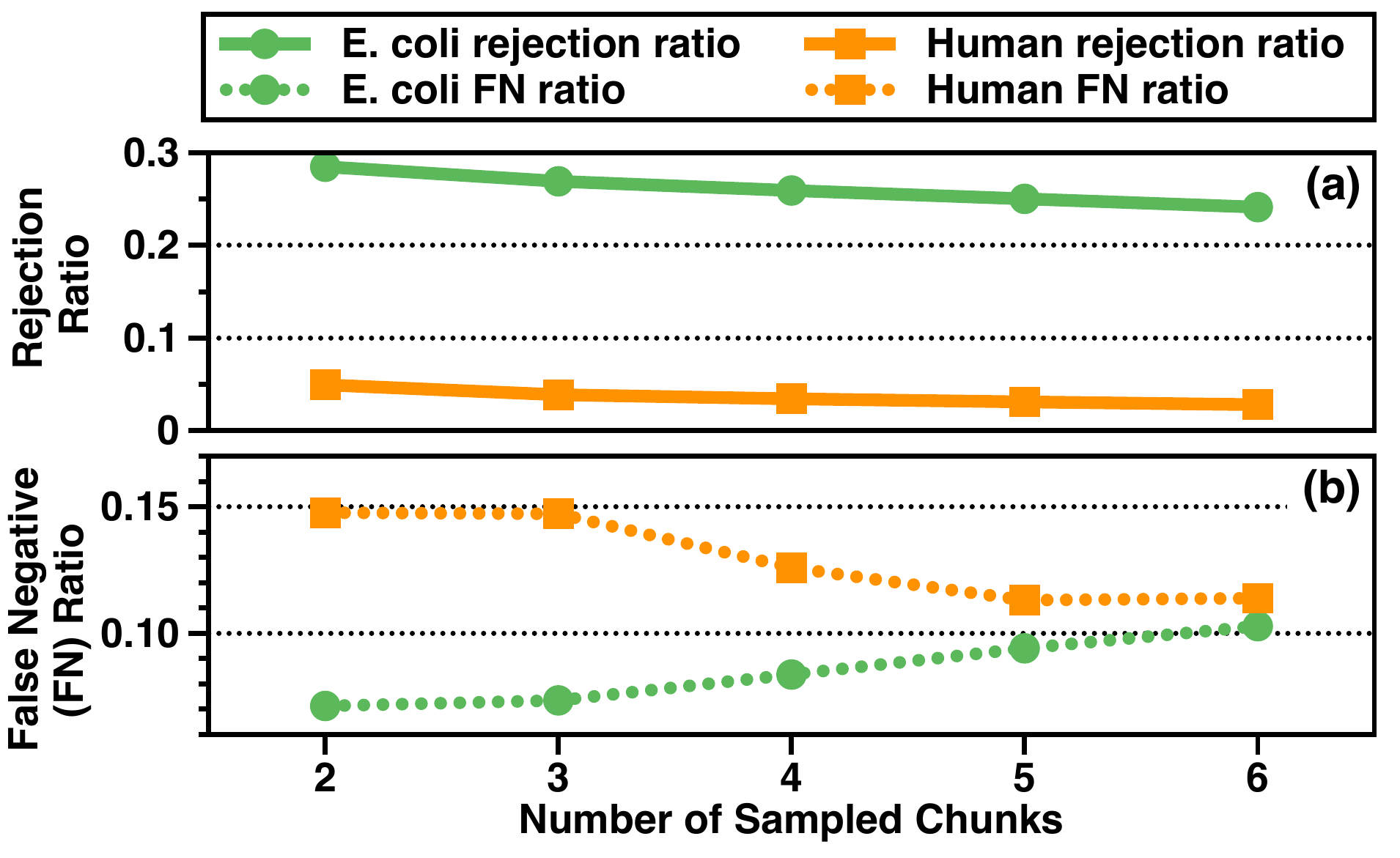}
    \caption{Effect of the number of sampled chunks on \earlyrej-QSR's (a) rejection ratio and (b) false negative ratio.}
    \label{fig:qs}
\end{figure}

To study the effect of \omsix{the} number of \omsix{sampled} chunks on the effectiveness of \earlyrej-$QSR$, we calculate the \emph{rejection ratio} and \emph{false negative ratio} metrics while varying the number of \omsix{sampled} chunks from 2 to 6.  We identify a rejection as false negative (FN) if \earlyrej-$QSR$ rejects the read while the average read quality score of the \emph{entire} read is above the read quality score threshold. Figure~\ref{fig:qs}(a-b) shows \omsix{the} rejection ratio and \omsix{the} false negative ratio for \earlyrej-$QSR$ using the E. coli and human datasets, respectively. We make three key observations. First, the rejection ratio slightly decreases \omsix{as} the number of \omsix{sampled} chunks \omsix{increases} for both \omsix{the} E. coli and \omsix{the} human datasets. This is because there are \hysix{many} short reads in both datasets that consist of \omsix{only} a few chunks (e.g., 3 chunks); \omsix{increasing} the number of sampled chunks reduces the likelihood of early rejection of such short reads. 
Second, increasing the number of sampled chunks \omsix{decreases the} false negative ratio for \omsix{the} human dataset \omsix{but increases the} false negative ratio for the E. coli dataset. 
\hysev{For \omsev{the} human dataset, increasing the number of sampled chunks provides better read quality prediction accuracy, which leads to \omsev{a} lower false negative ratio. For the E. coli dataset, there are many regions with low-quality chunks although the average quality of reads is high. Using \ome{more} sampled chunks leads to using more of these low-quality chunks in the read quality prediction, which is the main cause of \ome{the} false negative predictions \ome{in this dataset}.}
Third, the false negative ratio of the human dataset is slightly larger than that of the E. coli dataset. 
Such large FN ratios are still acceptable since the average alignment score of these incorrectly-rejected reads ($14.4$) is closer to the average alignment score of low-quality reads ($3.9$) than the average alignment score of all reads ($52.5$).
Thus, these incorrectly-rejected reads are unlikely to provide high-quality mapping in the read mapping \dc{step}~\cite{li2018minimap2}.

Based on our sensitivity analysis, we use two and five sampled chunks for the E. coli and human datasets, respectively, which provides a good balance between achieving a high rejection ratio and achieving a low false negative ratio. 

\subsubsection{Effect of \omsev{the} Number of \omsix{Sampled} Chunks on \earlyrej-CMR}
\label{sec:cmr-analysis}
\begin{figure}[h!]
    \centering
    \includegraphics[width=\linewidth]{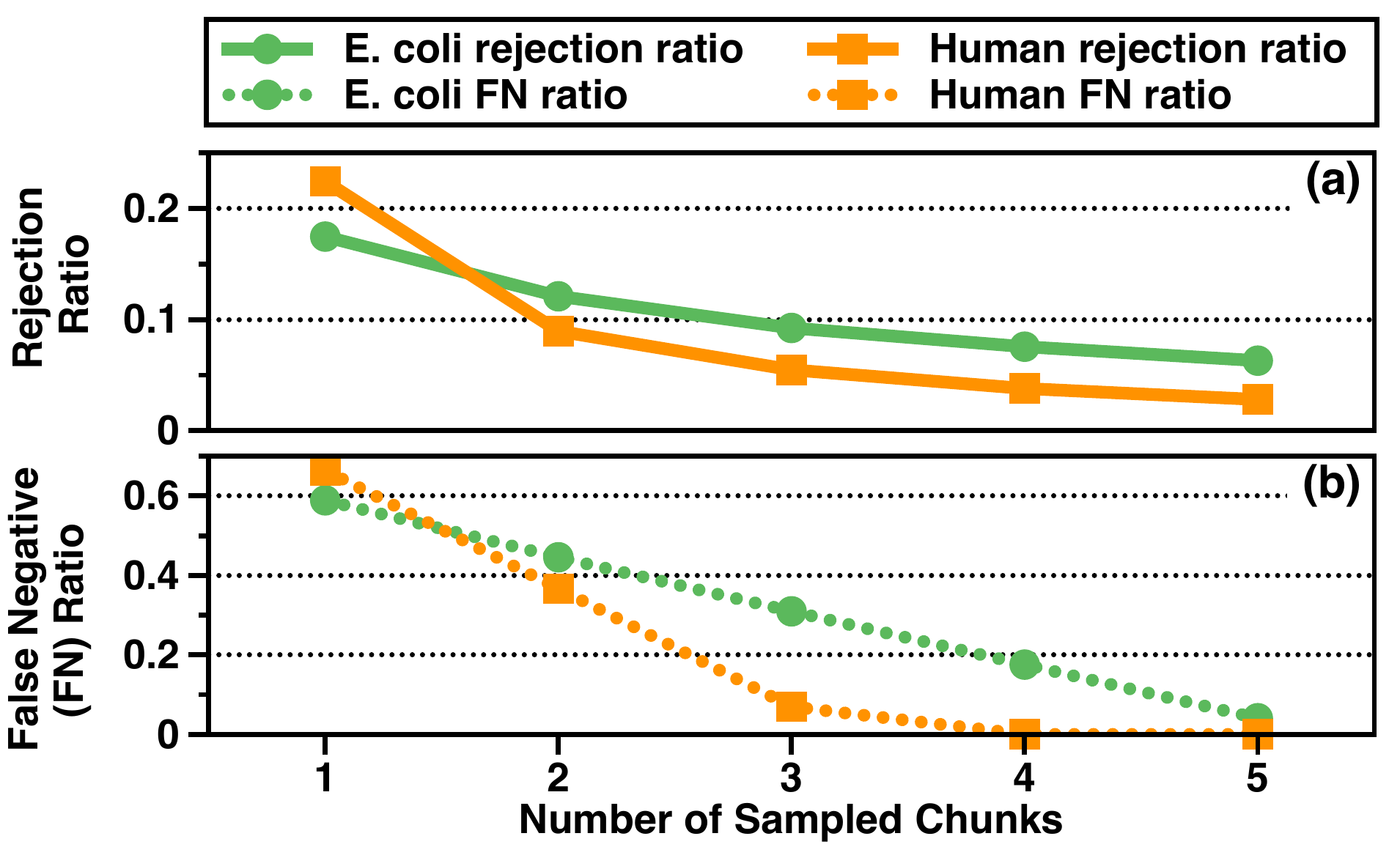}
    \vspace{-1.5em}
    \caption{Effect of the number of sampled chunks on
ER-CMR’s (a) rejection ratio and (b) false negative ratio.}
    \label{fig:cmrrfn}
\end{figure}

To study the effect of \omsix{the} number of \omsix{sampled} chunks on the effectiveness of \earlyrej-$CMR$, we calculate the \emph{rejection ratio} and \emph{false negative ratio} metrics while varying the number of \omsix{sampled} chunks from 1 to 5. We identify a rejection as FN if the read rejected by \earlyrej-$CMR$ (predicted as unmapped) can be mapped to the reference genome. Figure~\ref{fig:cmrrfn}(a-b) shows \omsix{the} rejection ratio and \omsix{the} false negative ratio for \earlyrej-$CMR$ using the E.coli and human datasets, \omsix{respectively}. 
We make two key observations.
First, the rejection ratio decreases \omsix{as} the number of \omsix{sampled} chunks \omsix{increases} for both \omsix{the} E. coli and \omsix{the} human datasets. 
This is due to two main reasons.
1) There are \hysix{many} short reads in both datasets that consist of a few chunks (e.g., 3 chunks). Increasing the number of sampled chunks reduces the likelihood of early rejection of such short reads.
2) Increasing the number of \omsix{sampled} chunks increases the accuracy of \earlyrej-$CMR$, \omsix{which leads to the rejection of} fewer reads. 
Second, the false negative ratio decreases \omsix{as} the number of sampled chunks \omsix{increases} for both \omsix{the} E. coli and  \omsix{the} human datasets. This is because using a larger number of \omsix{sampled} \omsix{chunks} results in a larger assembled chunk that is likely more representative of the entire read.

Based on our sensitivity analysis, we use five and three samples for the E. coli and human datasets, respectively, because the false negative ratios \omsix{they provide} are close to zero while the rejection ratios are \omsix{reasonable (i.e., $6.3\%$ and $5.5\%$, respectively)}. 

\subsection{Area and Power Analysis}
\label{sec:area}
\hyfv{To study the area and power \omsix{overheads} of \name, we 1)~measure the area and power of the new components designed for \name (Figure~\ref{fig:overview}\circled{2}\circled{3}\circled{4}\omsev{\circledletter{c}}) and 2)~use the \omsix{area and power} data of \omsix{other} components \omsix{as reported in previous \omsev{works}} (\circled{1}~\cite{lou2020helix}, \circled{5}~\cite{chen2020parc}).}
Table~\ref{table:configuration} shows the area and power breakdown of \hyfv{\name's components in three modules, the basecalling module (\circledletter{a}), the read mapping module (\circledletter{b}), and the \name controller (\circledletter{c})}. 
\hyfv{\name occupies 163.8 $mm^2$ chip area and consumes 147.2W power \omsix{at \omsev{the} 32nm technology node}. Our analysis shows that the read mapping module is the most expensive \hysix{module} in terms of area and power consumption, as it accounts for 56.9\% \omsev{of \ome{the} \name total area} and 77.8\% of \omsev{\ome{the} \name total} power consumption. 
}

\vspace{4mm}

\begin{table}[h!]
\begin{center}
\caption{Area and power breakdown of \name.}
\vspace{-2mm}
\label{table:configuration}
  \resizebox{1\linewidth}{!}{
\begin{tabular}{|c|c||c|c|}
\hline
  \textbf{Component} & \textbf{Specification} & \textbf{Power W} & \textbf{Area mm$^2$} \\ \hline\hline

\multirow{2}{*}{PIM Basecaller \ding{202}} & 168 Tiles  & \multirow{2}{*}{27.1}    & \multirow{2}{*}{49.2}  \\ 
& 4MB eDRAM &  &  \\ \hline
\multirow{2}{*}{PIM-CQS \ding{203}} & SOT-MRAM PIM  &  \multirow{2}{*}{0.307}  &  \multirow{2}{*}{ 0.0256}  \\
 & Array size: 16x1024  &     &     \\ \hline
\multicolumn{2}{|c||}{\textbf{Basecalling Module \raisebox{.5pt}{\textcircled{\raisebox{-.5pt} {a}}} Total} }                & \textbf{27.4}     & \textbf{49.2}     \\ \hline
\hline
\multirow{6}{*}{Seeding \ding{204}}                 & 4096 seeding units           &   \multirow{6}{*}{28.2}           &   \multirow{6}{*}{76.68}        \\
  & ReRAM-based CAM-RAM           &              &             \\ 
  & 8 $32 \times 128$ CAMs per unit           &              &             \\
  & 1 QSG per CAM           &                &              \\
  & 8 16KB RAMs per unit           &              &             \\ 
  & 1 4KB eDRAM per unit           &                &              \\ \hline
 RMC \ding{205}                 & 4 MB eDRAM           &   1.346             &   5.472      \\  \hline
DP \ding{206}       & 1024 units  & 85     & 10.9       \\ \hline
\multicolumn{2}{|c||}{\textbf{Read Mapping Module \raisebox{.5pt}{\textcircled{\raisebox{-.5pt} {b}}} Total} }              & \textbf{114.5}    & \textbf{93.1}      \\ \hline
\hline

\multirow{4}{*}{\textbf{Module \raisebox{.5pt}{\textcircled{\raisebox{-.5pt} {c}}}}}      & 12 MB eDRAM  &  \multirow{5}{*}{\textbf{5.3}}   &   \multirow{5}{*}{ \textbf{21.5}}    \\ 
 & AQS calculator  &     &        \\
 & ER-QSR controller  &     &        \\
 & ER-CMR controller  &     &        \\ \cline{1-2}
\multicolumn{2}{|c||}{\textbf{\name Controller Module \raisebox{.5pt}{\textcircled{\raisebox{-.5pt} {c}}} Total}  }         &    &       \\ \hline \hline
\multicolumn{2}{|c||}{\textbf{\name Total}  }             & \textbf{147.2}    & \textbf{163.8}    \\ \hline
\end{tabular}
}
\end{center}
\end{table}



\section{Related Work} ~\label{sec:related}
To our knowledge, \name is the \emph{first} \omsix{processing-in-memory} (PIM) accelerator for the genome analysis pipeline that tightly integrates the two key steps of genome analysis (basecalling and read mapping) to minimize 1)~the data movement by eliminating the need to store intermediate results and 2)~useless computation \omsix{due to} low-quality and unmapped reads. We have already compared \name extensively to the state-of-the-art CPU-based, GPU-based, and PIM-based systems in Section~\ref{sec:result}. 
In this section, we describe other related works in four categories: (1)~PIM acceleration of genome analysis, (2)~non-PIM acceleration of basecalling, (3)~non-PIM acceleration of read mapping, and (4)~\omsix{basecalling-free genome analysis}.

\head{PIM Acceleration of Genome Analysis} Previous 
PIM works focus on the acceleration of either basecalling~\cite{lou2020helix,lou2018brawl} or read mapping~\cite{zokaee2018aligner, gupta2019rapid, chen2020parc, khatamifard2021genvom,kaplan2017resistive,kaplan2018rassa,zokaee2019finder,angizi2020exploring,angizi2020pim,kaplan2020bioseal,laguna2020seed, kim2018grim, mansouri2022genstore, khalifa_filtpim_2021, chowdhury_dna_2020, huangfu_radar_2018, diab2022high, diab2022framework, shahroodi2022demeter, li2021pim, angizi2019aligns}. 
For basecalling, previous PIM accelerators~\cite{lou2020helix, lou2018brawl} accelerate the neural networks of basecallers using non-volatile memory. These accelerators can significantly reduce the performance and energy overheads associated with frequently moving the data for neural networks by implementing these neural networks in-memory~\cite{shafiee2016isaac}. 
For read mapping, PIM accelerators~\cite{zokaee2018aligner, gupta2019rapid, chen2020parc, khatamifard2021genvom, kaplan2017resistive, kaplan2018rassa, zokaee2019finder, angizi2020exploring, angizi2020pim, kaplan2020bioseal, laguna2020seed, kim2018grim, mansouri2022genstore, khalifa_filtpim_2021, chowdhury_dna_2020, huangfu_radar_2018, diab2022high, diab2022framework, shahroodi2022demeter, li2021pim, angizi2019aligns} accelerate several computationally-costly steps in read mapping (e.g., chaining and sequence alignment). 
\omsix{Many of these works \omsix{provide} \omsix{large-scale in-memory} parallelism while reducing the data movement overheads of mapping reads to a reference genome.}

\omsix{T}hese works suffer from two main issues. 
First, none of these PIM accelerators are designed to accelerate \omsev{\emph{both}} basecalling and read mapping, which requires storing and moving \omsix{a large amount of data (long reads)} after the basecalling step instead of streaming these reads directly to the read mapping step \omsix{in a pipelined manner that maximizes concurrency}. 
\omsix{Second, even though read quality control filters out the low-quality reads, these reads have already
been processed by the expensive basecalling step (because basecalling happens earlier in a separate PIM accelerator).}
Compared to these existing approaches, \name effectively and efficiently orchestrates basecalling and read mapping steps to \omsix{1)~reduce the data movement overhead between the basecalling and read mapping steps and enable fine-grained overlapping between these two steps, and 2)}~eliminate the redundant computations in both basecalling and read mapping by \omsev{quickly} rejecting low-quality reads and unmapped reads.

\head{Non-PIM Acceleration of Basecalling} SquiggleFilter~\cite{dunn2021squigglefilter} accelerates the basecalling step by filtering raw \omsix{electrical} signals before basecalling based on their similarity to a \emph{certain} genome (e.g., a viral DNA). SquiggleFilter~\cite{dunn2021squigglefilter} targets \omsix{a metagenomics use case}, where there are large numbers of reads from different species.
Many prior works accelerate the basecalling step by implementing the basecaller using GPUs \omsix{(e.g.,~\cite{xu2021fast, Guppy, bonito, perevsini2021nanopore, lv_end--end_2020, zeng_causalcall_2020, yeh_msrcall_2022, huang_sacall_2022, konishi_halcyon_2021, boza_deepnano_2017}) and FPGAs (e.g.,~\cite{ramachandra_ont-x_2021, hammad_scalable_2021, wu_fpga_2022, wu_fpga-accelerated_2020})}. 

\head{Non-PIM Acceleration of Read Mapping} There are several works that focus on accelerating different steps of read mapping, such as pre-alignment filtering \omsix{(e.g.,~\cite{xin2013accelerating, xin2015shifted, alser2017gatekeeper, alser2019shouji, alser2020sneakysnake, singh2021fpga, alser2017magnet, bingol2021gatekeeper}), chaining (e.g.,~\cite{guo_hardware_2019, sadasivan_accelerating_2022}), and sequence alignment (e.g.,~\cite{chen2013hybrid, khatamifard2017non, turakhia2018darwin, s_d_goenka_segalign_2020, nag2019gencache, aguado-puig_accelerating_2022, aguado-puig_wfa-gpu_2022, haghi_fpga_2021, cali2020genasm, lindegger2022algorithmic, lindegger2022scrooge, cali_segram_2022, fujiki2018genax, madhavan2014race, cheng2018bitmapper2, houtgast2018hardware, houtgast2017efficient, zeni2020logan, ahmed2019gasal2, nishimura2017accelerating, de2016cudalign, liu2015gswabe, liu2013cudasw++, liu2009cudasw++, liu2010cudasw++, wilton2015arioc, goyal2017ultra, chen2016spark, chen2014accelerating, chen2021high, fujiki2020seedex, banerjee2018asap, fei2018fpgasw, waidyasooriya2015hardware, chen2015novel, rucci2018swifold, li2021pipebsw, wu2019fpga, yan_accel-align_2021, vasimuddin_efficient_2019, daily_parasail_2016, kalikar_accelerating_2022, marco-sola_fast_2021, eizenga_improving_2022, marco-sola_optimal_2022}).} 
\hysix{\name is different from these works as none of the \omsev{prior} read mapping accelerators tightly integrate the basecalling and read mapping steps to reduce the data movement and useless computations in the \entire.}

\hysix{\head{Basecalling-free Genome Analysis} Several works avoid the computationally-costly basecalling step from the genome analysis pipeline by \omsev{directly} mapping raw electrical signals to genomic sequences such as reference genomes (e.g.,~\cite{kovaka2021targeted, zhang2021real, dunn2021squigglefilter, loose_real-time_2016, edwards_real-time_2019, payne_readfish_2021, de_maio_boss-runs_2020, danilevsky_adaptive_2022, joppich_sequ-into_2020}. These works change the representation of the genomic sequence from the base (i.e., DNA character) space into the electrical signal space and perform analysis fully in the signal space. \omsev{As such, they} can accelerate the genome analysis pipeline by reducing or eliminating the need for basecalling for certain use cases (e.g., targeted sequencing~\cite{kovaka2021targeted}). \name is different from these works as} \omsev{it uses the basecalling step and performs genome analysis in the base space, which can be integrated into \emph{any} genome analysis use case.}

\section{Conclusion}
Nanopore genome analysis pipeline has two main computationally-costly processing steps, basecalling and read mapping, which are executed separately \omsix{on different machines} in conventional systems. We observe that \omsix{the} separate execution of these two critical steps results in (1) significant data movement and (2) useless computations on the low-quality and unmapped reads, slowing down the genome analysis pipeline and wasting significant energy.
To effectively overcome these two limitations, we propose \name, an in-memory genome analysis accelerator that tightly integrates basecalling and read mapping. 
\name ~\omsix{uses} two key mechanisms: (1) a chunk-based pipeline, \chuncpp, to collaboratively execute the major genome analysis steps in parallel, and (2) a new early-rejection technique, \earlyrej, to timely terminate the analysis on low-quality and unmapped reads. 
Our experimental results show that \name achieves significant performance improvement and energy saving\omsix{s} compared to \omsev{prior} \omsix{genome analysis} accelerators.
We envision \name to be best implemented inside the sequencing machine to \omsix{maximize the efficiency of} genome \omsix{sequence} analysis.
\omsix{We hope that our work inspires further rethinking of the construction and acceleration of the genome analysis pipeline in a holistic manner.}


\section*{Acknowledgments}
We thank the anonymous reviewers of MICRO 2022 and ISCA 2022 for their valuable feedback. We thank the SAFARI Research Group members for their valuable feedback
and \omsix{the} stimulating intellectual environment they provide.
We acknowledge the generous gifts provided by our industrial partners, including Google, Huawei, Intel, Microsoft, and VMware.
This research is partially supported by the Semiconductor Research Corporation and the ETH Future Computing Laboratory.
This work is partially supported by the European Union’s Horizon programme for research and innovation under grant agreement No 101047160, project BioPIM (Processing-in-memory architectures and programming libraries for bioinformatics algorithms)



%

\bibliographystyle{unsrt}
\bibliography{references}

\begin{thebibliography}{100}

\bibitem{amarasinghe2020opportunities}
Shanika~L Amarasinghe, Shian Su, Xueyi Dong, Luke Zappia, Matthew~E Ritchie,
  and Quentin Gouil.
\newblock {Opportunities and challenges in long-read sequencing data analysis}.
\newblock {\em Genome Biology}, 2020.

\bibitem{murigneux2020comparison}
Valentine Murigneux, Subash~Kumar Rai, Agnelo Furtado, Timothy~JC Bruxner, Wei
  Tian, Ivon Harliwong, Hanmin Wei, Bicheng Yang, Qianyu Ye, Ellis Anderson,
  et~al.
\newblock {Comparison of long-read methods for sequencing and assembly of a
  plant genome}.
\newblock {\em GigaScience}, 2020.

\bibitem{wang2019efficient}
Ou~Wang, Robert Chin, Xiaofang Cheng, Michelle Ka~Yan Wu, Qing Mao, Jingbo
  Tang, Yuhui Sun, Ellis Anderson, Han~K Lam, Dan Chen, et~al.
\newblock {Efficient and unique cobarcoding of second-generation sequencing
  reads from long DNA molecules enabling cost-effective and accurate
  sequencing, haplotyping, and de Novo assembly}.
\newblock {\em Genome Research}, 2019.

\bibitem{cali2017nanopore}
Damla Senol~Cali, Jeremie~S Kim, Saugata Ghose, Can Alkan, and Onur Mutlu.
\newblock {Nanopore sequencing technology and tools for genome assembly:
  computational analysis of the current state, bottlenecks and future
  directions}.
\newblock {\em Briefings in Bioinformatics}, 2018.

\bibitem{clark2019diagnosis}
Michelle~M Clark, Amber Hildreth, Sergey Batalov, Yan Ding, Shimul Chowdhury,
  Kelly Watkins, Katarzyna Ellsworth, Brandon Camp, Cyrielle~I Kint, Calum
  Yacoubian, et~al.
\newblock {Diagnosis of genetic diseases in seriously ill children by rapid
  whole-genome sequencing and automated phenotyping and interpretation}.
\newblock {\em Science Translational Medicine}, 2019.

\bibitem{farnaes2018rapid}
Lauge Farnaes, Amber Hildreth, Nathaly~M Sweeney, Michelle~M Clark, Shimul
  Chowdhury, Shareef Nahas, Julie~A Cakici, Wendy Benson, Robert~H Kaplan,
  Richard Kronick, et~al.
\newblock {Rapid whole-genome sequencing decreases infant morbidity and cost of
  hospitalization}.
\newblock {\em NPJ Genomic Medicine}, 2018.

\bibitem{Ashley2016}
Euan~A Ashley.
\newblock {Towards precision medicine}.
\newblock {\em Nature Reviews Genetics}, 2016.

\bibitem{flores2013p4}
Mauricio Flores, Gustavo Glusman, Kristin Brogaard, Nathan~D Price, and Leroy
  Hood.
\newblock {P4 Medicine: How systems medicine will transform the healthcare
  sector and society}.
\newblock {\em Personalized Medicine}, 2013.

\bibitem{chin2011cancer}
Lynda Chin, Jannik~N Andersen, and P~Andrew Futreal.
\newblock {Cancer genomics: from discovery science to personalized medicine}.
\newblock {\em Nature Medicine}, 2011.

\bibitem{alkan2009personalized}
Can Alkan, Jeffrey~M Kidd, Tomas Marques-Bonet, Gozde Aksay, Francesca
  Antonacci, Fereydoun Hormozdiari, Jacob~O Kitzman, Carl Baker, Maika Malig,
  Onur Mutlu, S~Cenk Sahinalp, Richard~A Gibbs, and Evan~E Eichler.
\newblock {Personalized copy number and segmental duplication maps using
  next-generation sequencing}.
\newblock {\em Nature Genetics}, 2009.

\bibitem{ginsburg2009genomic}
Geoffrey~S Ginsburg and Huntington~F Willard.
\newblock {Genomic and personalized medicine: foundations and applications}.
\newblock {\em Translational Research}, 2009.

\bibitem{alvarez2017next}
Maria~Jesus Alvarez-Cubero, Maria Saiz, Bel{\'e}n Mart{\'\i}nez-Garc{\'\i}a,
  Sara~M Sayalero, Carmen Entrala, Jose~Antonio Lorente, and Luis~Javier
  Martinez-Gonzalez.
\newblock {Next generation sequencing: An application in forensic sciences?}
\newblock {\em Annals of Human Biology}, 2017.

\bibitem{borsting2015next}
Claus B{\o}rsting and Niels Morling.
\newblock {Next generation sequencing and its applications in forensic
  genetics}.
\newblock {\em FSI Genetics}, 2015.

\bibitem{Prohaska2019}
Ana Prohaska, Fernando Racimo, Andrew~J Schork, Martin Sikora, Aaron~J Stern,
  Melissa Ilardo, Morten~Erik Allentoft, Lasse Folkersen, Alfonso Buil,
  J~Víctor Moreno-Mayar, Thorfinn Korneliussen, Daniel Geschwind, Andrés
  Ingason, Thomas Werge, Rasmus Nielsen, and Eske Willerslev.
\newblock {Human disease variation in the light of population genomics}.
\newblock {\em Cell}, 2019.

\bibitem{ellegren2016determinants}
Hans Ellegren and Nicolas Galtier.
\newblock {Determinants of genetic diversity}.
\newblock {\em Nature Reviews Genetics}, 2016.

\bibitem{hoban2016finding}
Sean Hoban, Joanna~L Kelley, Katie~E Lotterhos, Michael~F Antolin, Gideon
  Bradburd, David~B Lowry, Mary~L Poss, Laura~K Reed, Andrew Storfer, and
  Michael~C Whitlock.
\newblock {Finding the genomic basis of local adaptation: pitfalls, practical
  solutions, and future directions}.
\newblock {\em The American Naturalist}, 2016.

\bibitem{ellegren2014genome}
Hans Ellegren.
\newblock {Genome sequencing and population genomics in non-model organisms}.
\newblock {\em Trends in Ecology \& Evolution}, 2014.

\bibitem{romiguier2014comparative}
J~Romiguier, Philippe Gayral, Marion Ballenghien, Arnaud Bernard, Vincent
  Cahais, A~Chenuil, Ylenia Chiari, R~Dernat, L~Duret, Nicolas Faivre, et~al.
\newblock {Comparative population genomics in animals uncovers the determinants
  of genetic diversity}.
\newblock {\em Nature}, 2014.

\bibitem{Prado-Martinez2013}
Javier Prado-Martinez, Peter~H. Sudmant, Jeffrey~M. Kidd, Heng Li, Joanna~L.
  Kelley, Belen Lorente-Galdos, Krishna~R. Veeramah, August~E. Woerner,
  Timothy~D. O'Connor, Gabriel Santpere, Alexander Cagan, Christoph Theunert,
  Ferran Casals, Hafid Laayouni, Kasper Munch, Asger Hobolth, Anders~E.
  Halager, Maika Malig, Jessica Hernandez-Rodriguez, Irene Hernando-Herraez,
  Kay Pr{\"{u}}fer, Marc Pybus, Laurel Johnstone, Michael Lachmann, Can Alkan,
  Dorina Twigg, Natalia Petit, Carl Baker, Fereydoun Hormozdiari, Marcos
  Fernandez-Callejo, Marc Dabad, Michael~L. Wilson, Laurie Stevison, Cristina
  Camprub{\'{\i}}, Tiago Carvalho, Aurora Ruiz-Herrera, Laura Vives, Marta
  Mele, Teresa Abello, Ivanela Kondova, Ronald~E. Bontrop, Anne Pusey, Felix
  Lankester, John~A. Kiyang, Richard~A. Bergl, Elizabeth Lonsdorf, Simon Myers,
  Mario Ventura, Pascal Gagneux, David Comas, Hans Siegismund, Julie Blanc,
  Lidia Agueda-Calpena, Marta Gut, Lucinda Fulton, Sarah~A. Tishkoff, James~C.
  Mullikin, Richard~K. Wilson, Ivo~G. Gut, Mary~Katherine Gonder, Oliver~A.
  Ryder, Beatrice~H. Hahn, Arcadi Navarro, Joshua~M. Akey, Jaume Bertranpetit,
  David Reich, Thomas Mailund, Mikkel~H. Schierup, Christina Hvilsom, Aida~M.
  Andr{\'{e}}s, Jeffrey~D. Wall, Carlos~D. Bustamante, Michael~F. Hammer,
  Evan~E. Eichler, and Tomas Marques-Bonet.
\newblock {Great ape genetic diversity and population history}.
\newblock {\em Nature}, 2013.

\bibitem{bloom2021massively}
Joshua~S Bloom, Laila Sathe, Chetan Munugala, Eric~M Jones, Molly Gasperini,
  Nathan~B Lubock, Fauna Yarza, Erin~M Thompson, Kyle~M Kovary, Jimin Park,
  et~al.
\newblock {Massively scaled-up testing for SARS-COV-2 RNA via next-generation
  sequencing of pooled and barcoded nasal and saliva samples}.
\newblock {\em Nature Biomedical Engineering}, 2021.

\bibitem{yelagandula2021multiplexed}
Ramesh Yelagandula, Aleksandr Bykov, Alexander Vogt, Robert Heinen, Ezgi
  {\"O}zkan, Marcus~Martin Strobl, Juliane~Christina Baar, Kristina Uzunova,
  Bence Hajdusits, Darja Kordic, et~al.
\newblock {Multiplexed detection of SARS-CoV-2 and other respiratory infections
  in high throughput by SARSeq}.
\newblock {\em Nature Communications}, 2021.

\bibitem{wang2020initial}
Fang Wang, Shujia Huang, Rongsui Gao, Yuwen Zhou, Changxiang Lai, Zhichao Li,
  Wenjie Xian, Xiaobo Qian, Zhiyu Li, Yushan Huang, et~al.
\newblock {Initial whole-genome sequencing and analysis of the host genetic
  contribution to Covid-19 severity and susceptibility}.
\newblock {\em Cell Discovery}, 2020.

\bibitem{nikolayevskyy2016whole}
Vlad Nikolayevskyy, Katharina Kranzer, Stefan Niemann, and Francis Drobniewski.
\newblock {Whole genome sequencing of mycobacterium tuberculosis for detection
  of recent transmission and rtacing outbreaks: A systematic review}.
\newblock {\em Tuberculosis}, 2016.

\bibitem{qiu2015whole}
Shaofu Qiu, Peng Li, Hongbo Liu, Yong Wang, Nan Liu, Chengyi Li, Shenlong Li,
  Ming Li, Zhengjie Jiang, Huandong Sun, et~al.
\newblock {Whole-genome sequencing for tracing the transmission link between
  two ard outbreaks caused by a novel hadv serotype 7 variant, China}.
\newblock {\em Scientific Reports}, 2015.

\bibitem{gilchrist2015whole}
Carol~A Gilchrist, Stephen~D Turner, Margaret~F Riley, William~A Petri, and
  Erik~L Hewlett.
\newblock {Whole-genome sequencing in outbreak analysis}.
\newblock {\em Clinical Microbiology Reviews}, 2015.

\bibitem{meyer2022critical}
Fernando Meyer, Adrian Fritz, Zhi-Luo Deng, David Koslicki, Till~Robin Lesker,
  Alexey Gurevich, Gary Robertson, Mohammed Alser, Dmitry Antipov, Francesco
  Beghini, et~al.
\newblock {Critical assessment of metagenome interpretation: the second round
  of challenges}.
\newblock {\em Nature Methods}, 2022.

\bibitem{gire2014genomic}
Stephen~K Gire, Augustine Goba, Kristian~G Andersen, Rachel~SG Sealfon,
  Daniel~J Park, Lansana Kanneh, Simbirie Jalloh, Mambu Momoh, Mohamed Fullah,
  Gytis Dudas, et~al.
\newblock {Genomic surveillance elucidates ebola virus origin and transmission
  during the 2014 outbreak}.
\newblock {\em Science}, 2014.

\bibitem{alser2022covidhunter}
Mohammed Alser, Jeremie~S Kim, Nour~Almadhoun Alserr, Stefan~W Tell, and Onur
  Mutlu.
\newblock {COVIDHunter: COVID-19 pandemic wave prediction and mitigation via
  seasonality aware modeling}.
\newblock {\em Frontiers in public health}, 2022.

\bibitem{le2013selected}
Vien Thi~Minh Le and Binh~An Diep.
\newblock {Selected insights from application of whole genome sequencing for
  outbreak investigations}.
\newblock {\em Current Opinion in Critical Care}, 2013.

\bibitem{lapierre2019micop}
Nathan LaPierre, Serghei Mangul, Mohammed Alser, Igor Mandric, Nicholas~C Wu,
  David Koslicki, and Eleazar Eskin.
\newblock {MiCoP: microbial community profiling method for detecting viral and
  fungal organisms in metagenomic samples}.
\newblock {\em BMC genomics}, 2019.

\bibitem{wang2021nanopore}
Yunhao Wang, Yue Zhao, Audrey Bollas, Yuru Wang, and Kin~Fai Au.
\newblock {Nanopore sequencing technology, bioinformatics and applications}.
\newblock {\em Nature Biotechnology}, 2021.

\bibitem{segerman2020most}
Bo~Segerman.
\newblock {The most frequently used sequencing technologies and assembly
  methods in different time segments of the bacterial surveillance and refseq
  genome databases}.
\newblock {\em Frontiers in Cellular and Infection Microbiology}, 2020.

\bibitem{jain2016oxford}
Miten Jain, Hugh~E Olsen, Benedict Paten, and Mark Akeson.
\newblock {The Oxford Nanopore Minion: Delivery of nanopore sequencing to the
  genomics community}.
\newblock {\em Genome Biology}, 2016.

\bibitem{de2021towards}
Wouter De~Coster, Matthias~H Weissensteiner, and Fritz~J Sedlazeck.
\newblock {Towards population-scale long-read sequencing}.
\newblock {\em Nature Reviews Genetics}, 2021.

\bibitem{shafin2020nanopore}
Kishwar Shafin, Trevor Pesout, Ryan Lorig-Roach, Marina Haukness, Hugh~E Olsen,
  Colleen Bosworth, Joel Armstrong, Kristof Tigyi, Nicholas Maurer, Sergey
  Koren, et~al.
\newblock {Nanopore sequencing and the shasta toolkit enable efficient de Novo
  assembly of eleven human genomes}.
\newblock {\em Nature Biotechnology}, 2020.

\bibitem{logsdon2020long}
Glennis~A Logsdon, Mitchell~R Vollger, and Evan~E Eichler.
\newblock {Long-read human genome sequencing and its applications}.
\newblock {\em Nature Reviews Genetics}, 2020.

\bibitem{payne2019bulkvis}
Alexander Payne, Nadine Holmes, Vardhman Rakyan, and Matthew Loose.
\newblock {BulkVis: a graphical viewer for Oxford nanopore bulk FAST5 files}.
\newblock {\em Bioinformatics}, 2019.

\bibitem{van2018third}
Erwin~L van Dijk, Yan Jaszczyszyn, Delphine Naquin, and Claude Thermes.
\newblock {The third revolution in sequencing technology}.
\newblock {\em Trends in Genetics}, 2018.

\bibitem{ardui2018single}
Simon Ardui, Adam Ameur, Joris~R Vermeesch, and Matthew~S Hestand.
\newblock {Single molecule real-time SMRT sequencing comes of age: applications
  and utilities for medical diagnostics}.
\newblock {\em Nucleic Acids Research}, 2018.

\bibitem{de2018nanopack}
Wouter De~Coster, Svenn D’hert, Darrin~T Schultz, Marc Cruts, and Christine
  Van~Broeckhoven.
\newblock {NanoPack: visualizing and processing long-read sequencing data}.
\newblock {\em Bioinformatics}, 2018.

\bibitem{jain2018nanopore}
Miten Jain, Sergey Koren, Karen~H Miga, Josh Quick, Arthur~C Rand, Thomas~A
  Sasani, John~R Tyson, Andrew~D Beggs, Alexander~T Dilthey, Ian~T Fiddes,
  et~al.
\newblock {Nanopore sequencing and assembly of a human genome with ultra-long
  reads}.
\newblock {\em Nature Biotechnology}, 2018.

\bibitem{rang2018squiggle}
Franka~J Rang, Wigard~P Kloosterman, and Jeroen de~Ridder.
\newblock {From squiggle to basepair: Computational approaches for improving
  nanopore sequencing read accuracy}.
\newblock {\em Genome Biology}, 2018.

\bibitem{belser2018chromosome}
Caroline Belser, Benjamin Istace, Erwan Denis, Marion Dubarry, Franc-Christophe
  Baurens, Cyril Falentin, Mathieu Genete, Wahiba Berrabah, Anne-Marie
  Ch{\`e}vre, R{\'e}gine Delourme, et~al.
\newblock {Chromosome-scale assemblies of plant genomes using nanopore long
  reads and optical maps}.
\newblock {\em Nature Plants}, 2018.

\bibitem{pollard2018long}
Martin~O Pollard, Deepti Gurdasani, Alexander~J Mentzer, Tarryn Porter, and
  Manjinder~S Sandhu.
\newblock {Long Reads: Their purpose and place}.
\newblock {\em Human Molecular Genetics}, 2018.

\bibitem{kchouk2017generations}
Mehdi Kchouk, Jean-Francois Gibrat, and Mourad Elloumi.
\newblock {Generations of sequencing technologies: From first to next
  generation}.
\newblock {\em Biology and Medicine}, 2017.

\bibitem{weirather2017comprehensive}
Jason~L Weirather, Mariateresa de~Cesare, Yunhao Wang, Paolo Piazza, Vittorio
  Sebastiano, Xiu-Jie Wang, David Buck, and Kin~Fai Au.
\newblock {Comprehensive comparison of Pacific Biosciences and Oxford nanopore
  technologies and their applications to transcriptome analysis}.
\newblock {\em F1000Research}, 2017.

\bibitem{jain2017minion}
Miten Jain, John~R Tyson, Matthew Loose, Camilla~LC Ip, David~A Eccles, Justin
  O'Grady, Sunir Malla, Richard~M Leggett, Ola Wallerman, Hans~J Jansen, et~al.
\newblock {Minion analysis and reference consortium: Phase 2 data release and
  analysis of R9.0 chemistry}.
\newblock {\em F1000Research}, 2017.

\bibitem{giordano2017novo}
Francesca Giordano, Louise Aigrain, Michael~A Quail, Paul Coupland, James~K
  Bonfield, Robert~M Davies, German Tischler, David~K Jackson, Thomas~M Keane,
  Jing Li, et~al.
\newblock {De novo yeast genome assemblies from MinION, PacBio and MiSeq
  platforms}.
\newblock {\em Scientific Reports}, 2017.

\bibitem{clarke2009continuous}
James Clarke, Hai-Chen Wu, Lakmal Jayasinghe, Alpesh Patel, Stuart Reid, and
  Hagan Bayley.
\newblock {Continuous base identification for single-molecule nanopore DNA
  sequencing}.
\newblock {\em Nature Nanotechnology}, 2009.

\bibitem{alser2022molecules}
Mohammed Alser, Joel Lindegger, Can Firtina, Nour Almadhoun, Haiyu Mao,
  Gagandeep Singh, Juan Gomez-Luna, and Onur Mutlu.
\newblock {From molecules to genomic variations: accelerating genome analysis
  via intelligent algorithms and architectures}.
\newblock {\em CSBJ}, 2022.

\bibitem{bonito}
{A Pytorch basecaller for Oxford nanopore reads}.
\newblock \url{https://github.com/nanoporetech/bonito}.

\bibitem{Guppy}
Oxford~Nanopore Technology.
\newblock {Guppy}.
\newblock
  \url{https://denbi-nanopore-training-course.readthedocs.io/en/latest/basecalling/basecalling.html
  }.

\bibitem{minion}
{Minion}.
\newblock \url{https://nanoporetech.com/products/minion}.

\bibitem{mansouri2022genstore}
Nika Mansouri~Ghiasi, Jisung Park, Harun Mustafa, Jeremie Kim, Ataberk Olgun,
  Arvid Gollwitzer, Damla Senol~Cali, Can Firtina, Haiyu Mao, Nour
  Almadhoun~Alserr, Rachata Ausavarungnirun, Nandita Vijaykumar, Mohammed
  Alser, and Onur Mutlu.
\newblock {GenStore: A high-performance in-storage processing system for genome
  sequence analysis}.
\newblock In {\em ASPLOS}, 2022.

\bibitem{firtina_blend_2021}
Can Firtina, Jisung Park, Jeremie~S Kim, Mohammed Alser, Damla~Senol Cali, Taha
  Shahroodi, Nika~Mansouri Ghiasi, Gagandeep Singh, Konstantinos Kanellopoulos,
  Can Alkan, et~al.
\newblock {BLEND: A fast, memory-efficient, and accurate mechanism to find
  fuzzy seed matches}.
\newblock {\em arXiv}, 2021.

\bibitem{alser2020accelerating}
Mohammed Alser, Zulal Bingöl, Damla Senol~Cali, Jeremie Kim, Saugata Ghose,
  Can Alkan, and Onur Mutlu.
\newblock {Accelerating genome analysis: A primer on an ongoing journey}.
\newblock {\em IEEE Micro}, 2020.

\bibitem{cali2020genasm}
Damla Senol~Cali, Gupreet Kalsi, Zulal Bing{\"o}l, Lavanya Subramanian, Can
  Firtina, Jeremie Kim, Rachata Ausavarungnirun, Mohammed Alser, Anant Nori,
  Juan Luna, et~al.
\newblock {GenASM: A high-performance, low-power approximate string matching
  acceleration framework for genome sequence analysis}.
\newblock In {\em MICRO}, 2020.

\bibitem{alser2020technology}
Mohammed Alser, Jeremy Rotman, Dhrithi Deshpande, Kodi Taraszka, Huwenbo Shi,
  Pelin~Icer Baykal, Harry~Taegyun Yang, Victor Xue, Sergey Knyazev,
  Benjamin~D. Singer, Brunilda Balliu, David Koslicki, Pavel Skums, Alex
  Zelikovsky, Can Alkan, Onur Mutlu, and Serghei Mangul.
\newblock {Technology dictates algorithms: recent developments in read
  alignment}.
\newblock {\em Genome Biology}, 2021.

\bibitem{kim_fastremap_2022}
Jeremie~S Kim, Can Firtina, Meryem~Banu Cavlak, Damla Senol~Cali, Can Alkan,
  and Onur Mutlu.
\newblock {FastRemap: A tool for quickly remapping reads between genome
  assemblies}.
\newblock {\em Bioinformatics}, 2022.

\bibitem{firtina_aphmm_2022}
Can Firtina, Kamlesh Pillai, Gurpreet~S. Kalsi, Bharathwaj Suresh, Damla~Senol
  Cali, Jeremie Kim, Taha Shahroodi, Meryem~Banu Cavlak, Joel Lindegger,
  Mohammed Alser, Juan~Gómez Luna, Sreenivas Subramoney, and Onur Mutlu.
\newblock {ApHMM: Accelerating profile hidden markov models for fast and
  energy-efficient genome analysis}.
\newblock {\em arXiv}, 2022.

\bibitem{cali_segram_2022}
Damla Senol~Cali, Konstantinos Kanellopoulos, Joël Lindegger, Zülal Bingöl,
  Gurpreet~S. Kalsi, Ziyi Zuo, Can Firtina, Meryem~Banu Cavlak, Jeremie Kim,
  Nika~Mansouri Ghiasi, Gagandeep Singh, Juan Gómez-Luna, Nour~Almadhoun
  Alserr, Mohammed Alser, Sreenivas Subramoney, Can Alkan, Saugata Ghose, and
  Onur Mutlu.
\newblock {SeGraM: A universal hardware accelerator for genomic
  sequence-to-graph and sequence-to-sequence mapping}.
\newblock In {\em ISCA}, 2022.

\bibitem{singh2021fpga}
Gagandeep Singh, Mohammed Alser, Damla Senol~Cali, Diamantopoulos
  Diamantopoulos, Juan Gómez-Luna, Henk Corporaal, and Onur Mutlu.
\newblock {FPGA-based near-memory acceleration of modern data-intensive
  applications}.
\newblock {\em IEEE Micro}, 2021.

\bibitem{lou2020helix}
Qian Lou, Sarath~Chandra Janga, and Lei Jiang.
\newblock {Helix: Algorithm/architecture co-design for accelerating nanopore
  genome base-calling}.
\newblock In {\em PACT}, 2020.

\bibitem{firtina2020apollo}
Can Firtina, Jeremie~S Kim, Mohammed Alser, Damla Senol~Cali, A~Ercument Cicek,
  Can Alkan, and Onur Mutlu.
\newblock {Apollo: A sequencing-technology-independent, scalable and accurate
  assembly polishing algorithm}.
\newblock {\em Bioinformatics}, 2020.

\bibitem{s_d_goenka_segalign_2020}
Sneha~D. Goenka, Yatish Turakhia, Benedict Paten, and Mark Horowitz.
\newblock {SegAlign: A scalable GPU-based whole genome aligner}.
\newblock In {\em SC}, 2020.

\bibitem{angizi2020pim}
Shaahin Angizi, Jiao Sun, Wei Zhang, and Deliang Fan.
\newblock {PIM-Aligner: A processing-in-MRAM platform for biological sequence
  alignment}.
\newblock In {\em DATE}, 2020.

\bibitem{alser2020sneakysnake}
Mohammed Alser, Taha Shahroodi, Juan G{\'o}mez-Luna, Can Alkan, and Onur Mutlu.
\newblock {SneakySnake: A fast and accurate universal genome pre-alignment
  filter for CPUs, GPUs and FPGAs}.
\newblock {\em Bioinformatics}, 2020.

\bibitem{nag2019gencache}
Anirban Nag, C.~N. Ramachandra, Rajeev Balasubramonian, Ryan Stutsman, Edouard
  Giacomin, Hari Kambalasubramanyam, and Pierre-Emmanuel Gaillardon.
\newblock {GenCache: Leveraging in-Cache operators for efficient sequence
  alignment}.
\newblock In {\em MICRO}, 2019.

\bibitem{kim2019airlift}
Jeremie~S Kim, Can Firtina, Meryem~Banu Cavlak, Damla~Senol Cali, Mohammed
  Alser, Nastaran Hajinazar, Can Alkan, and Onur Mutlu.
\newblock {AirLift: A fast and comprehensive technique for remapping alignments
  between reference genomes}.
\newblock {\em arXiv}, 2019.

\bibitem{kim2018grim}
Jeremie~S. Kim, Damla Senol~Cali, Hongyi Xin, Donghyuk Lee, Saugata Ghose,
  Mohammed Alser, Hasan Hassan, Oguz Ergin, Can Alkan, and Onur Mutlu.
\newblock {GRIM-Filter: Fast seed location filtering in DNA read mapping using
  processing-in-memory technologies}.
\newblock {\em BMC Genomics}, 2018.

\bibitem{turakhia2018darwin}
Yatish Turakhia, Gill Bejerano, and William~J. Dally.
\newblock {Darwin: A genomics co-processor provides up to 15,000x acceleration
  on long read assembly}.
\newblock In {\em ASPLOS}, 2018.

\bibitem{alser2017gatekeeper}
Mohammed Alser, Hasan Hassan, Hongyi Xin, O{\u{g}}uz Ergin, Onur Mutlu, and Can
  Alkan.
\newblock {GateKeeper: A new hardware architecture for accelerating
  pre-alignment in DNA short read mapping}.
\newblock {\em Bioinformatics}, 2017.

\bibitem{xin2015shifted}
Hongyi Xin, John Greth, John Emmons, Gennady Pekhimenko, Carl Kingsford, Can
  Alkan, and Onur Mutlu.
\newblock {Shifted Hamming Distance: A fast and accurate simd-friendly filter
  to accelerate alignment verification in read mapping}.
\newblock {\em Bioinformatics}, 2015.

\bibitem{xin2013accelerating}
Hongyi Xin, Donghyuk Lee, Farhad Hormozdiari, Samihan Yedkar, Onur Mutlu, and
  Can Alkan.
\newblock {Accelerating read mapping with fasthash}.
\newblock {\em BMC Genomics}, 2013.

\bibitem{smith1981identification}
Temple~F Smith, Michael~S Waterman, et~al.
\newblock {Identification of common molecular subsequences}.
\newblock {\em JMB}, 1981.

\bibitem{needleman1970general}
Saul~B Needleman and Christian~D Wunsch.
\newblock {A general method applicable to the search for similarities in the
  amino acid sequence of two proteins}.
\newblock {\em JMB}, 1970.

\bibitem{lou2018brawl}
Qian Lou and Lei Jiang.
\newblock {Brawl: A spintronics-based portable basecalling-in-memory
  architecture for nanopore genome sequencing}.
\newblock {\em CAL}, 2018.

\bibitem{wu2018fpga}
ZhongPan Wu, Karim Hammad, Robinson Mittmann, Sebastian Magierowski, Ebrahim
  Ghafar-Zadeh, and Xiaoyong Zhong.
\newblock {FPGA-based DNA basecalling hardware acceleration}.
\newblock In {\em MWSCAS}, 2018.

\bibitem{khatamifard2021genvom}
S~Karen Khatamifard, Zamshed Chowdhury, Nakul Pande, Meisam Razaviyayn, Chris~H
  Kim, and Ulya~R Karpuzcu.
\newblock {GeNVoM: Read mapping near non-volatile memory}.
\newblock {\em TCBB}, 2021.

\bibitem{jain2019accelerating}
Chirag Jain, Sanchit Misra, Haowen Zhang, Alexander Dilthey, and Srinivas
  Aluru.
\newblock {Accelerating sequence alignment to graphs}.
\newblock In {\em IPDPS}, 2019.

\bibitem{feng2019accelerating}
Zonghao Feng, Shuang Qiu, Lipeng Wang, and Qiong Luo.
\newblock {Accelerating long read alignment on three processors}.
\newblock In {\em ICPP}, 2019.

\bibitem{gupta2019rapid}
Saransh Gupta, Mohsen Imani, Behnam Khaleghi, Venkatesh Kumar, and Tajana
  Rosing.
\newblock {RAPID: A reRAM processing in-memory architecture for DNA sequence
  alignment}.
\newblock In {\em ISLPED}, 2019.

\bibitem{zokaee2018aligner}
Farzaneh Zokaee, Hamid~R Zarandi, and Lei Jiang.
\newblock {AligneR: A process-in-Memory architecture for short read alignment
  in ReRAMs}.
\newblock {\em CAL}, 2018.

\bibitem{diab2022framework}
Safaa Diab, Amir Nassereldine, Mohammed Alser, Juan G{\'o}mez-Luna, Onur Mutlu,
  and Izzat~El Hajj.
\newblock {A framework for high-throughput sequence alignment using real
  processing-in-memory systems}.
\newblock {\em arXiv}, 2022.

\bibitem{bowden2019sequencing}
Rory Bowden, Robert~W Davies, Andreas Heger, Alistair~T Pagnamenta, Mariateresa
  de~Cesare, Laura~E Oikkonen, Duncan Parkes, Colin Freeman, Fatima Dhalla,
  Smita~Y Patel, et~al.
\newblock {Sequencing of human genomes with nanopore technology}.
\newblock {\em Nature Communications}, 2019.

\bibitem{lapierre2020metalign}
Nathan LaPierre, Mohammed Alser, Eleazar Eskin, David Koslicki, and Serghei
  Mangul.
\newblock {Metalign: Efficient alignment-based metagenomic profiling via
  containment min hash}.
\newblock {\em Genome Biology}, 2020.

\bibitem{ecoli}
Nicholas~J. Loman.
\newblock {Nanopore R9 rapid run data release}, 2016.
\newblock \url{http://lab.loman.net/2016/07/30/nanopore-r9-data-release/}.

\bibitem{chen2020parc}
Fan Chen, Linghao Song, Yiran Chen, et~al.
\newblock {PARC: A processing-in-CAM architecture for genomic long read
  pairwise alignment using ReRAM}.
\newblock In {\em ASP-DAC}, 2020.

\bibitem{Flappie}
Oxford Nanopore~Technology (ONT).
\newblock {Flappie}.
\newblock \url{https://github.com/nanoporetech/flappie}.

\bibitem{Scrappie}
Oxford~Nanopore Technology.
\newblock {Scrappie}.
\newblock \url{https://github.com/nanoporetech/scrappie }.

\bibitem{10.1093/gigascience/giy037}
Haotian Teng, Minh~Duc Cao, Michael~B Hall, Tania Duarte, Sheng Wang, and
  Lachlan J~M Coin.
\newblock {Chiron: Translating nanopore raw signal directly into nucleotide
  sequence using deep learning}.
\newblock {\em GigaScience}, 2018.

\bibitem{li2018minimap2}
Heng Li.
\newblock {Minimap2: Pairwise alignment for nucleotide sequences}.
\newblock {\em Bioinformatics}, 2018.

\bibitem{roberts_reducing_2004}
Michael Roberts, Wayne Hayes, Brian~R. Hunt, Stephen~M. Mount, and James~A.
  Yorke.
\newblock {Reducing storage requirements for biological sequence comparison}.
\newblock {\em Bioinformatics}, 2004.

\bibitem{schleimer2003winnowing}
Saul Schleimer, Daniel~S Wilkerson, and Alex Aiken.
\newblock {Winnowing: Local algorithms for document fingerprinting}.
\newblock In {\em SIGMOD}, 2003.

\bibitem{alser2021technology}
Mohammed Alser, Jeremy Rotman, Dhrithi Deshpande, Kodi Taraszka, Huwenbo Shi,
  Pelin~Icer Baykal, Harry~Taegyun Yang, Victor Xue, Sergey Knyazev, Benjamin~D
  Singer, et~al.
\newblock {Technology dictates algorithms: recent developments in read
  alignment}.
\newblock {\em Genome Biology}, 2021.

\bibitem{eddy2004dynamic}
Sean~R Eddy.
\newblock {What is dynamic programming?}
\newblock {\em Nature Biotechnology}, 2004.

\bibitem{fukasawa2020longqc}
Yoshinori Fukasawa, Luca Ermini, Hai Wang, Karen Carty, and Min-Sin Cheung.
\newblock {LongQC: A quality control tool for third generation sequencing long
  read data}.
\newblock {\em G3}, 2020.

\bibitem{leger2019pycoqc}
Adrien Leger and Tommaso Leonardi.
\newblock {pycoQC, interactive quality control for Oxford nanopore sequencing}.
\newblock {\em JOSS}, 2019.

\bibitem{minionrequirements}
{Minion Mk1b IT requirements}.
\newblock
  \url{https://community.nanoporetech.com/requirements_documents/minion-it-reqs.pdf}.

\bibitem{kingsmore2020measurement}
Stephen~F Kingsmore, Audrey Henderson, Mallory~J Owen, Michelle~M Clark,
  Christian Hansen, David Dimmock, Christina~D Chambers, Laura~L
  Jeliffe-Pawlowski, and Charlotte Hobbs.
\newblock {Measurement of genetic diseases as a cause of mortality in infants
  receiving whole genome sequencing}.
\newblock {\em NPJ Genomic Medicine}, 2020.

\bibitem{ankit2019puma}
Aayush Ankit, Izzat~El Hajj, Sai~Rahul Chalamalasetti, Geoffrey Ndu, Martin
  Foltin, R~Stanley Williams, Paolo Faraboschi, Wen-mei~W Hwu, John~Paul
  Strachan, Kaushik Roy, et~al.
\newblock {PUMA: A programmable ultra-efficient memristor-based accelerator for
  machine learning inference}.
\newblock In {\em ASPLOS}, 2019.

\bibitem{ankit2020panther}
Aayush Ankit, Izzat El~Hajj, Sai~Rahul Chalamalasetti, Sapan Agarwal, Matthew
  Marinella, Martin Foltin, John~Paul Strachan, Dejan Milojicic, Wen-Mei Hwu,
  and Kaushik Roy.
\newblock {PANTHER: A programmable architecture for neural network training
  harnessing energy-efficient ReRAM}.
\newblock {\em IEEE TC}, 2020.

\bibitem{yuan2021forms}
Geng Yuan, Payman Behnam, Zhengang Li, Ali Shafiee, Sheng Lin, Xiaolong Ma,
  Hang Liu, Xuehai Qian, Mahdi~Nazm Bojnordi, Yanzhi Wang, et~al.
\newblock {Forms: Fine-grained polarized ReRAM-based in-situ computation for
  mixed-signal DNN accelerator}.
\newblock {\em arXiv}, 2021.

\bibitem{huang2021mixed}
Sitao Huang, Aayush Ankit, Plinio Silveira, Rodrigo Antunes, Sai~Rahul
  Chalamalasetti, Izzat El~Hajj, Dong~Eun Kim, Glaucimar Aguiar, Pedro Bruel,
  Sergey Serebryakov, et~al.
\newblock {Mixed precision quantization for ReRAM-based DNN inference
  accelerators}.
\newblock In {\em ASP-DAC}, 2021.

\bibitem{longlive}
Yi~Cai, Yujun Lin, Lixue Xia, Xiaoming Chen, Song Han, Yu~Wang, and Huazhong
  Yang.
\newblock {Long Live Time: Improving lifetime for training-in-memory engines by
  structured gradient sparsification}.
\newblock In {\em DAC}, 2018.

\bibitem{song2018graphr}
Linghao Song, Youwei Zhuo, Xuehai Qian, Hai Li, and Yiran Chen.
\newblock {GraphR: Accelerating graph processing using ReRAM}.
\newblock In {\em HPCA}, 2018.

\bibitem{imani2019floatpim}
Mohsen Imani, Saransh Gupta, Yeseong Kim, and Tajana Rosing.
\newblock {FloatPIM: In-memory acceleration of deep neural network training
  with high precision}.
\newblock In {\em ISCA}, 2019.

\bibitem{mittal2019survey}
Sparsh Mittal.
\newblock {A survey of ReRAM-based architectures for processing-in-memory and
  neural networks}.
\newblock {\em MAKE}, 2019.

\bibitem{gupta2019nnpim}
Saransh Gupta, Mohsen Imani, Harveen Kaur, and Tajana~Simunic Rosing.
\newblock {NNPIM: A processing in-memory architecture for neural network
  acceleration}.
\newblock {\em TC}, 2019.

\bibitem{lergan}
Haiyu Mao, Mingcong Song, Tao Li, Yuting Dai, and Jiwu Shu.
\newblock {LerGAN: A zero-free, low data movement and PIM-based GAN
  architecture}.
\newblock In {\em MICRO}, 2018.

\bibitem{mao2020lrgan}
Haiyu Mao, Jiwu Shu, Mingcong Song, and Tao Li.
\newblock {LrGAN: A compact and energy efficient PIM-Based architecture for GAN
  training}.
\newblock {\em TC}, 2020.

\bibitem{karam2015emerging}
Robert Karam, Ruchir Puri, Swaroop Ghosh, and Swarup Bhunia.
\newblock {Emerging trends in design and applications of memory-based computing
  and content-addressable memories}.
\newblock {\em Proceedings of the IEEE}, 2015.

\bibitem{imani2016masc}
Mohsen Imani, Shruti Patil, and Tajana~S Rosing.
\newblock {MASC: Ultra-low energy multiple-access single-charge TCAM for
  approximate computing}.
\newblock In {\em DATE}, 2016.

\bibitem{li2016nvsim}
Shuangchen Li, Liu Liu, Peng Gu, Cong Xu, and Yuan Xie.
\newblock {NVSim-CAM: A circuit-level simulator for emerging nonvolatile memory
  based content-addressable memory}.
\newblock In {\em ICCAD}, 2016.

\bibitem{deng2016multi}
Erya Deng, Lorena Anghel, Guillaume Prenat, and Weisheng Zhao.
\newblock {Multi-context non-volatile content addressable memory using magnetic
  tunnel junctions}.
\newblock In {\em NANOARCH}, 2016.

\bibitem{chang20173t1r}
Meng-Fan Chang, Chien-Chen Lin, Albert Lee, Yen-Ning Chiang, Chia-Chen Kuo,
  Geng-Hau Yang, Hsiang-Jen Tsai, Tien-Fu Chen, and Shyh-Shyuan Sheu.
\newblock {A 3T1R nonvolatile TCAM using MLC ReRAM for frequent-Off instant-on
  filters in IoT and big-data processing}.
\newblock {\em JSSC}, 2017.

\bibitem{yantir2017approximate}
Hasan~Erdem Yantir, Ahmed~M Eltawil, and Fadi~J Kurdahi.
\newblock {Approximate memristive in-memory computing}.
\newblock {\em TECS}, 2017.

\bibitem{yin2017design}
Xunzhao Yin, Michael Niemier, and X~Sharon Hu.
\newblock {Design and benchmarking of ferroelectric FET-based TCAM}.
\newblock In {\em DATE}, 2017.

\bibitem{imani2017multi}
Mohsen Imani, Abbas Rahimi, Pietro Mercati, and Tajana~Simunic Rosing.
\newblock {Multi-stage tunable approximate search in resistive associative
  memory}.
\newblock {\em TMSCS}, 2017.

\bibitem{yin2018ultra}
Xunzhao Yin, Kai Ni, Dayane Reis, Suman Datta, Michael Niemier, and
  Xiaobo~Sharon Hu.
\newblock {An ultra-dense 2FeFET TCAM design based on a multi-domain FeFET
  model}.
\newblock {\em TCAS}, 2018.

\bibitem{wang2018novel}
Chengzhi Wang, Deming Zhang, Lang Zeng, Erya Deng, Jie Chen, and Weisheng Zhao.
\newblock {A novel MTJ-based non-volatile ternary content-addressable memory
  for high-speed, low-power, and high-reliable search operation}.
\newblock {\em TCAS}, 2018.

\bibitem{gnawali2018low}
Krishna~Prasad Gnawali, Seyed~Nima Mozaffari, and Spyros Tragoudas.
\newblock {Low power spintronic ternary content addressable memory}.
\newblock {\em TNANO}, 2018.

\bibitem{behnam2018r}
Payman Behnam, Arjun~Pal Chowdhury, and Mahdi~Nazm Bojnordi.
\newblock {R-Cache: A highly set-associative in-package cache using memristive
  arrays}.
\newblock In {\em ICCD}, 2018.

\bibitem{tan2019experimental}
Ava~J Tan, Korok Chatterjee, Jiuren Zhou, Daewoong Kwon, Yu-Hung Liao, Suraj
  Cheema, Chenming Hu, and Sayeef Salahuddin.
\newblock {Experimental demonstration of a ferroelectric HfO2-based content
  addressable memory cell}.
\newblock {\em EDL}, 2019.

\bibitem{arakawa2019multi}
Ren Arakawa, Naoya Onizawa, Jean-Philippe Diguet, and Takahiro Hanyu.
\newblock {Multi-context TCAM based selective computing architecture for a
  low-power NN}.
\newblock In {\em ICECS}, 2019.

\bibitem{Zhao2019RFAccA3}
Lei Zhao, Quan Deng, Youtao Zhang, and J.~Yang.
\newblock {RFAcc: a 3D ReRAM associative array based random forest
  accelerator}.
\newblock {\em SC}, 2019.

\bibitem{halawani2019reram}
Yasmin Halawani, Baker Mohammad, Muath~Abu Lebdeh, Mahmoud Al-Qutayri, and
  Said~F Al-Sarawi.
\newblock {ReRAM-based in-Memory computing for search engine and neural network
  applications}.
\newblock {\em JETCAS}, 2019.

\bibitem{li2020analog}
Can Li, Catherine~E Graves, Xia Sheng, Darrin Miller, Martin Foltin, Giacomo
  Pedretti, and John~Paul Strachan.
\newblock {Analog content-addressable memories with memristors}.
\newblock {\em Nature Communications}, 2020.

\bibitem{graves2020memory}
Catherine~E Graves, Can Li, Xia Sheng, Darrin Miller, Jim Ignowski, Lennie
  Kiyama, and John~Paul Strachan.
\newblock {In-memory computing with memristor content addressable memories for
  pattern matching}.
\newblock {\em Advanced Materials}, 2020.

\bibitem{yin2020fecam}
Xunzhao Yin, Chao Li, Qingrong Huang, Li~Zhang, Michael Niemier, Xiaobo~Sharon
  Hu, Cheng Zhuo, and Kai Ni.
\newblock {FeCAM: A universal compact digital and analog content addressable
  memory using ferroelectric}.
\newblock {\em TED}, 2020.

\bibitem{xiu2021capacitive}
Nuo Xiu, Yiming Chen, Guodong Yin, Xiaoyang Ma, Huazhong Yang, Sumitha George,
  and Xueqing Li.
\newblock {Capacitive content-addressable memory: A highly reliable and
  scalable approach to energy-efficient parallel pattern matching
  applications}.
\newblock In {\em GLSVLSI}, 2021.

\bibitem{chi2016prime}
Ping Chi, Shuangchen Li, Cong Xu, Tao Zhang, Jishen Zhao, Yongpan Liu, Yu~Wang,
  and Yuan Xie.
\newblock {PRIME: A novel processing-in-memory architecture for neural network
  computation in ReRAM-based main memory}.
\newblock {\em CAN}, 2016.

\bibitem{gpu}
{NVIDIA RTX GeForce 2080 Ti 11GB}, 2018.
\newblock
  \url{https://www.pny.com.tw/jp/upload/download_files/jp_download_list_19b22_wr2i3c5qid.pdf}.

\bibitem{cpu}
{Introduction to Intel architecture}, 2016.
\newblock
  \url{https://www.intel.com/content/dam/www/public/us/en/documents/white-papers/ia-introduction-basics-paper.pdf}.

\bibitem{edram}
Sparsh Mittal, Jeffrey~S Vetter, and Dong Li.
\newblock {A survey of architectural approaches for managing embedded DRAM and
  non-volatile on-chip caches}.
\newblock {\em TPDS}, 2014.

\bibitem{shafiee2016isaac}
A.~Shafiee, Anirban Nag, N.~Muralimanohar, Rajeev Balasubramonian, J.~Strachan,
  Miao Hu, R.~Williams, and Vivek Srikumar.
\newblock {ISAAC: A convolutional neural network accelerator with in-situ
  analog arithmetic in crossbars}.
\newblock In {\em ISCA}, 2016.

\bibitem{kurup2012logic}
Pran Kurup and Taher Abbasi.
\newblock {\em {Logic synthesis using Synopsys{\textregistered}}}.
\newblock Springer Science \& Business Media, 2012.

\bibitem{dong2012nvsim}
Xiangyu Dong, Cong Xu, Yuan Xie, and Norman~P Jouppi.
\newblock {NVSIM: A circuit-level performance, energy, and area model for
  emerging nonvolatile memory}.
\newblock {\em TCAD}, 2012.

\bibitem{muralimanohar2007optimizing}
Naveen Muralimanohar, Rajeev Balasubramonian, and Norm Jouppi.
\newblock {Optimizing NUCA organizations and wiring alternatives for large
  caches with CACTI 6.0}.
\newblock In {\em MICRO}, 2007.

\bibitem{ena}
{European Nucleotide Archive (ENA)}.
\newblock \url{https://www.ebi.ac.uk/ena/browser/home}.

\bibitem{ncbi}
{National cCenter for Biotechnology Information (NCBI)}.
\newblock \url{https://www.ncbi.nlm.nih.gov/}.

\bibitem{senol2017nanopore}
Damla Senol, Jeremie Kim, Saugata Ghose, Can Alkan, and Onur Mutlu.
\newblock {Nanopore sequencing technology and tools: cComputational analysis of
  the current state, bottlenecks and future directions}.
\newblock In {\em PSB}, 2017.

\bibitem{sereika2022oxford}
Mantas Sereika, Rasmus~Hansen Kirkegaard, S{\o}ren~Michael Karst, Thomas~Yssing
  Michaelsen, Emil~Aarre S{\o}rensen, Rasmus~Dam Wollenberg, and Mads
  Albertsen.
\newblock {Oxford nanopore R10.4 long-read sequencing enables the generation of
  near-finished bacterial genomes from pure cultures and metagenomes without
  short-read or reference polishing}.
\newblock {\em Nature Methods}, 2022.

\bibitem{ecoli_dataset}
{Nanopore R9 rapid run data release}.
\newblock \url{http://lab.loman.net/2016/07/30/nanopore-r9-data-release/}.

\bibitem{human_dataset}
{European Nucleotide Archive - Sequencing of human genomes with nanopore
  technology}.
\newblock \url{https://www.ebi.ac.uk/ena/browser/view/PRJEB30620}.

\bibitem{kaplan2017resistive}
Roman Kaplan, Leonid Yavits, Ran Ginosar, and Uri Weiser.
\newblock {A resistive CAM processing-in-storage architecture for DNA sequence
  alignment}.
\newblock {\em IEEE Micro}, 2017.

\bibitem{kaplan2018rassa}
Roman Kaplan, Leonid Yavits, and Ran Ginosar.
\newblock {RASSA: Resistive pre-alignment accelerator for approximate DNA long
  read mapping}.
\newblock {\em IEEE Micro}, 2018.

\bibitem{zokaee2019finder}
Farzaneh Zokaee, Mingzhe Zhang, and Lei Jiang.
\newblock {FindeR: Accelerating FM-index-based exact pattern matching in
  genomic sequences through ReRAM technology}.
\newblock In {\em PACT}, 2019.

\bibitem{angizi2020exploring}
Shaahin Angizi, Wei Zhang, and Deliang Fan.
\newblock {Exploring DNA alignment-in-memory leveraging emerging SOT-MRAM}.
\newblock In {\em GLSVLSI}, 2020.

\bibitem{kaplan2020bioseal}
Roman Kaplan, Leonid Yavits, and Ran Ginosasr.
\newblock {BioSEAL: In-memory biological sequence alignment accelerator for
  large-scale genomic data}.
\newblock In {\em SYSTOR}, 2020.

\bibitem{laguna2020seed}
Ann~Franchesca Laguna, Hasindu Gamaarachchi, Xunzhao Yin, Michael Niemier, Sri
  Parameswaran, and X~Sharon Hu.
\newblock {Seed-and-Vote based in-memory accelerator for DNA read mapping}.
\newblock In {\em ICCAD}, 2020.

\bibitem{khalifa_filtpim_2021}
Marcel Khalifa, Rotem Ben-Hur, Ronny Ronen, Orian Leitersdorf, Leonid Yavits,
  and Shahar Kvatinsky.
\newblock {FiltPIM: In-memory filter for DNA sequencing}.
\newblock In {\em ICECS}, 2021.

\bibitem{chowdhury_dna_2020}
Zamshed~I. Chowdhury, Masoud Zabihi, S.~Karen Khatamifard, Zhengyang Zhao,
  Salonik Resch, Meisam Razaviyayn, Jian-Ping Wang, Sachin~S. Sapatnekar, and
  Ulya~R. Karpuzcu.
\newblock {A DNA read alignment accelerator based on computational RAM}.
\newblock {\em JXCDC}, 2020.

\bibitem{huangfu_radar_2018}
Wenqin Huangfu, Shuangchen Li, Xing Hu, and Yuan Xie.
\newblock {RADAR: A 3D-ReRAM based DNA alignment accelerator architecture}.
\newblock In {\em DAC}, 2018.

\bibitem{diab2022high}
Safaa Diab, Amir Nassereldine, Mohammed Alser, Juan~G{\'o}mez Luna, Onur Mutlu,
  and Izzat~El Hajj.
\newblock {High-throughput pairwise alignment with the wavefront algorithm
  using processing-in-memory}.
\newblock {\em arXiv}, 2022.

\bibitem{shahroodi2022demeter}
Taha Shahroodi, Mahdi Zahedi, Can Firtina, Mohammed Alser, Stephan Wong, Onur
  Mutlu, and Said Hamdioui.
\newblock Demeter: {A} {fast} and {energy}-{efficient} {food} {profiler}
  {using} {hyperdimensional} {computing} in {memory}.
\newblock {\em IEEE Access}, 2022.

\bibitem{li2021pim}
Xue-Qi Li, Guang-Ming Tan, and Ning-Hui Sun.
\newblock {PIM-Align: A processing-in-memory architecture for FM-Index search
  algorithm}.
\newblock {\em JCST}, 2021.

\bibitem{angizi2019aligns}
Shaahin Angizi, Jiao Sun, Wei Zhang, and Deliang Fan.
\newblock {Aligns: A processing-in-memory accelerator for DNA short read
  alignment leveraging SOT-MRAM}.
\newblock In {\em DAC}, 2019.

\bibitem{dunn2021squigglefilter}
Tim Dunn, Harisankar Sadasivan, Jack Wadden, Kush Goliya, Kuan-Yu Chen, David
  Blaauw, Reetuparna Das, and Satish Narayanasamy.
\newblock {SquiggleFilter: An accelerator for portable virus detection}.
\newblock In {\em MICRO}, 2021.

\bibitem{xu2021fast}
Zhimeng Xu, Yuting Mai, Denghui Liu, Wenjun He, Xinyuan Lin, Chi Xu, Lei Zhang,
  Xin Meng, Joseph Mafofo, Walid~Abbas Zaher, et~al.
\newblock {Fast-bonito: A faster deep learning based basecaller for nanopore
  sequencing}.
\newblock {\em Artificial Intelligence in the Life Sciences}, 2021.

\bibitem{perevsini2021nanopore}
Peter Pere{\v{s}}{\'\i}ni, Vladim{\'\i}r Bo{\v{z}}a, Bro{\v{n}}a Brejov{\'a},
  and Tom{\'a}{\v{s}} Vina{\v{r}}.
\newblock Nanopore base calling on the edge.
\newblock {\em Bioinformatics}, 2021.

\bibitem{lv_end--end_2020}
Xuan Lv, Zhiguang Chen, Yutong Lu, and Yuedong Yang.
\newblock {An end-to-end Oxford nanopore basecaller using convolution-augmented
  transformer}.
\newblock In {\em BIBM}, 2020.

\bibitem{zeng_causalcall_2020}
Jingwen Zeng, Hongmin Cai, Hong Peng, Haiyan Wang, Yue Zhang, and Tatsuya
  Akutsu.
\newblock {Causalcall: Nanopore basecalling using a temporal convolutional
  network}.
\newblock {\em Frontiers in Genetics}, 2020.

\bibitem{yeh_msrcall_2022}
Yang-Ming Yeh and Yi-Chang Lu.
\newblock {MSRCall: A multi-scale deep neural network to basecall Oxford
  nanopore sequences}.
\newblock {\em Bioinformatics}, 2022.

\bibitem{huang_sacall_2022}
Neng Huang, Fan Nie, Peng Ni, Feng Luo, and Jianxin Wang.
\newblock {SACall: A neural network basecaller for Oxford nanopore sequencing
  data based on self-attention mechanism}.
\newblock {\em TCBB}, 2022.

\bibitem{konishi_halcyon_2021}
Hiroki Konishi, Rui Yamaguchi, Kiyoshi Yamaguchi, Yoichi Furukawa, and Seiya
  Imoto.
\newblock {Halcyon: an accurate basecaller exploiting an encoder-decoder model
  with monotonic attention}.
\newblock {\em Bioinformatics}, 2021.

\bibitem{boza_deepnano_2017}
Vladimír Boža, Broňa Brejová, and Tomáš Vinař.
\newblock {DeepNano: Deep recurrent neural networks for base calling in MinION
  nanopore reads}.
\newblock {\em PLOS One}, 2017.

\bibitem{ramachandra_ont-x_2021}
C~N Ramachandra, Anirban Nag, Rajeev Balasubramonion, Gurpreet Kalsi, Kamlesh
  Pillai, and Sreenivas Subramoney.
\newblock {ONT-X: An FPGA approach to real-time portable genomic analysis}.
\newblock In {\em FCCM}, 2021.

\bibitem{hammad_scalable_2021}
Karim Hammad, Zhongpan Wu, Ebrahim Ghafar-Zadeh, and Sebastian Magierowski.
\newblock {A scalable hardware accelerator for mobile DNA sequencing}.
\newblock {\em TVLSI}, 2021.

\bibitem{wu_fpga_2022}
Zhongpan Wu, Karim Hammad, Abel Beyene, Yunus Dawji, Ebrahim Ghafar-Zadeh, and
  Sebastian Magierowski.
\newblock {An FPGA implementation of a portable DNA sequencing device based on
  RISC-V}.
\newblock In {\em Newcas}, 2022.

\bibitem{wu_fpga-accelerated_2020}
Zhongpan Wu, Karim Hammad, Ebrahim Ghafar-Zadeh, and Sebastian Magierowski.
\newblock {FPGA-accelerated 3rd generation DNA sequencing}.
\newblock {\em TBCS}, 2020.

\bibitem{alser2019shouji}
Mohammed Alser, Hasan Hassan, Akash Kumar, Onur Mutlu, and Can Alkan.
\newblock {Shouji: A fast and efficient pre-alignment filter for sequence
  alignment}.
\newblock {\em Bioinformatics}, 2019.

\bibitem{alser2017magnet}
Mohammed Alser, Onur Mutlu, and Can Alkan.
\newblock {MAGNET: Understanding and improving the accuracy of genome
  pre-Alignment filtering}.
\newblock {\em arXiv}, 2017.

\bibitem{bingol2021gatekeeper}
Z{\"u}lal Bing{\"o}l, Mohammed Alser, Onur Mutlu, Ozcan Ozturk, and Can Alkan.
\newblock {GateKeeper-GPU: Fast and accurate pre-alignment filtering in short
  read mapping}.
\newblock In {\em IPDPSW}. IEEE, 2021.

\bibitem{guo_hardware_2019}
Licheng Guo, Jason Lau, Zhenyuan Ruan, Peng Wei, and Jason Cong.
\newblock {Hardware acceleration of long read pairwise overlapping in genome
  sequencing: a race between FPGA and GPU}.
\newblock In {\em FCCM}, 2019.

\bibitem{sadasivan_accelerating_2022}
Harisankar Sadasivan, Milos Maric, Eric Dawson, Vishanth Iyer, Johnny Israeli,
  and Satish Narayanasamy.
\newblock {Accelerating {Minimap2} for accurate long read alignment on {GPUs}}.
\newblock {\em bioRxiv}, 2022.

\bibitem{chen2013hybrid}
Yupeng Chen, Bertil Schmidt, and Douglas~L Maskell.
\newblock {A hybrid short read mapping accelerator}.
\newblock {\em BMC Bioinformatics}, 2013.

\bibitem{khatamifard2017non}
S~Karen Khatamifard, Zamshed Chowdhury, Nakul Pande, Meisam Razaviyayn, Chris
  Kim, and Ulya~R Karpuzcu.
\newblock {Read mapping near non-volatile memory}.
\newblock {\em arXiv}, 2017.

\bibitem{aguado-puig_accelerating_2022}
Quim Aguado-Puig, Santiago Marco-Sola, Juan~Carlos Moure, David Castells-Rufas,
  Lluc Alvarez, Antonio Espinosa, and Miquel Moreto.
\newblock {Accelerating edit-distance sequence alignment on GPU using the
  wavefront algorithm}.
\newblock {\em IEEE Access}, 2022.

\bibitem{aguado-puig_wfa-gpu_2022}
Quim Aguado-Puig, Santiago Marco-Sola, Juan~Carlos Moure, Christos Matzoros,
  David Castells-Rufas, Antonio Espinosa, and Miquel Moreto.
\newblock {WFA-GPU: Gap-affine pairwise alignment using GPUs}.
\newblock {\em bioRxiv}, 2022.

\bibitem{haghi_fpga_2021}
Abbas Haghi, Santiago Marco-Sola, Lluc Alvarez, Dionysios Diamantopoulos,
  Christoph Hagleitner, and Miquel Moreto.
\newblock {An FPGA accelerator of the wavefront algorithm for genomics pairwise
  alignment}.
\newblock In {\em FPL}, 2021.

\bibitem{lindegger2022algorithmic}
Jo{\"e}l Lindegger, Damla~Senol Cali, Mohammed Alser, Juan G{\'o}mez-Luna, and
  Onur Mutlu.
\newblock {Algorithmic improvement and GPU acceleration of the GenASM
  algorithm}.
\newblock {\em arXiv}, 2022.

\bibitem{lindegger2022scrooge}
Jo{\"e}l Lindegger, Damla~Senol Cali, Mohammed Alser, Juan G{\'o}mez-Luna,
  Nika~Mansouri Ghiasi, and Onur Mutlu.
\newblock {Scrooge: A fast and memory-frugal genomic sequence aligner for CPUs,
  GPUs, and ASICs}.
\newblock {\em arXiv}, 2022.

\bibitem{fujiki2018genax}
Daichi Fujiki, Arun Subramaniyan, Tianjun Zhang, Yu~Zeng, Reetuparna Das, David
  Blaauw, and Satish Narayanasamy.
\newblock {GenAx: A genome sequencing accelerator}.
\newblock In {\em ISCA}, 2018.

\bibitem{madhavan2014race}
Advait Madhavan, Timothy Sherwood, and Dmitri Strukov.
\newblock {Race Logic: A hardware acceleration for dynamic programming
  algorithms}.
\newblock {\em CAN}, 2014.

\bibitem{cheng2018bitmapper2}
Haoyu Cheng, Yong Zhang, and Yun Xu.
\newblock {Bitmapper2: A GPU-accelerated all-mapper based on the sparse Q-gram
  index}.
\newblock {\em TCBB}, 2018.

\bibitem{houtgast2018hardware}
Ernst~Joachim Houtgast, Vlad-Mihai Sima, Koen Bertels, and Zaid Al-Ars.
\newblock {Hardware acceleration of BWA-MEM genomic short read mapping for
  longer read lengths}.
\newblock {\em Computational Biology and Chemistry}, 2018.

\bibitem{houtgast2017efficient}
Ernst~Joachim Houtgast, VladMihai Sima, Koen Bertels, and Zaid AlArs.
\newblock {An efficient GPU-accelerated implementation of genomic short read
  mapping with BWA-MEM}.
\newblock {\em CAN}, 2017.

\bibitem{zeni2020logan}
Alberto Zeni, Giulia Guidi, Marquita Ellis, Nan Ding, Marco~D Santambrogio,
  Steven Hofmeyr, Ayd{\i}n Bulu{\c{c}}, Leonid Oliker, and Katherine Yelick.
\newblock {Logan: High-performance GPU-based X-drop long-read alignment}.
\newblock In {\em IPDPS}, 2020.

\bibitem{ahmed2019gasal2}
Nauman Ahmed, Jonathan L{\'e}vy, Shanshan Ren, Hamid Mushtaq, Koen Bertels, and
  Zaid Al-Ars.
\newblock {GASAL2: A GPU accelerated sequence alignment library for
  high-throughput NGS data}.
\newblock {\em BMC Bioinformatics}, 2019.

\bibitem{nishimura2017accelerating}
Takahiro Nishimura, Jacir~L Bordim, Yasuaki Ito, and Koji Nakano.
\newblock {Accelerating the Smith-waterman algorithm using bitwise parallel
  bulk computation technique on GPU}.
\newblock In {\em IPDPSW}, 2017.

\bibitem{de2016cudalign}
Edans~Flavius de~Oliveira~Sandes, Guillermo Miranda, Xavier Martorell, Eduard
  Ayguade, George Teodoro, and Alba Cristina~Magalhaes Melo.
\newblock {CUDAlign 4.0: Incremental speculative traceback for exact
  chromosome-wide alignment in GPU clusters}.
\newblock {\em TPDS}, 2016.

\bibitem{liu2015gswabe}
Yongchao Liu and Bertil Schmidt.
\newblock {GSWABE: Faster GPU-accelerated sequence alignment with optimal
  alignment retrieval for short DNA sequences}.
\newblock {\em Concurrency and Computation: Practice and Experience}, 2015.

\bibitem{liu2013cudasw++}
Yongchao Liu, Adrianto Wirawan, and Bertil Schmidt.
\newblock {CUDASW++ 3.0: Accelerating Smith-Waterman protein database search by
  coupling CPU and GPU SIMD instructions}.
\newblock {\em BMC Bioinformatics}, 2013.

\bibitem{liu2009cudasw++}
Yongchao Liu, Douglas~L Maskell, and Bertil Schmidt.
\newblock {CUDASW++: Optimizing Smith-Waterman sequence database searches for
  CUDA-enabled graphics processing units}.
\newblock {\em BMC Research Notes}, 2009.

\bibitem{liu2010cudasw++}
Yongchao Liu, Bertil Schmidt, and Douglas~L Maskell.
\newblock {CUDASW++ 2.0: Enhanced Smith-Waterman protein database search on
  CUDA-enabled GPUs based on SIMT and virtualized SIMD abstractions}.
\newblock {\em BMC Research Notes}, 2010.

\bibitem{wilton2015arioc}
Richard Wilton, Tamas Budavari, Ben Langmead, Sarah~J Wheelan, Steven~L
  Salzberg, and Alexander~S Szalay.
\newblock {Arioc: High-throughput read alignment with GPU-accelerated
  exploration of the seed-and-extend search space}.
\newblock {\em PeerJ}, 2015.

\bibitem{goyal2017ultra}
Amit Goyal, Hyuk~Jung Kwon, Kichan Lee, Reena Garg, Seon~Young Yun, Yoon~Hee
  Kim, Sunghoon Lee, and Min~Seob Lee.
\newblock {Ultra-fast next generation human genome sequencing data processing
  using $DRAGEN^{TM}$ Bio-IT processor for precision medicine}.
\newblock {\em OJGen}, 2017.

\bibitem{chen2016spark}
Yu-Ting Chen, Jason Cong, Zhenman Fang, Jie Lei, and Peng Wei.
\newblock {When Spark Meets FPGAs: A case study for next-generation DNA
  sequencing acceleration}.
\newblock In {\em HotCloud}, 2016.

\bibitem{chen2014accelerating}
Peng Chen, Chao Wang, Xi~Li, and Xuehai Zhou.
\newblock {Accelerating the next generation long read mapping with the
  FPGA-based system}.
\newblock {\em TCBB}, 2014.

\bibitem{chen2021high}
Yen-Lung Chen, Bo-Yi Chang, Chia-Hsiang Yang, and Tzi-Dar Chiueh.
\newblock {A high-throughput FPGA accelerator for short-read mapping of the
  whole human genome}.
\newblock {\em TPDS}, 2021.

\bibitem{fujiki2020seedex}
Daichi Fujiki, Shunhao Wu, Nathan Ozog, Kush Goliya, David Blaauw, Satish
  Narayanasamy, and Reetuparna Das.
\newblock {SeedEx: A genome sequencing accelerator for optimal alignments in
  subminimal space}.
\newblock In {\em MICRO}, 2020.

\bibitem{banerjee2018asap}
Subho~Sankar Banerjee, Mohamed El-Hadedy, Jong~Bin Lim, Zbigniew~T Kalbarczyk,
  Deming Chen, Steven~S Lumetta, and Ravishankar~K Iyer.
\newblock {ASAP: Accelerated short-read alignment on programmable hardware}.
\newblock {\em TC}, 2019.

\bibitem{fei2018fpgasw}
Xia Fei, Zou Dan, Lu~Lina, Man Xin, and Zhang Chunlei.
\newblock {FPGASW: Accelerating large-scale Smith--Waterman sequence alignment
  application with backtracking on FPGA linear systolic array}.
\newblock {\em Interdisciplinary Sciences: Computational Life Sciences}, 2018.

\bibitem{waidyasooriya2015hardware}
Hasitha~Muthumala Waidyasooriya and Masanori Hariyama.
\newblock {Hardware-acceleration of short-read alignment based on the
  Burrows-wheeler transform}.
\newblock {\em TPDS}, 2015.

\bibitem{chen2015novel}
Yu-Ting Chen, Jason Cong, Jie Lei, and Peng Wei.
\newblock {A novel high-throughput acceleration engine for read alignment}.
\newblock In {\em FCCM}, 2015.

\bibitem{rucci2018swifold}
Enzo Rucci, Carlos Garcia, Guillermo Botella, Armando De~Giusti, Marcelo
  Naiouf, and Manuel Prieto-Matias.
\newblock {SWIFOLD: Smith-Waterman implementation on FPGA with OpenCL for long
  DNA sequences}.
\newblock {\em BMC Systems Biology}, 2018.

\bibitem{li2021pipebsw}
Luyi Li, Jun Lin, and Zhongfeng Wang.
\newblock {PipeBSW: A two-stage pipeline structure for banded Smith-Waterman
  algorithm on FPGA}.
\newblock In {\em ISVLSI}, 2021.

\bibitem{wu2019fpga}
Lisa Wu, David Bruns-Smith, Frank~A Nothaft, Qijing Huang, Sagar Karandikar,
  Johnny Le, Andrew Lin, Howard Mao, Brendan Sweeney, Krste Asanovi{\'c},
  et~al.
\newblock {FPGA accelerated indel realignment in the cloud}.
\newblock In {\em HPCA}, 2019.

\bibitem{yan_accel-align_2021}
Yiqing Yan, Nimisha Chaturvedi, and Raja Appuswamy.
\newblock {Accel-{Align}: a fast sequence mapper and aligner based on the
  seed–embed–extend method}.
\newblock {\em BMC Bioinformatics}, 2021.

\bibitem{vasimuddin_efficient_2019}
Md. Vasimuddin, Sanchit Misra, Heng Li, and Srinivas Aluru.
\newblock {Efficient {architecture}-{aware} {acceleration} of {BWA}-{MEM} for
  {multicore} {systems}}.
\newblock In {\em IPDPS}, 2019.

\bibitem{daily_parasail_2016}
Jeff Daily.
\newblock {Parasail: {SIMD} {C} library for global, semi-global, and local
  pairwise sequence alignments}.
\newblock {\em BMC Bioinformatics}, 2016.

\bibitem{kalikar_accelerating_2022}
Saurabh Kalikar, Chirag Jain, Md~Vasimuddin, and Sanchit Misra.
\newblock {Accelerating minimap2 for long-read sequencing applications on
  modern {CPUs}}.
\newblock {\em Nature Computational Science}, 2022.

\bibitem{marco-sola_fast_2021}
Santiago Marco-Sola, Juan~Carlos Moure, Miquel Moreto, and Antonio Espinosa.
\newblock {Fast gap-affine pairwise alignment using the wavefront algorithm}.
\newblock {\em Bioinformatics}, 2021.

\bibitem{eizenga_improving_2022}
Jordan~M. Eizenga and Benedict Paten.
\newblock {Improving the time and space complexity of the {WFA} algorithm and
  generalizing its scoring}.
\newblock {\em bioRxiv}, 2022.

\bibitem{marco-sola_optimal_2022}
Santiago Marco-Sola, Jordan~M. Eizenga, Andrea Guarracino, Benedict Paten, Erik
  Garrison, and Miquel Moreto.
\newblock {Optimal gap-affine alignment in {O}(s) space}.
\newblock {\em bioRxiv}, 2022.

\bibitem{kovaka2021targeted}
Sam Kovaka, Yunfan Fan, Bohan Ni, Winston Timp, and Michael~C Schatz.
\newblock {Targeted nanopore sequencing by real-time mapping of raw electrical
  signal with uncalled}.
\newblock {\em Nature Biotechnology}, 2021.

\bibitem{zhang2021real}
Haowen Zhang, Haoran Li, Chirag Jain, Haoyu Cheng, Kin~Fai Au, Heng Li, and
  Srinivas Aluru.
\newblock {Real-time mapping of nanopore raw signals}.
\newblock {\em Bioinformatics}, 2021.

\bibitem{loose_real-time_2016}
Matthew Loose, Sunir Malla, and Michael Stout.
\newblock {Real-time selective sequencing using nanopore technology}.
\newblock {\em Nature Methods}, 2016.

\bibitem{edwards_real-time_2019}
Harrison~S. Edwards, Raga Krishnakumar, Anupama Sinha, Sara~W. Bird, Kamlesh~D.
  Patel, and Michael~S. Bartsch.
\newblock Real-{Time} {selective} {sequencing} with {RUBRIC}: {Read} {Until}
  with {basecall} and {reference}-{informed} {criteria}.
\newblock {\em Scientific Reports}, 2019.

\bibitem{payne_readfish_2021}
Alexander Payne, Nadine Holmes, Thomas Clarke, Rory Munro, Bisrat~J. Debebe,
  and Matthew Loose.
\newblock {Readfish enables targeted nanopore sequencing of gigabase-sized
  genomes}.
\newblock {\em Nature Biotechnology}, 2021.

\bibitem{de_maio_boss-runs_2020}
Nicola De~Maio, Charlotte Manser, Rory Munro, Ewan Birney, Matthew Loose, and
  Nick Goldman.
\newblock {{BOSS}-{RUNS}: a flexible and practical dynamic read sampling
  framework for nanopore sequencing}.
\newblock {\em bioRxiv}, 2020.

\bibitem{danilevsky_adaptive_2022}
Artem Danilevsky, Avital~Luba Polsky, and Noam Shomron.
\newblock {Adaptive sequencing using nanopores and deep learning of
  mitochondrial {DNA}}.
\newblock {\em Briefings in Bioinformatics}, 2022.

\bibitem{joppich_sequ-into_2020}
Markus Joppich, Margaryta Olenchuk, Julia~M. Mayer, Quirin Emslander, Luisa~F.
  Jimenez-Soto, and Ralf Zimmer.
\newblock {{SEQU}-{INTO}: {Early} detection of impurities, contamination and
  off-targets ({ICOs}) in long read/{MinION} sequencing}.
\newblock {\em CSBJ}, 2020.

\end{thebibliography}

\end{document}